\documentclass[journal=jcisd8,manuscript=note]{achemso}
\usepackage[T1]{fontenc} 
\usepackage{url}
\usepackage{longtable}
\usepackage{graphicx} 
\usepackage{color} 
\usepackage{svg}

\newcommand{\etal}{\textit{et al}.}

\newcommand{\dbname}{B3LYP/6-31G*/\!/PM6}

\usepackage{adjustbox}
\usepackage{amsmath}
\usepackage{fancyvrb}
\newenvironment{myverb}{%
 \VerbatimEnvironment
 \begin{adjustbox}{max width=\linewidth}
 \begin{BVerbatim}
  }{
  \end{BVerbatim}
 \end{adjustbox}
}

\author{Maho Nakata}
\affiliation[RIKEN ACCC]{
RIKEN Cluster for Pioneering Research, 2-1 Hirosawa, Wako-City, Saitama 351-0198, JAPAN}
\email{maho@riken.jp}
\author{Toshiyuki Maeda}
\affiliation[STAIR]
{Software Technology and Artificial Intelligence Research Laboratory,\\Chiba Institute of Technology, 2-17-1 Tsudanuma, Narashino, Chiba 275-0016, JAPAN.}

\title{PubChemQC B3LYP/6-31G*/\!/PM6 dataset: the Electronic Structures of 86 Million Molecules using B3LYP/6-31G* calculations}

\SectionNumbersOn

\begin{document}

\begin{abstract}
This article introduces the extensive ``PubChemQC \dbname{}'' dataset, which contains electronic properties, including orbitals, orbital energies, total energies, dipole moments, and other relevant electronic properties for 85,938,443 molecules. The dataset covers a wide range of molecules, from small but essential compounds to large biomolecules weighing up to 1000 molecular weight, covering 94.0\% of the original PubChem Compound catalog as retrieved on August 29, 2016. The electronic properties were calculated using the B3LYP/6-31G* method, one of the first principle methods, and the PM6 method, an empirical method derived from molecular geometries. The dataset is available in three formats: (i) the comprehensive input/output of the GAMESS quantum chemistry program, (ii) selected data from output files in JSON format, and (iii) a PostgreSQL database, which allows researchers to query molecular properties. Additionally, five sub-datasets were created to provide more specific and targeted data. The first subset includes molecules containing C, H, O, and N, with a molecular weight of less than 300 and no salt, while the second subset encompasses molecules containing C, H, O, and N, with a molecular weight of less than 500 and no salt. The third subset includes molecules containing C, H, N, O, P, S, F, and Cl, with a molecular weight of less than 300 and no salt. In contrast, the fourth subset encompasses molecules containing C, H, N, O, P, S, F, and Cl, with a molecular weight of less than 500 and no salt. Finally, the fifth subset contains molecules containing C, H, N, O, P, S, F, Cl, Na, K, Mg, and Ca, with a molecular weight of less than 500. Notably, our analysis revealed coefficients of determination ranging from 0.892 (CHON500) to 0.803 (whole) for the HOMO-LUMO energy gap, among other findings. To the best of our knowledge, these comparisons represent the most extensive investigations of their kind. By developing machine learning models, researchers can use the comprehensive dataset for various applications, such as drug discovery and material science. The datasets are available under the Creative Commons Attribution 4.0 International license and can be found at \protect{\url{https://nakatamaho.riken.jp/pubchemqc.riken.jp/b3lyp_pm6_datasets.html}}.
\end{abstract}

\section{Introduction}
The availability of diverse molecular databases is critical in the field of chemistry. Numerous public databases, such as PubChem~\cite{10.1093/nar/gky1033,10.1093/nar/gkac956}, ZINC~\cite{zinc,doi:10.1021/acs.jcim.2c01253}, CAS~\cite{CAS}, ChEMBL~\cite{chembl,10.1093/nar/gky1075}, ChemSpider~\cite{chemspider}, DrugBank~\cite{10.1093/nar/gkx1037}, Open Materials Database~\cite{OpenMaterialsdatabase}, Crystallography Open Database~\cite{Grazulis2009,Grazulis2012}, and the Cambridge Structural Database~\cite{Bruno14:__cryst,CCD}, among others~\cite{listofchemicaldatabases}, serve as valuable sources of molecules.

These databases play a pivotal role in numerous domains, such as drug discovery~\cite{CHENG20141751,Miller02:_chemic,PMID:28602100}, where molecules are designed to target specific proteins, considering factors like conformers, interaction energies, toxicity, solubility, stability, regulatory approval, and patents~\cite{10.1093/nar/gky1033,10.1093/nar/gkac956}, and materials discovery and development~\cite{advs.201900808,TRIPATHI20201245,D0NA00388C}, which includes the search for novel organic compounds for applications in OLEDs~\cite{C9QM00716D,adfm.201808803} and high-energy-density materials for batteries~\cite{Zhao20:_design,Bruno14:__cryst}. In addition, these databases provide a wealth of information on the properties and characteristics of different molecules, including their electronic and physical properties, which can design and develop novel materials for various applications.

Including quantum chemical calculations~\cite{Helgaker2000,doi:10.1063/1.4869598,C7CP04913G,doi:10.1142/4910} can substantially increase the utility of diverse molecular databases. In the field of drug discovery, for instance, quantum chemical calculations can aid in studying molecular interactions~\cite{Bissantz10:_a_medi}, identifying new drug targets, designing novel drug molecules~\cite{qua.25678}, understanding reaction mechanisms~\cite{Fukui70:_formul,qua.24757,doi:10.1142/9789812839664_0016,doi:10.1063/1.1329672}, analyzing drug stability~\cite{Douroudgari21:__compu}, developing more reliable QSAR (Quantitative Structure-Activity Relationship) models~\cite{DEBENEDETTI20141921,Karelson96:_quantu,qsar.19870060406}, and drug metabolism~\cite{doi:10.1021/acs.chemrev.0c00901,Kirchmair15:_predic}, among other applications. In the context of organic light-emitting diode (OLED) materials, the Highest Occupied Molecular Orbital energy (HOMO) minus the Lowest Unoccupied Molecular Orbital energy (LUMO), or the HOMO-LUMO energy gap plays a crucial role in determining their optoelectronic properties. A narrower HOMO-LUMO energy gap promotes fluorescence emission, as excited molecules can readily undergo radiative decay. On the other hand, a wider HOMO-LUMO energy gap favors non-fluorescent mechanisms such as triplet-triplet upconversion. Therefore, it is essential to consider the HOMO-LUMO energy gap when designing and refining electroluminescent materials~\cite{OLED,OTFT,C9QM00716D,Yadav20:_role_o,Kaji2015,B512646K}. However, estimating the gap through experiments can be expensive. Incorporating quantum chemical data into these databases can facilitate the prediction of the HOMO-LUMO energy gap and other electronic properties, making it easier to identify and design OLED materials with optimal properties.

Although quantum chemical calculations are less expensive and more applicable for evaluating individual molecules compared to experiments~\cite{Helgaker2000,doi:10.1142/4910,Nist2005,C7CP04913G}, they are impractical for virtual high-throughput screening due to the time-consuming optimization of molecular geometry and finding the lowest energy conformer of the molecule, which can take several minutes to days, depending on the size of the molecule and levels of theory. As a result, the development of machine learning models for predicting molecular properties has become increasingly prevalent in recent years, as they can accelerate calculations and broaden their applications.

The prodigious number of potential drug candidate molecules, estimated to be approximately $10^{60}$\cite{LIPINSKI19973,chemicalspace}, limits the diversity of molecules in eminent quantum chemistry datasets, such as QM7, QM9\cite{QC134K,1703.00564}, ANI-1~\cite{Smith17:_-1_da}, ANI-1x and ANI-1ccx~\cite{Smith20:_the-1c}, and Qmugs~\cite{Isert22:_qmugs}.
The QM9, ANI-1, and Qmugs datasets provide valuable molecular insights; however, their restricted size and scope curtail the assortment of molecular and electronic configurations available for analysis. While QM9 comprises 134,000 molecules calculated using DFT (B3LYP/6-31G(2df,p)) optimized molecular geometries, ANI-1 consists of 57,462 small organic molecules with 17.2 million conformers determined by DFT ($\omega$B97-X/6-31G*), ANI-1x~\cite{doi:10.1021/jacs.2c13467} and contains ANI-1ccx~\cite{Smith19:_approa} subset of ANI-1 with better quantum chemical calculations. Qmugs encompasses 665,000 molecules with 200 million conformers ascertained utilizing the GFN2-xTB~\cite{GFN2-xtb} method and DFT ($\omega$B97X-D/def2-SVP) single-point electronic structure computations. Nevertheless, these numbers are trivial compared to the projected $10^{60}$ potential drug candidate molecules and even the PubChem database, which contains over $10^8$ unique molecules. Consequently, a dataset with a broader and more diverse collection of molecular structures is essential to overcome this challenge. 

In 2019, Glavatskikh \etal{} found that QM9 is limited in functional group diversity and suggested the need for complementary datasets~\cite{Glavatskikh19:_datase}. The study highlighted that combining QM9 with PubChemQC can improve the generalizability of the models, as PubChemQC has a significantly more diverse set of molecules. Additionally, the QM9 and PubChemQC datasets contain different chemical structures and functionalities, with PubChemQC having functional groups not present in QM9, such as radicals and triplets. Consequently, neural networks trained on PubChemQC may have higher generalization ability than those trained on QM9.

In 2021, Hu~\etal{} organized a competition to estimate HOMO-LUMO energy gaps~\cite{2103.09430} using previously published datasets~\cite{PubChemQC2017}. Relying exclusively on SMILES as molecular information and not utilizing molecular geometries, they estimated the HOMO-LUMO energy gap for a dataset comprising 3.8 million molecules. The reference data was provided, and the most accurate model employed GIN-virtual~\cite{xu2018how}, which demonstrated an MAE of 0.11 eV. Remarkably, even when employing a mere 10\% of the training data, they achieved an MAE of 0.15 eV with GIN-virtual. The esteemed research group led by Masters \etal{} garnered significant recognition with their GPS++ model~\cite{2212.02229}, which, through the innovative amalgamation of the Message Passing Neural Network (MPNN) and the Transformer model, managed to obtain an MAE of a mere 0.072 eV. It is crucial to note that, throughout these competitive endeavors, three-dimensional (3D) atomic coordinates served as the fundamental training data. Intriguingly, however, these coordinates were not employed during the prediction phase.

Therefore, the diversity of molecules appears to be significant. While it is not definitive, if other methods utilize QM9 as training molecules, there might be overfitting, and predictability and transferability could be limited outside of the set.

The PubChemQC project~\cite{PubChemQCsite,PubChemQC2015} aims to provide a dataset with a more extensive and diverse array to facilitate the inverse design of chemistry~\cite{doi:10.1021/jacs.2c13467,C9ME00039A} by offering a dataset of small molecules with quantum chemically calculated molecular properties, which may be used to generate superior machine learning models. It is widely recognized that the dataset size is crucial for the performance of machine learning models~\cite{4804817,8237359,7968387,1712.00409}. Thus, the extensive nature of the PubChemQC project's dataset holds significant value. Therefore, first, what small molecules are commonly known and what properties they have are provided by quantum chemical calculations. Since quantum chemical calculations are inexpensive and do not require experiments, we can provide accurate data without performing expensive, dangerous, and time-consuming experiments. By providing a standardized, accurate, and freely available molecular data set, we hope PubChemQC facilitates collaboration and accelerates progress toward critical scientific goals, including the inverse design of molecules. We selected PubChem~\cite{pubchemsite} as the molecular source because the PubChemQC project includes only molecules known to humans, ensuring molecular diversity, a wide range of atoms, and addressing the challenge of selecting molecules. Our previous research presented the PubChemQC dataset~\cite{PubChemQC2017}, which includes geometry optimization for approximately 3.8 million molecules at the B3LYP/6-31G* level and employs time-dependent density functional theory (TD-DFT) with the B3LYP functional and the 6-31+G* basis set to calculate ten low-lying excited states for more than two million molecules. Subsequently, we introduced the PubChemQC PM6 dataset~\cite{PubChemQC2020}, which encompasses electronic structures, optimized geometries, and molecular vibration frequencies for 86,213,135 neutral molecules, 51,555,911 cations, 45,581,750 anions, and 37,839,619 spin-flipped states. The PM6 method~\cite{Stewart2007} optimized the molecular geometry, resulting in a close approximation to B3LYP/6-31G*-optimized molecules, with a median difference in bond length of 0.016 \AA{} and a median bond angle difference of 1.7 degrees. This suggests that PM6-optimized geometries for molecules are reasonably accurate compared to B3LYP/6-31G* optimized geometries, except for dihedral angles, which were not investigated in that paper.

This paper presents the PubChemQC \dbname{} dataset, comprising 85,938,443 molecules optimized at the PM6 level (semi-empirical), employing the B3LYP/6-31G* level ({\it ab initio} DFT) of theory. The dataset is available as input and output files for the GAMESS quantum chemistry program~\cite{GAMESS} and is over 50 TB in size. Through the effort, we could compare HOMO, LUMO, HOMO-LUMO energy gaps, and dipole moments between PM6/\!/PM6 and B3LYP/6-31G*/\!/PM6 calculations. Although discrepancies exist between the values, the coefficient of determinations exhibited for $0.877$ to $0.907$ for the HOMO,
and $0.821$ to $0.911$ for the LUMO, $0.803$ to $0.892$ for the HOMO-LUMO energy gap, and $0.929$ to $0.945$ for the dipole moment.

The discovery implicates that, given identical molecular geometry, there is a high probability of achieving precise estimations for parameters such as HOMO, LUMO, HOMO-LUMO energy gaps, and dipole moments, employing machine learning methodologies. Moreover, to the best of our knowledge, no previous study has extensively compared between the PM6/\!/PM6 and the B3LYP/6-31G*/\!/PM6 method.

To the best of our knowledge, PubChemQC \dbname{} currently constitutes the largest and most diverse quantum chemical dataset available. The database serves as a resource for the development of enhanced machine-learning models. 

The remaining sections of the paper are structured as follows. Section~\ref{sec:pubchem} describes the relation between PubChem and PubChemQC project. In Section~\ref{sec:common_abbrev}, we explain several mnemonic representations employed in quantum chemical calculations. In Section~\ref{sec:relatedwork}, we discuss related works. In Section~\ref{sec:workflow}, we elaborate on the workflow we followed to develop our database. In Section~\ref{sec:statistics}, we present statistics related to the PubChemQC \dbname{} dataset, including the number of molecules and molecular weight distribution. Furthermore, in Section~\ref{sec:comparison}, we comapre the data obtained using the \dbname{} and PM6/\!/PM6 methods. In Section~\ref{sec:howtouse}, we describe how to use the PubChemQC \dbname{} dataset. Finally, in Section~\ref{sec:futurework}, we discuss the future directions of our research.

The datasets are available under the Creative Commons Attribution 4.0 International license and can be found at \protect{\url{https://nakatamaho.riken.jp/pubchemqc.riken.jp/b3lyp_pm6_datasets.html}}.

\section{PubChem and the goal of the PubChemQC Project}
\label{sec:pubchem}
PubChem~\cite{pubchemsite,10.1093/nar/gky1033,10.1093/nar/gkac956} is a freely accessible online database overseen by the National Institutes of Health (NIH) that contains molecular information. Since 2004, it has been regularly updated with records from diverse sources, including universities, government agencies, scientific papers, and other data collection initiatives. The database is composed of three sub-databases: Substances, Compounds, and BioAssays. Compounds contain standardized compound data extracted from Substances. As of August 16, 2016, PubChem Compounds contained 91,679,247 compounds, each with unique PubChem CID, InChI, and SMILES encodings, molecular formula, and other information.

Despite PubChem's comprehensive coverage of chemical structures, identifiers, chemical and physical properties, biological activities, patents, health, safety, toxicity data, and more, it, unfortunately, lacks comprehensive quantum chemical data, which could be beneficial in advancing the inverse design of molecules. To address the limitation, the PubChemQC project~\cite{PubChemQCsite,PubChemQC2015,PubChemQC2017,PubChemQC2020} generated a dataset of optimized molecular structures with electronic properties calculated at the B3LYP/6-31G* level of theory. Unlike molecular force field-optimized PubChem3D, our data contains molecular orbitals and orbital energies by DFT calculation and semi-empirical PM6-optimized geometry. The HOMO, LUMO, and HOMO-LUMO energies, in particular, are crucial for making organic light-emitting diodes (OLEDs), which are not provided in the original PubChem.


The PubChemQC project aims to accomplish several objectives and surmount various challenges, including:

\begin{itemize}
\item Providing a diverse and comprehensive dataset of molecular information by capitalizing on PubChem as the molecular source.
\item Conducting conformer searches of molecules using quantum chemical calculations: investigating the unique arrangements of atoms that can exist for a given molecular formula, particularly those with the lowest energy.
\item The position of hydrogens in a molecule was determined, and tautomerism was considered to enhance the accuracy of the calculations.
\item Determining molecular vibrations in quantum chemistry, which involves identifying the frequencies and modes of vibrations for a specific molecule.
\item Calculating excited state energies of molecules, which are essential for understanding optical spectroscopy and photochemistry.
\item Additionally, conduct the same calculations in solvents such as water or ethanol to further investigate the system's behavior.
\item Assessing NMR shifts through quantum chemical calculations to determine the molecular structure and dynamics of a compound.
\item Predicting the aforementioned molecular properties using machine learning techniques.
\item Our approach aims to facilitate the inverse design of molecules, which involves the strategic design of a molecule with specific properties or functions, as opposed to the traditional method of synthesizing a molecule and then evaluating its properties. This is achieved through the implementation of machine learning methodologies.
\item We intend to contribute to the original PubChem project by uploading the data generated through our approach.
\end{itemize}

We hope the PubChemQC \dbname{} dataset is a valuable for accurately predicting molecular properties and developing machine learning models for drug discovery and materials science. It is available under a Creative Commons Attribution 4.0 International License and can be downloaded from \protect{\url{https://nakatamaho.riken.jp/pubchemqc.riken.jp/b3lyp_pm6_datasets.html}}. In the future, we plan to submit our data to the original PubChem project to recognize the significant effort made to make the data publicly available.

\section{Common Abbreviations in Quantum Chemical Calculations} \label{sec:common_abbrev}
In this section explains several conventional abbreviations employed in quantum chemical calculations. The notation ``A/B'' denotes a calculation of a molecule using method A and basis set B. For instance, B3LYP/6-31G* implies that the calculation has been executed using the B3LYP method with the 6-31G* basis set. Similarly, 
$\omega$B97M-V/cc-pVTZ indicates a calculation performed using the 
$\omega$B97M-V method with the cc-pVTZ basis set. Contrarily, for PM6 (or PM7) calculations, we merely denote it as PM6 (or PM7), as the method intrinsically includes the basis set.

The notation ``A/B/\!/C/D'' signifies a molecular electronic structure calculation conducted with method A and basis set B, premised on a moThis geometry that has been optimized via method C and basis set D. For example, the notation ``$\omega$B97M-V/cc-pVTZ/\!/B3LYP/6-31G*'' specifies that the optimization of molecular geometry was executed using the B3LYP/6-31G* method. Subsequently, the electronic structure calculation was performed using the $\omega$B97M-V/cc-pVTZ method, using the molecular geometry derived from the aforementioned B3LYP/6-31G* optimization.
Calculations following the protocol are frequently undertaken owing to the substantial computational demands associated with molecular geometry optimization, particularly when higher-quality basis sets such as cc-pV5Z are employed. Intriguingly, despite omitting these auxiliary functions, the optimized molecular geometry largely retains its characteristics, thereby facilitating a reduction in computational overhead (see, for example, Table 8.5 and 8.16 of Helgaker~\etal{}'s text book.~\cite{Helgaker2000}).

In the present investigation, we utilize the \dbname{} approach, which signifies the use of the PM6 method for molecular geometry optimization, followed by a single point calculation via the B3LYP/6-31G* method on the molecular geometry previously optimized by the PM6 method.

Finally, the basis sets 6-31G* and 6-31G(d) are the same.

\section{Related work}
\label{sec:relatedwork}
The NIST Chemistry WebBook contains the IR spectra of over 16,000 molecules and the UV spectra and quantum chemical calculations of 1,600 molecules as well as a lot of experimental data~\cite{Nist2005}.

The Harvard Clean Energy Project, reported in the literature~\cite{Hachmann11,C3EE42756K}, is considered one of the first and most significant applications of big quantum chemical data in chemistry. In 2011, the project aimed to study 2.3 million molecular motifs using 150 million density functional theory calculations, leading to over 400 TB of data. The primary focus of the project was to identify promising compounds suitable for use in organic photovoltaics rather than for general-purpose applications.

In 2014, the QM9 dataset~\cite{QC134K} emerged as a widely-used quantum chemical dataset for developing machine learning algorithms. The dataset includes 133,885 organic molecules containing up to nine heavy atoms, including C, H, O, N, and F. The QM9 dataset is derived from the GDB-17~\cite{GDB-17} database, catalogs organic molecules containing up to 17 atoms of C, N, O, S, and halogens, resulting in a database of over 166 billion organic molecules. The selection of the molecules in the GDB-17 database was based on their chemical relevance, ensuring that the enumerated molecules have chemical significance. However, it is noteworthy that the number and size of the molecules in QM9 are more minor than even PubChemQC~\cite{PubChemQC2017}, which contains 3.8 million molecules. Additionally, the molecular weights of the molecules in QM9 are less than 200.

The QM8 dataset, introduced in 2015~\cite{10.1063/1.4928757}, computes excited states (S0, S1, S2) for 20 kilo small organic molecules containing eight CONF atoms using RI-CC2 level with def2TZVP and LR-TDDFT level, employing the hybrid XC functional PBE0 with def2SVP basis set. In contrast, in 2017, PubChemQC~\cite{PubChemQC2017} calculated 2 million excited states of ten low-lying molecules using TDDFT/6-31+G*. While QM8 prioritized the quality of its calculations, PubChemQC focused on generating a more significant number of molecules.

The ANI-1 dataset~\cite{Smith17:_-1_da}, introduced in 2017, along with its extensions ANI-1x and ANI-1ccx~\cite{Smith20:_the-1c}, is a significant contribution to machine learning in quantum chemistry. These datasets comprise 17.2 million off-equilibrium conformations for 200k organic molecules, and their primary purpose is to facilitate the development of machine-learning models for predicting molecular energies and properties. However, it is crucial to note that the molecular diversity of these datasets is much lower than that of PubChem, with only 400 times fewer molecules.

In 2017, Aires-de-Sousa \etal{} reported the generation of over 111,000 molecules using the B3LYP/6-31G*/\!/PM7 or B3LYP/6-31G*/\!/PM6 methods, as well as the application of a machine learning approach for the prediction of HOMO-LUMO energy gaps~\cite{doi:10.1021/acs.jcim.6b00340}.

In 2020, Liang \etal{}~\cite{Liang20:symex} introduced the QM-symex database, which encompasses excited state information for 173,000 molecules composed of H, B, C, N, O, F, Cl, and Br atoms. The PubChemQC database~\cite{PubChemQC2017} contains the first ten singlet excited state energies for 3.8 million molecules, calculated using B3LYP/6-31G(d) for ground state optimization and 2.0 million molecules ut ilizing B3LYP/6-31+G(d) for excitation. In contrast, QM-symex provides data for both the first ten singlet and triplet excited state energies for 173,000 molecules, employing B3LYP/6-31G(2df,p) for ground state optimization and B3LYP/6-31G for excitation.

In 2022, a new quantum chemical dataset called QMugs~\cite{Isert22:_qmugs} emerged, which includes theory-based molecular properties at the GFN2-xTB and DFT ($\omega$B97X-D/def2-SVP) levels. The QMugs dataset contains quantum mechanical properties for approximately 665,000 different bioactive molecules. The molecular weight is up to approximately 1000. The molecular source for QMugs is the CheEMBL dataset, and conformer searching was performed using a combination of RDKit and GFN2-xTB-based conformational search with CREST~\cite{doi:10.1021/acs.jctc.9b00143}. No further geometry optimization was conducted.

When compared to PubChemQC \dbname{}, QMugs displays distinct molecular origins as it is sourced from the CheMBEL dataset, and its molecular structures surpass PubChemQC PM6, which only carried out geometry optimization with Open Babel-generated initial assumptions for molecules. Moreover, QMugs may exhibit slightly better final electronic structures than PubChemQC \dbname{}, given that it employs advanced density functionals but similar basis sets (B3LYP/6-31G* vs. $\omega$B97X-D/def2-SVP). Nonetheless, it is important to note that PubChemQC features a significantly larger repository of molecules than QMugs, with 130 times more molecules in total. Additionally, PubChemQC \dbname{} dataset contains a much more diverse set of molecules than QMugs, with both datasets containing molecules with a molecular weight of up to 1000 Da.
\section{Workflow}
\label{sec:workflow}

The PubChemQC \dbname{} dataset was generated by extracting “.xyz” files from the 20TB compressed archives of the PubChemQC PM6 database~\cite{PubChemQC2020}, which were processed using raw Gaussian~\cite{gaussian09} output files. The generation of GAMESS~\cite{GAMESS} input files for the 86,213,135 neutral molecules included in the dataset was facilitated by Open Babel~\cite{OpenBabel}.

To determine the electronic properties of the molecules, we employed the B3LYP density functional with the VWN1RPA exchange-correlation functional, which aligns with the selected electron gas formula~\cite{doi:10.1063/1.464913,PhysRevB.37.785,B3LYP1,doi:10.1021/j100096a001} and used the 6-31G* basis set for H-Kr~\cite{Wiberg1986AbIM,doi:10.1063/1.476673}. For Rb-Rn, the Def2-SV(P) basis set was utilized~\cite{B508541A}, while the Stuttgart RSC 1997 effective core potentials (ECPs)~\cite{Andrae90:_energy} were employed for Ce-Yb and Th-Lr.

The molecular calculations were conducted using the GAMESS quantum chemistry software, with the settings and parameters available for reference in the GitHub repository~\cite{b3lyp_pm6_scripts}. The tool GNU parallel~\cite{tange_ole_2018_1146014} was employed as applicable in the research. As shown in Table~\ref{tab:statistics}, out of the 86,213,135 calculations performed, 274,692 failed and were subsequently removed from the analysis.

All of the calculations were performed on the RIKEN HOKUSAI BigWave supercomputer (Intel Xeon Gold 6148 2.4GHz, 1,680 CPUs, 33,600 cores) and the QUEST cluster (Intel Core2 L7400 1.50 GHz, 700 nodes, 1,400 cores). The calculation time was 478 days, totaling 31.1 million core hours on HOKUSAI BigWave, and 274 days with 7.7 million on QUEST. The calculation commenced on May 11, 2018, and concluded on July 19, 2020. The summarization process was completed on July 14, 2022.

\section{Statistics}
\label{sec:statistics}
This section presents the statistical analysis of our calculations. Table~\ref{tab:statistics} provides an overview of the datasets used in the study. We retrieved the PubChem Compound database on August 29, 2016, which contained 91,679,247 molecules. Our previous dataset, PubChem PM6~\cite{PubChemQC2020}, consisted of 86,213,135 neutral molecules. From the dataset, we derived the PubChemQC \dbname{} set, which contains 85,938,443 molecules. We removed 274,692 molecules from the set due to failed calculations.

The CHNOPSFClNaKMgCa500 subset includes 69,280,174 molecules and contains the essential elements in the human body, except for fluorine, with a molecular weight of less than 500~\cite{ZORODDU2019120}. The CHNOPSFCl500 and CHNOPSFCl300 subsets consist of 69,280,174 and 29,774,707 molecules, respectively, and have a molecular weight of less than 500 or 300, respectively. The CHNOPS elements, including carbon, hydrogen, nitrogen, oxygen, phosphorus, and sulfur, are vital for biological molecules~\cite{10.5555/1198994}. Additionally, fluorine (F) and chlorine (Cl) have been incorporated.

The CHON300 and CHON500 subsets comprise 46,220,368 and 17,308,321 molecules, respectively, with a molecular weight of less than 500 or 300, respectively. These subsets include the most common elements present in living organisms~\cite{CHON}. Additionally, to ensure the exclusion of salts, we removed salts using the {\tt --r} option of Open Babel~\cite{OpenBabel}.

\begin{figure}
\caption{Histograms displaying the molecular weight distribution with a bin size 10 reveal distinct peaks around molecular weight 300. While it is expected to increase exponentially with increasing molecular weight, there is a decrease in the trend.} \label{mw_b3lyp_pm6}
\begin{center}
\includegraphics[width=14cm]{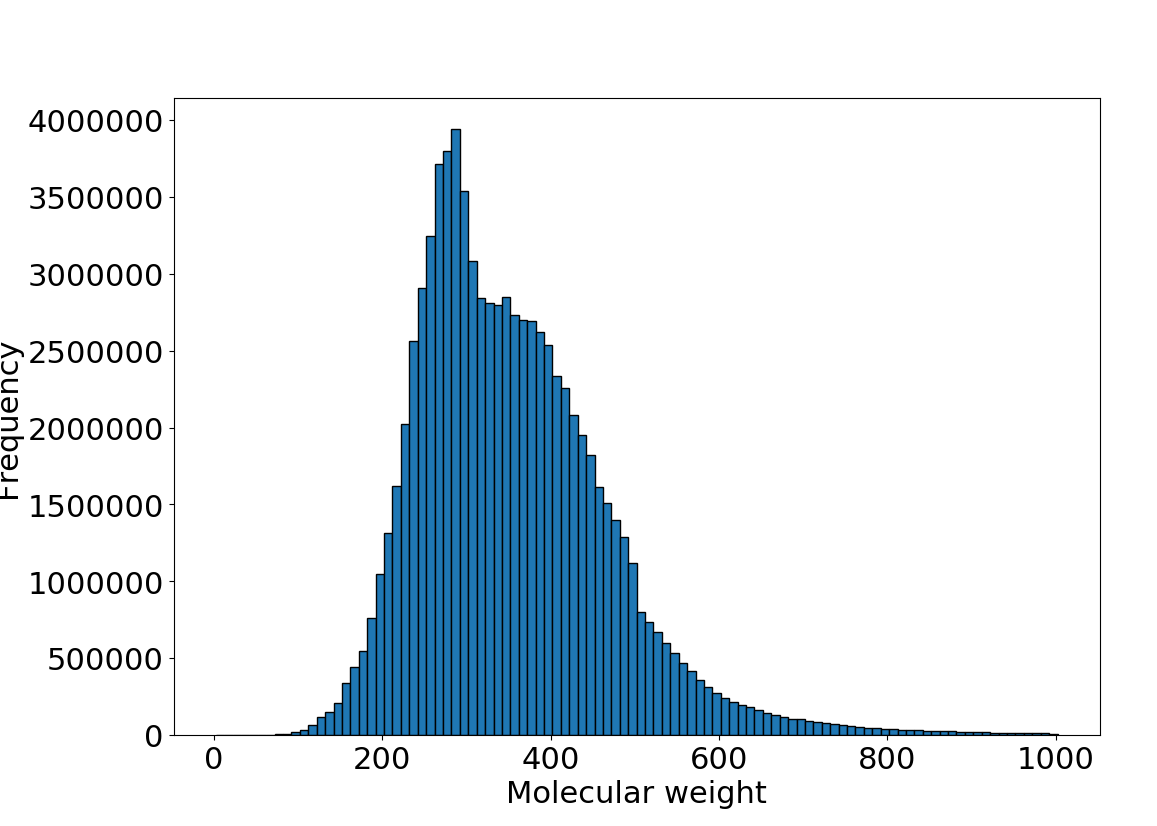}
\end{center}
\end{figure}

\begin{table}
\centering
\caption{The statistics of the PubChemQC \dbname{} dataset reveal that a total of 85,938,443 molecules were calculated, with 274,692 failures. All dataset data can be accessed at \protect\url{https://nakatamaho.riken.jp/pubchemqc.riken.jp/b3lyp_pm6_datasets.html}.}
\label{tab:statistics}
 \begin{tabular}{lrr}
 \hline \hline
 Results & Molecule count & percentage \\ \hline
 PubChem Compound (2016 Aug. 29th) & 91,679,247 & 100\% \\
 PubChem PM6 (neutral molecules only) & 86,213,135 &  94.0\% \\
 PubChemQC \dbname{} (whole set) & 85,938,443 & 93.7\%  \\
 CHNOPSFClNaKMgCa500 subset & 69,280,174 & 75.6\% \\
 CHNOPSFCl500 subset (no salt) & 67,720,082 & 73.9\% \\
 CHON500  subset (no salt) & 46,220,368 & 50.4\% \\
 CHNOPSFCl300 subset (no salt) & 29,774,707 & 32.5\% \\
 CHON300 subset (no salt) & 17,308,321 & 18.9\% \\ \hline
 Removed failed calculations & 274,692 & 0.32\% \\  \hline\hline
 \end{tabular}
 \label{statistics}
\end{table}
In Figure~\ref{mw_b3lyp_pm6}, the histogram delineates the molecular weight distribution with a bin size 10. Notably, a conspicuous peak is evident around molecular weights near 300, with a secondary shoulder near 400. Generally, the diversity of molecules increases with increasing molecular weight. However, for molecules cataloged in PubChem, a peak around molecular weight 300 suggests limited study and production of molecules beyond this threshold. Considering the diverse sources from which PubChem gathers data, such as government agencies, pharmaceutical companies, journal publishers, and more, the scarcity of molecules above a molecular weight of 300 is a fascinating observation. Although the number of molecules that can exist increases exponentially with increasing molecular weight, the number of molecules listed in PubChem decreases for molecular weights above 300, possibly indicating that research has not kept pace.

\section{Comparison to PM6 datasets}
\label{sec:comparison}
This section compares the HOMO energies, LUMO energies, HOMO-LUMO energy gaps, and dipole moments obtained from \dbname{} calculations with those from PM6/\!/PM6 calculations. 

\subsection{Comparison of HOMO Energies}

\begin{figure}[h!]
\caption{The figure presents a 2D histogram and heatmap of correlated HOMO energies calculated using PM6/\!/PM6 and \dbname{} methods. The horizontal axis signifies PM6/\!/PM6 HOMO energies, while the vertical axis exhibits the frequency distribution of \dbname{} HOMO values. It is noteworthy that the data point dispersion escalates with the molecule's atom count. The coefficients of determination are 0.877 to 0.907.}
    \centering
    \begin{tabular}{c c}
      \includegraphics[width=0.45\textwidth]{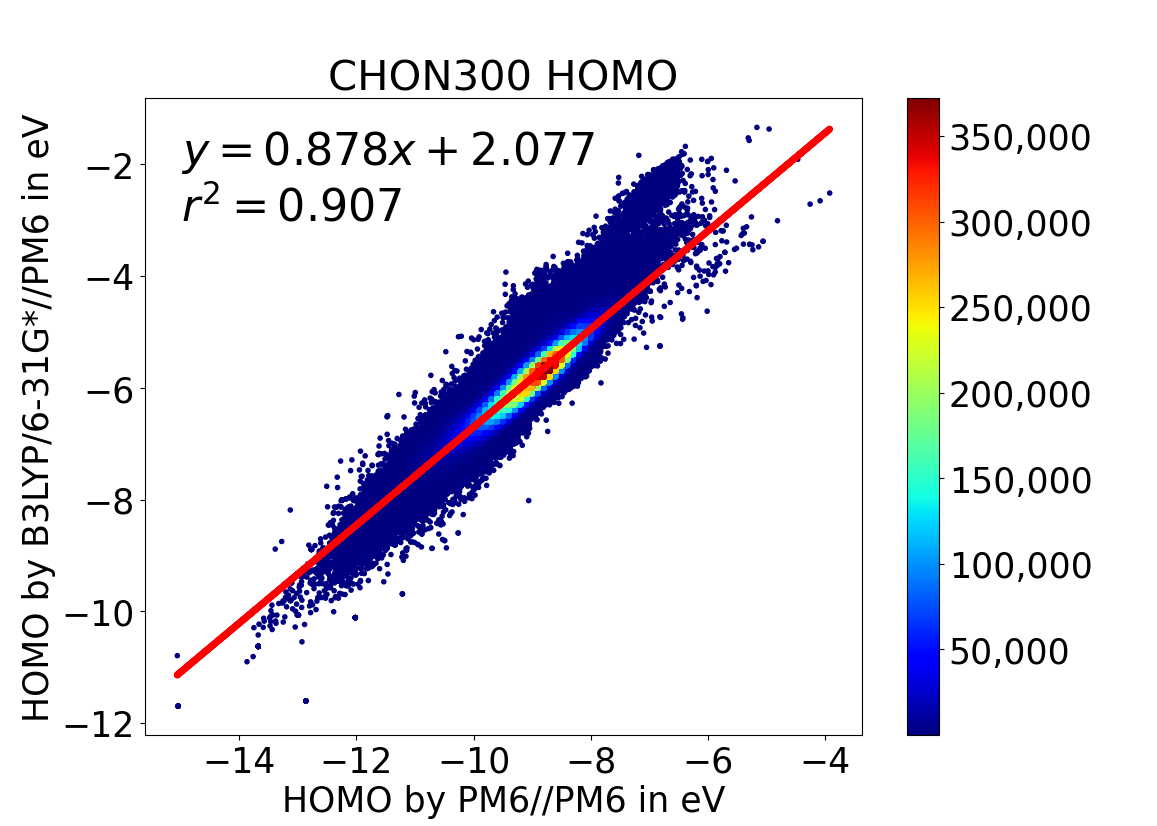} & \includegraphics[width=0.45\textwidth]{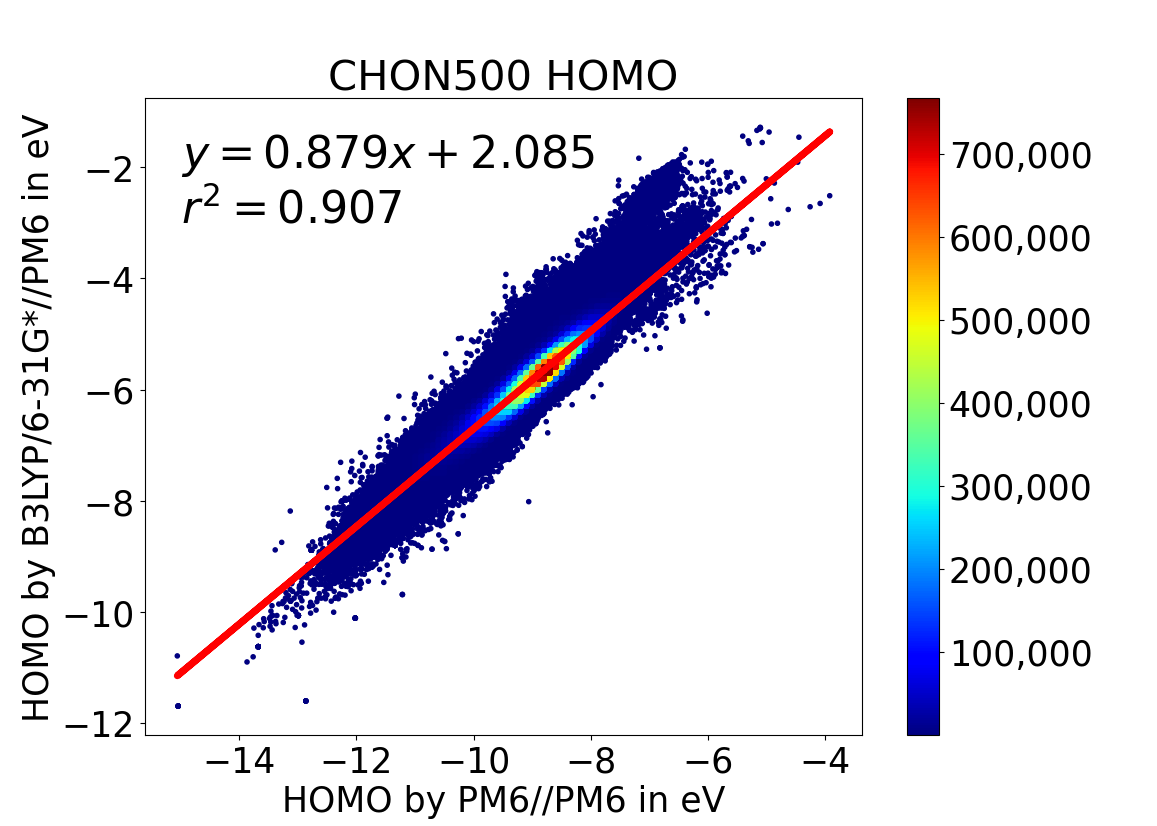} \\ 
      \includegraphics[width=0.45\textwidth]{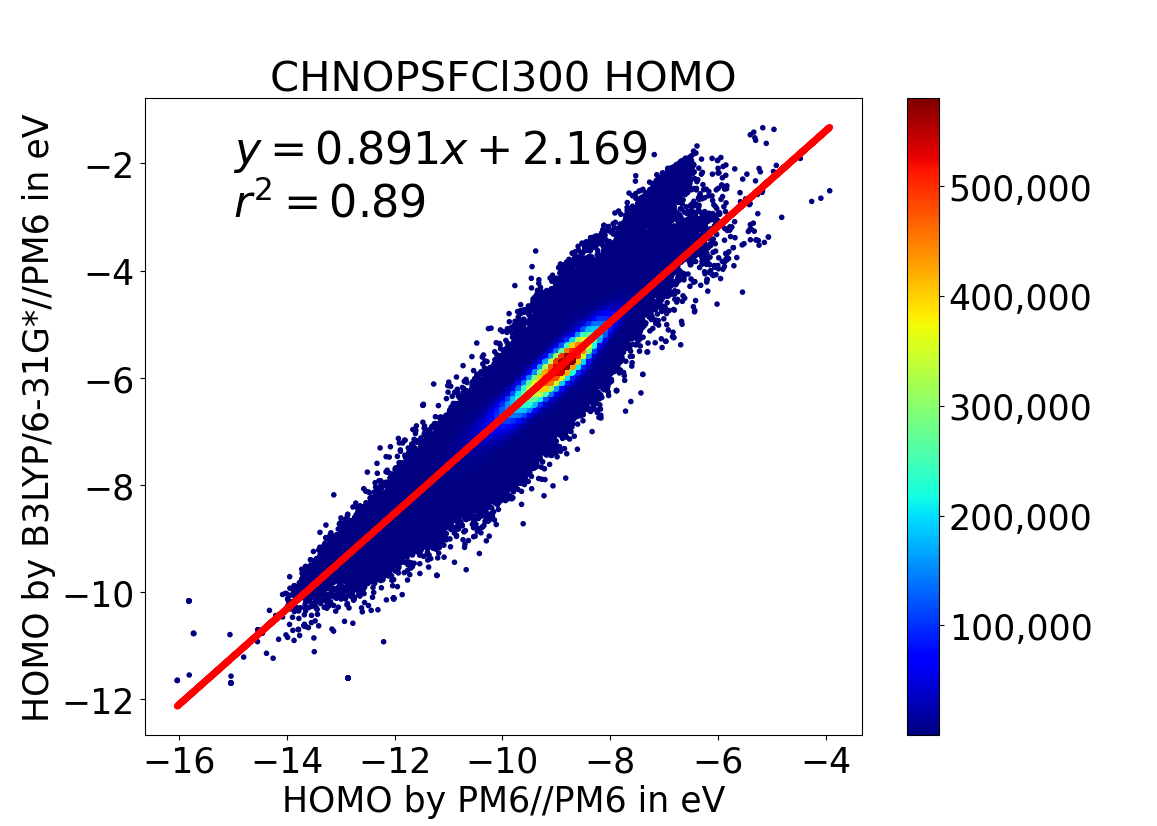} & \includegraphics[width=0.45\textwidth]{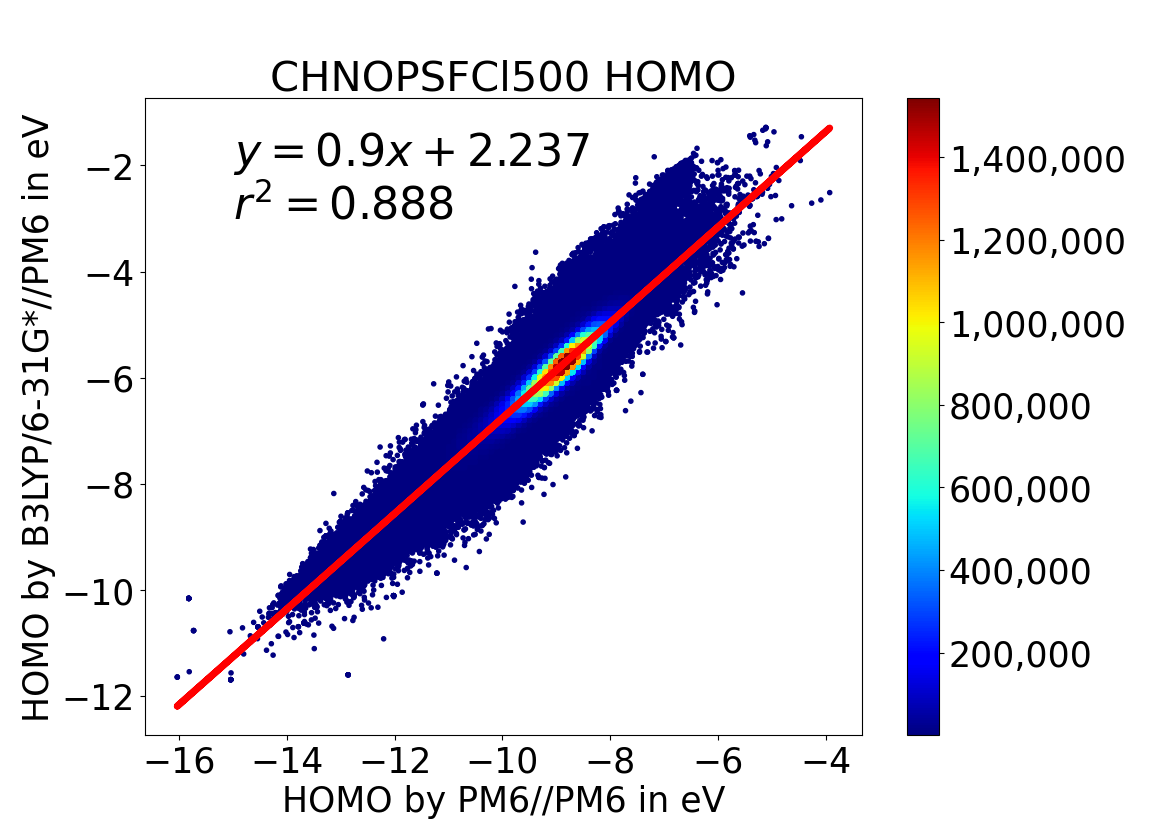} \\
      \includegraphics[width=0.45\textwidth]{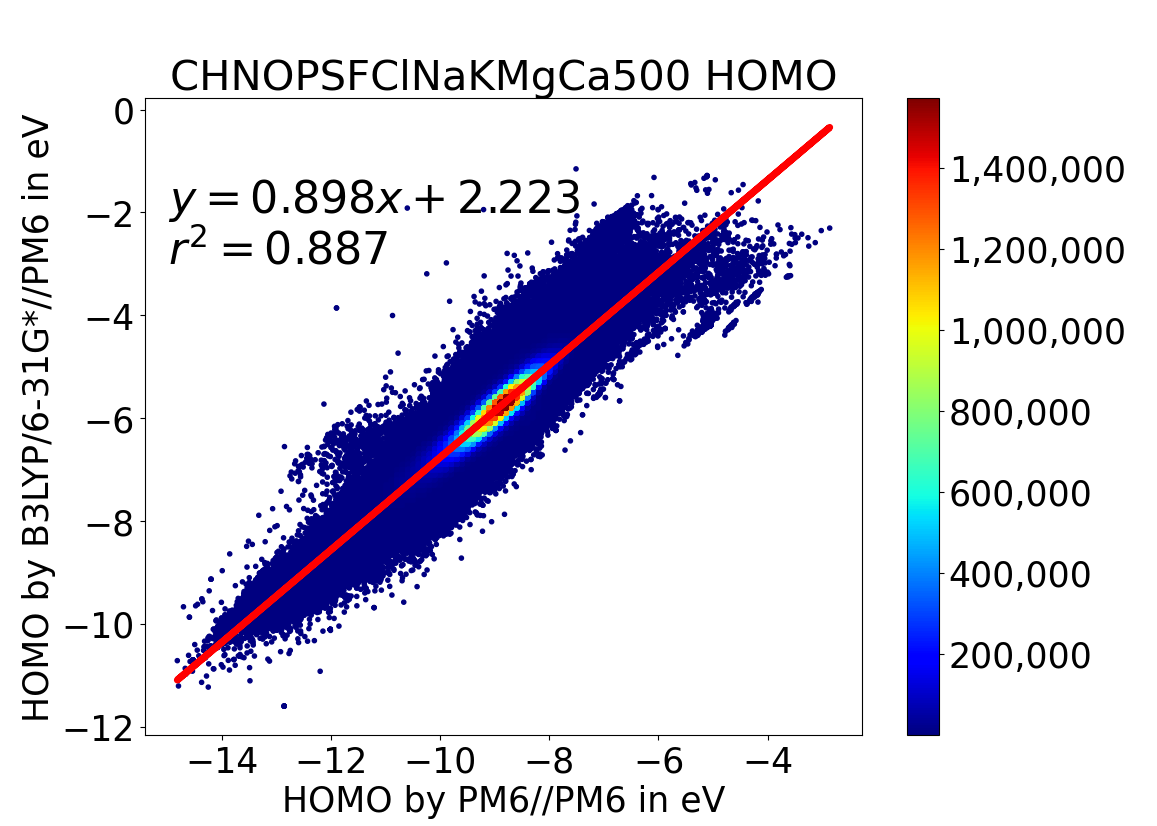} & 
      \includegraphics[width=0.45\textwidth]{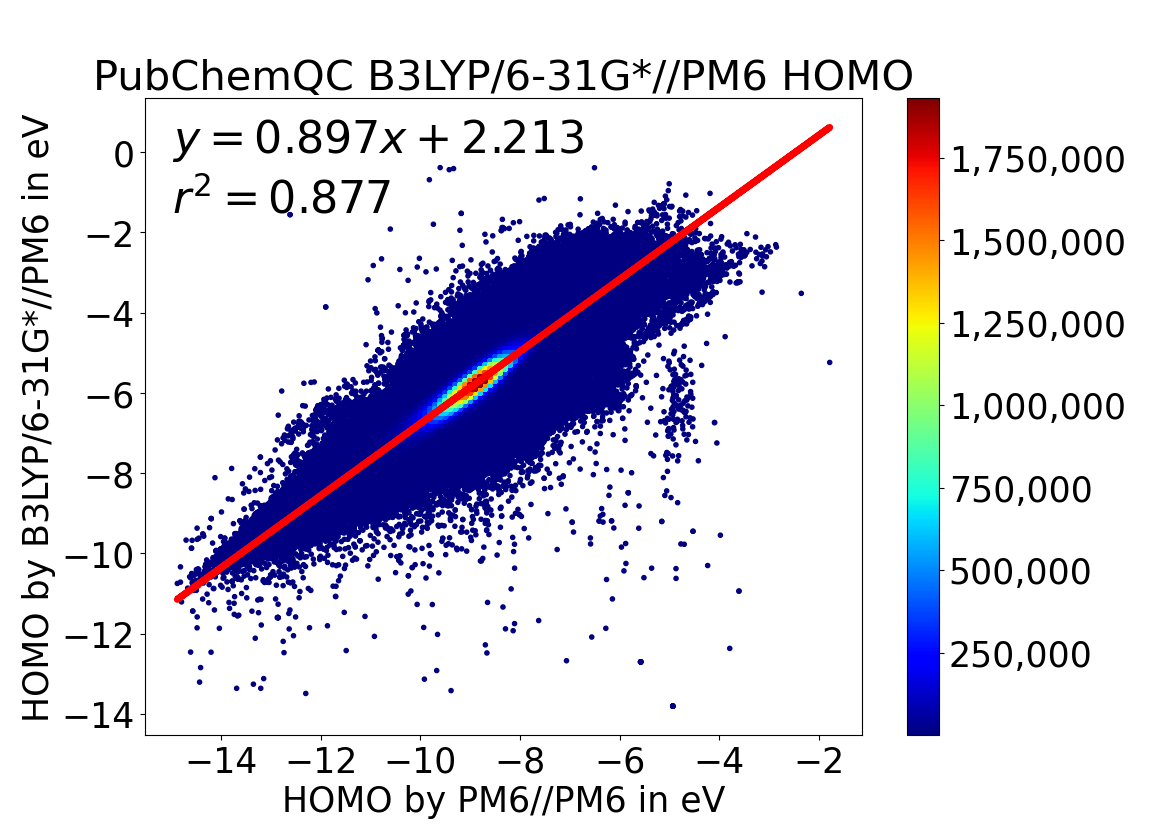} \\
    \end{tabular}
    \label{graph:homo}
\end{figure}

The comparative analysis of HOMO energies between the \dbname{} and PM6/\!/PM6 methods is illustrated in Fig.~\ref{graph:homo}, which presents a two-dimensional histogram and heatmap depicting the correlation of HOMO energies with a bin size of $0.1$ eV each. These energies were computed using both PM6/\!/PM6 and \dbname{} levels of theory for the entire set and all subsets, including CHON300, CHON500, CHNOPSFCl300, CHNOPSFCl500, CHNOPSFClNaKMgCa500, and the complete set.

For the subsets CHON300, CHON500, CHNOPSFCl300, and CHNOPSFCl500, we have plotted all the molecules and included them in the analysis. However, for the CHNOPSFClNaKMgCa500 subset and the complete molecular set, we have excluded molecules with physically and chemically implausible values. Such values include cases where the dipole moment exceeds $100$ debye, the HOMO-LUMO energy gap exceeds $100$ eV, the HOMO energy falls outside the range of $-5$ eV to $0.5$ eV, or the LUMO energy exceeds $10$ eV.

The horizontal axis represents HOMO energies by PM6/\!/PM6, while the vertical axis displays the frequency distribution of HOMO values by \dbname{}. Additionally, a linear regression analysis was conducted to examine the correlations.

The CHON300 dataset, located in the upper left corner, consists of molecules comprising C, H, O, and N elements without salt and exhibits a molecular weight below 300. Notably, a peak is observed at approximately $-9$ eV and $-6$ eV for PM6/\!/PM6 and \dbname{}, respectively. Furthermore, the coefficient of determination is 0.907, indicating an exceptionally high correlation based on the linear regression analysis.

Similarly, the CHON500 (upper right), CHNOPSFCl300 (middle left), CHNOPSFCl500 (middle right) subsets, CHNOPSFClNaKMgCa500 (lower left), and the complete set (lower right) exhibit high correlation coefficients of determination, which are 0.907, 0907, 0.890, 0.888, 0.876 and 0.877, respectively. These values demonstrate a strong correlation between the HOMO energies calculated using PM6/\!/PM6 and \dbname{} methods for all subsets and the complete set.

There were outliers in all subsets, but their shape was almost unchanged. The locations where peak values were present and their shape were almost identical in all subsets. However, the peak height increased with the number of molecules in that subset. Outliers were more common when all molecules were plotted. This is thought to be due to the increasing number of atom species and is not surprising as, in general, the heavier the atoms, the more complex the calculation becomes; there are not many molecules containing heavy atoms in PubChem. Therefore, the peak values and coefficients of determination were unchanged. In any case, close examination is necessary.

\begin{figure}[h!]
\caption{The figure is the same as Fig.~\ref{graph:homo}, viewed from a 45-degree angle. The plots demonstrate almost linear relationships, indicating that the number of molecules with substantial differences in HOMO values between the calculation methods, as observed in the two-dimensional plots, is minimal.}
    \label{graph:homo_3d}
    \centering
    \begin{tabular}{c c}
      \includegraphics[width=0.45\textwidth]{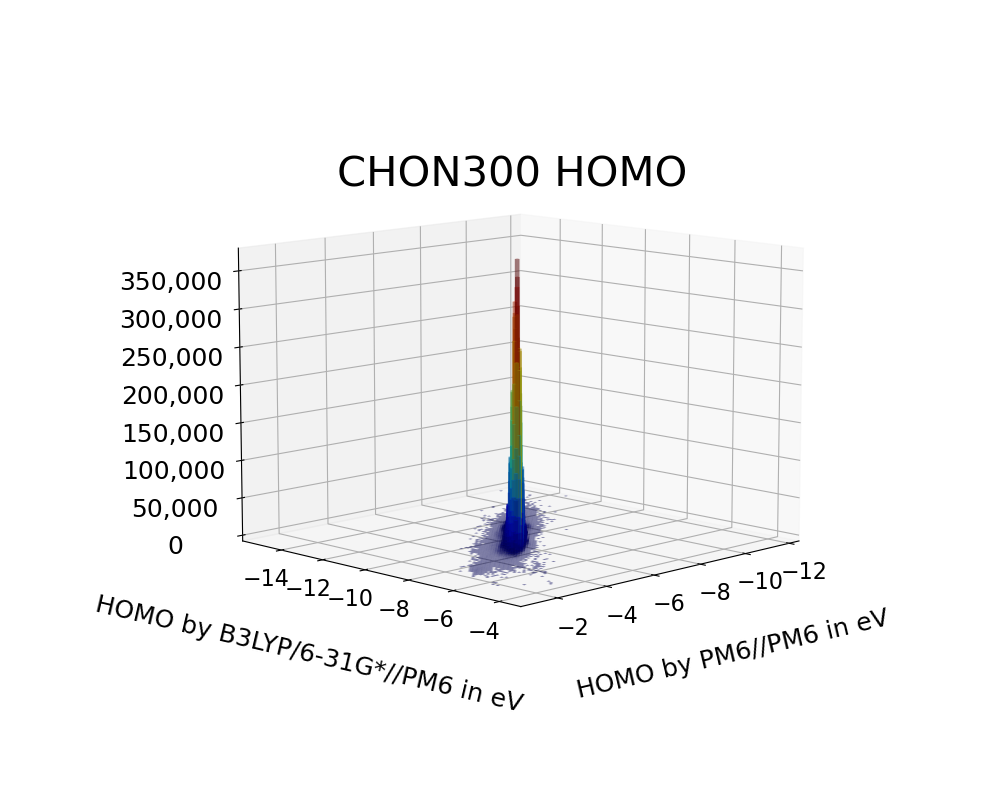} & \includegraphics[width=0.45\textwidth]{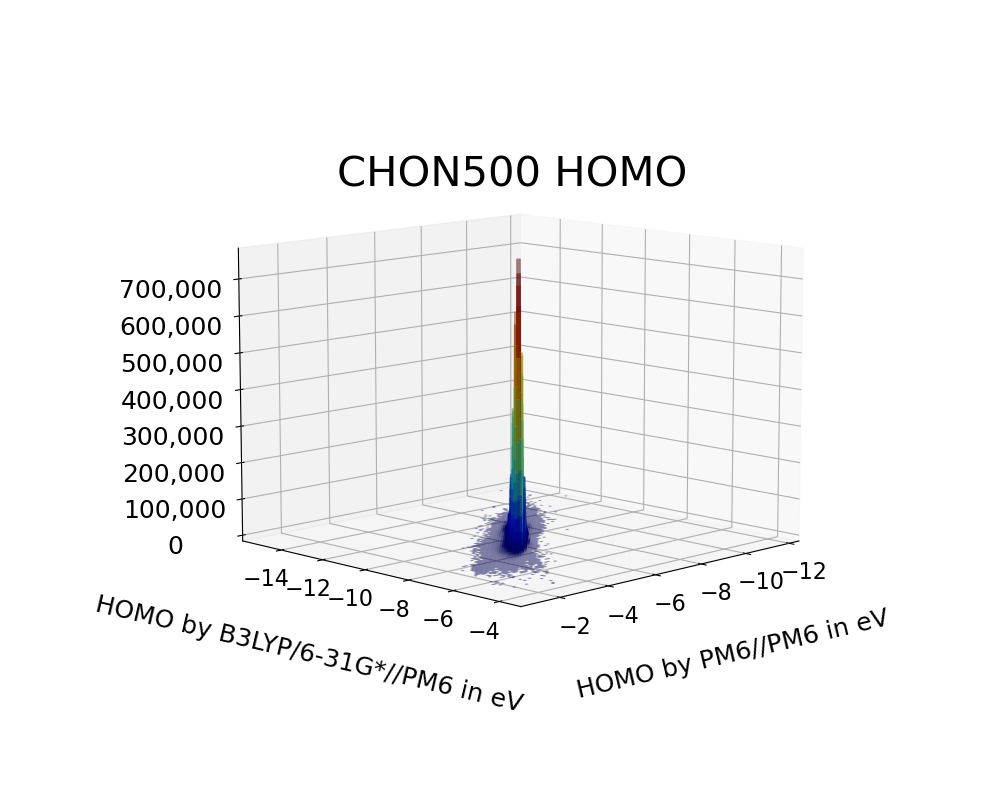} \\ 
      \includegraphics[width=0.45\textwidth]{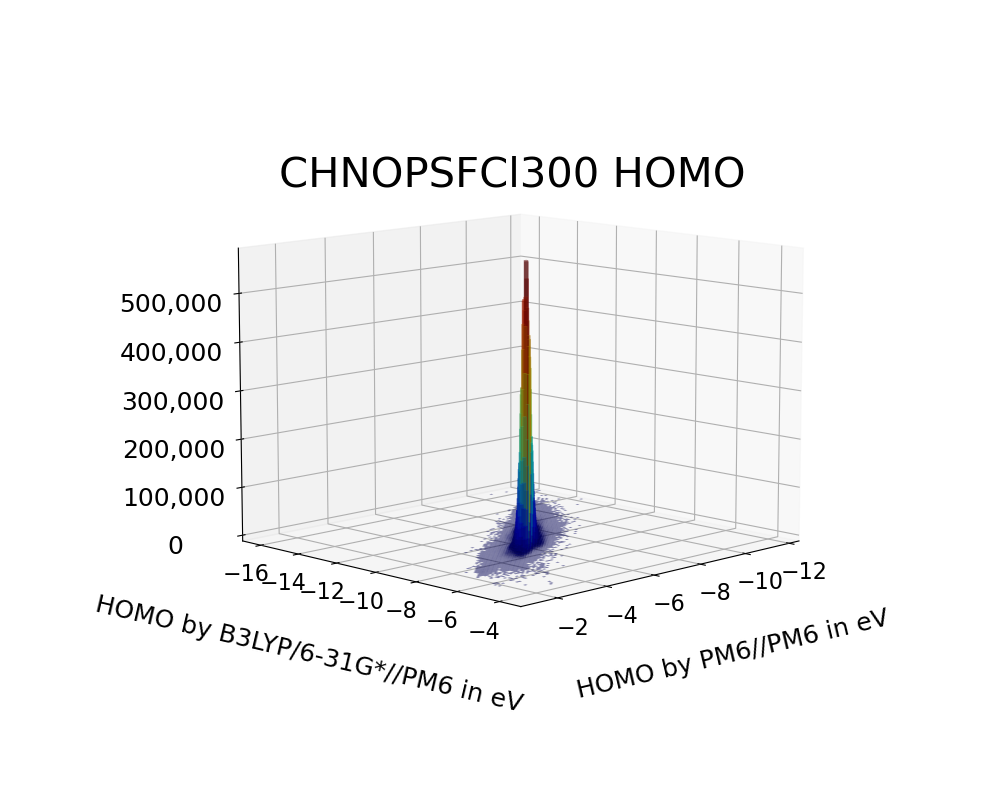} & \includegraphics[width=0.45\textwidth]{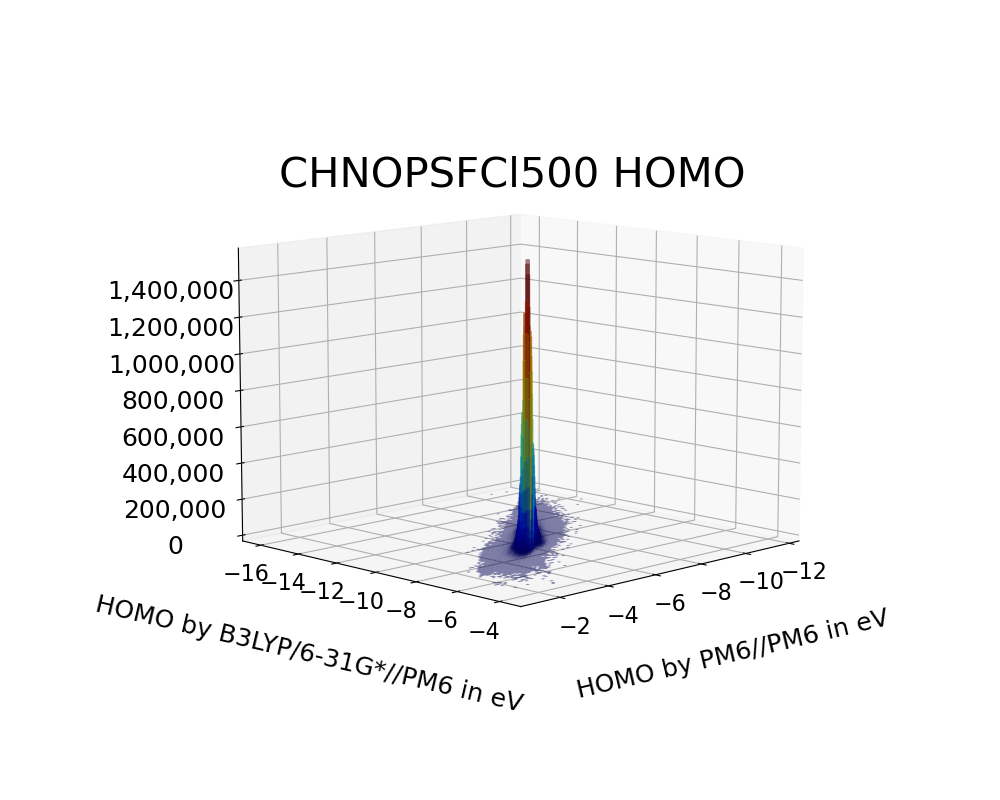} \\ 
      \includegraphics[width=0.45\textwidth]{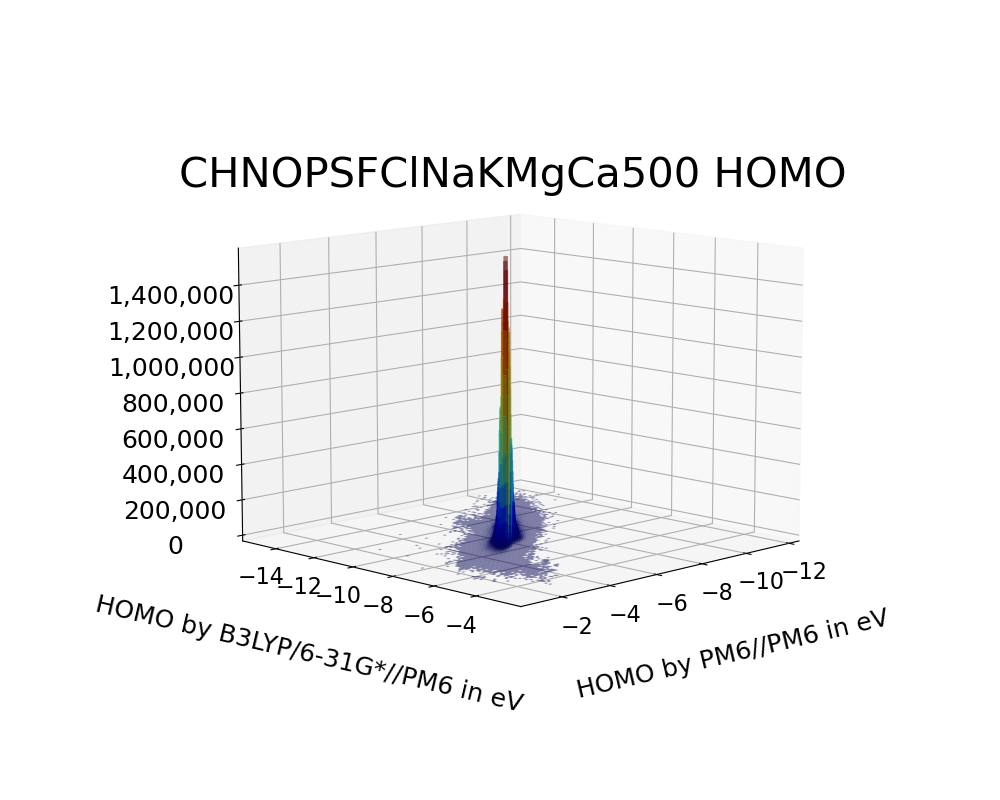} & 
      \includegraphics[width=0.45\textwidth]{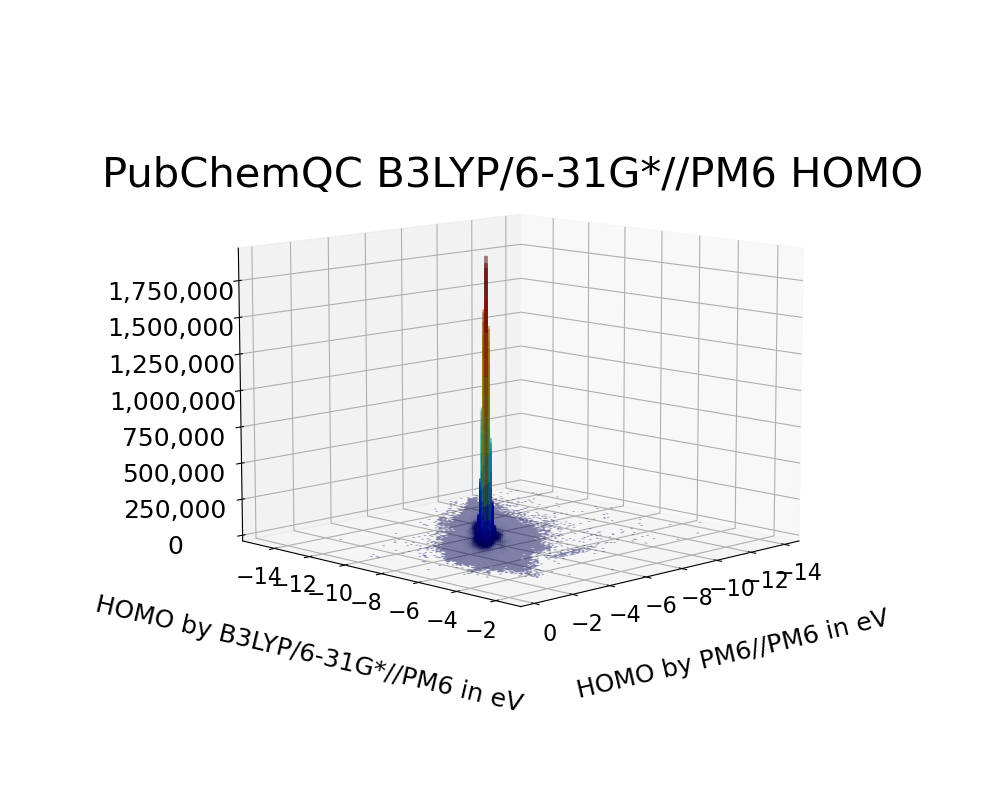} \\
    \end{tabular}
\end{figure}

Fig. \ref{graph:homo_3d} presents the data depicted in Fig. \ref{graph:homo}, offering an alternative perspective from a 45-degree angle. Notably, the visualization highlights a minimal presence of outliers and distinct peaks for all datasets.

When the results from both calculations exhibit agreement, a singular point is observed, with the linearity of the relationship being influenced by factors such as higher correlation coefficients or more minor variances.

\clearpage

\subsection{Comparison of LUMO Energies}
\begin{figure}[h!]
\caption{The figure presents a 2D histogram and heatmap of correlated LUMO energies calculated using PM6/\!/PM6 and \dbname{} methods. The horizontal axis signifies PM6/\!/PM6 LUMO energies, while the vertical axis exhibits the frequency distribution of B3LYP/6-31G*//PM6 LUMO values. It is noteworthy that the data point dispersion escalates with the molecule's atom count. The coefficients of determination are from 0.821 to 0.911}
    \centering
    \begin{tabular}{c c}
      \includegraphics[width=0.45\textwidth]{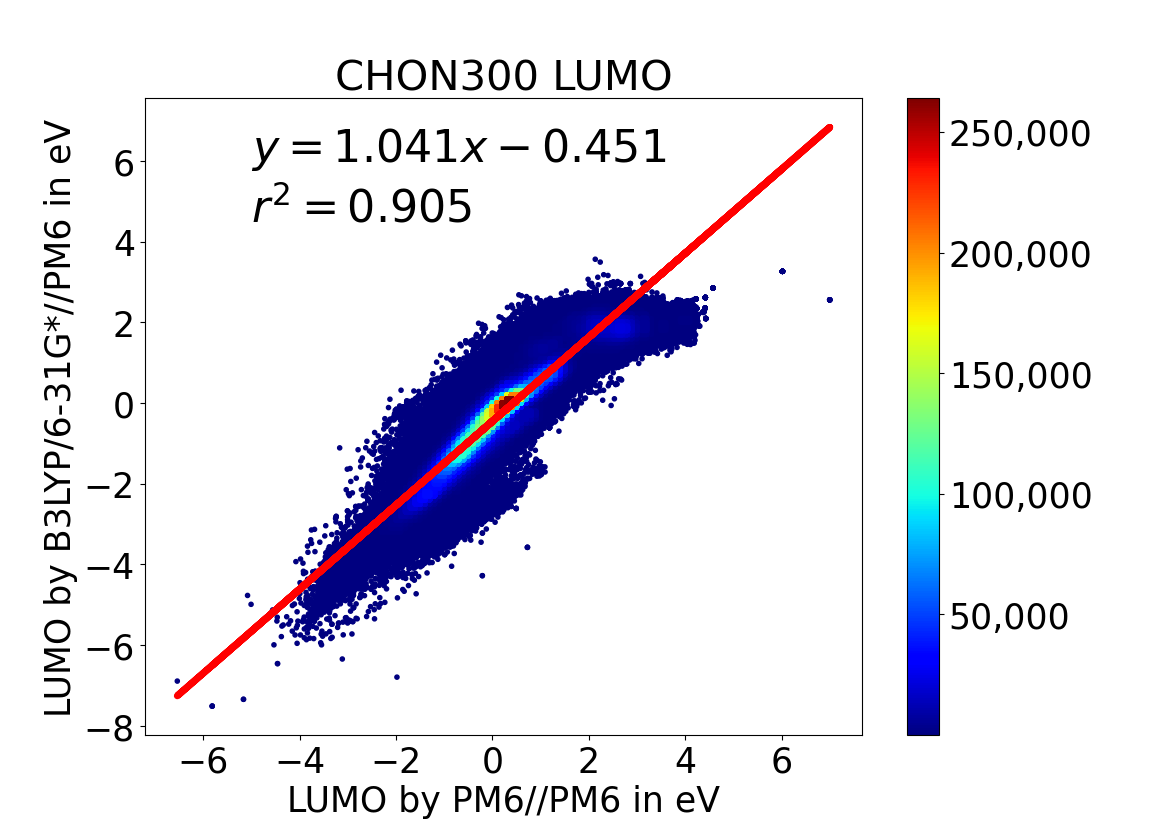} & \includegraphics[width=0.45\textwidth]{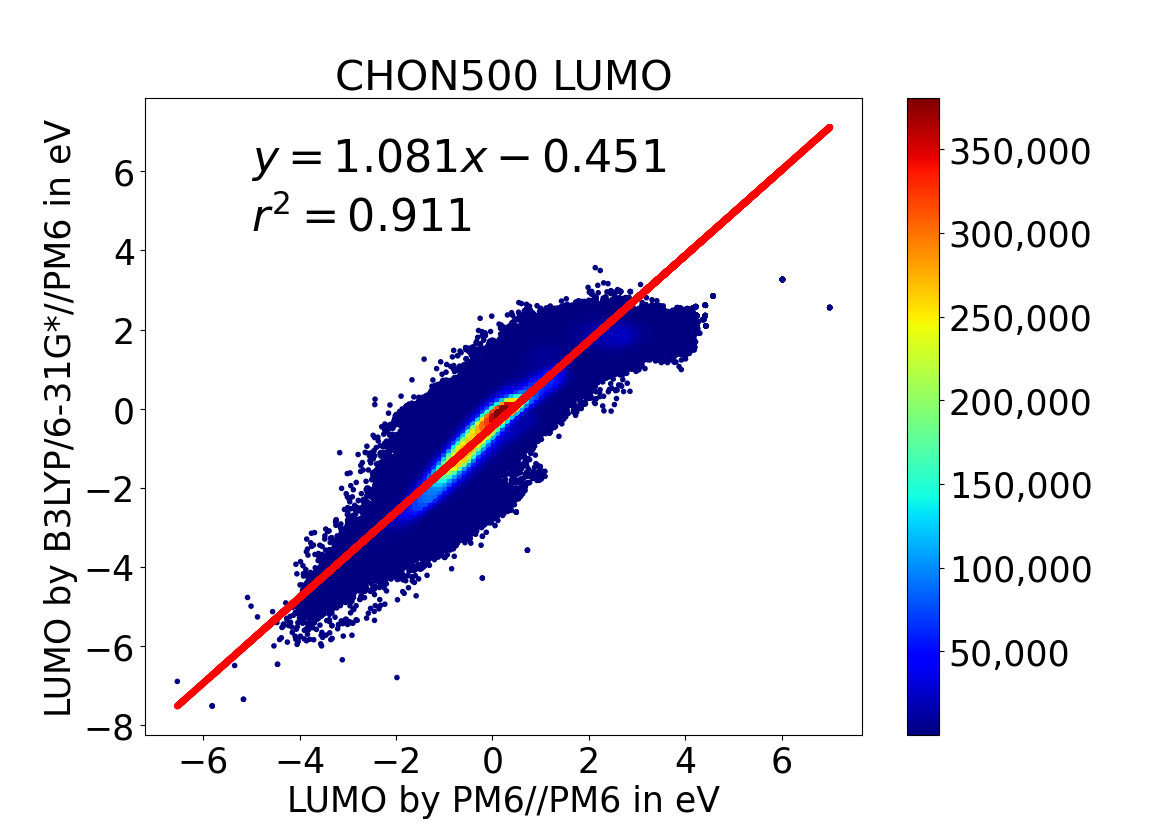} \\ 
      \includegraphics[width=0.45\textwidth]{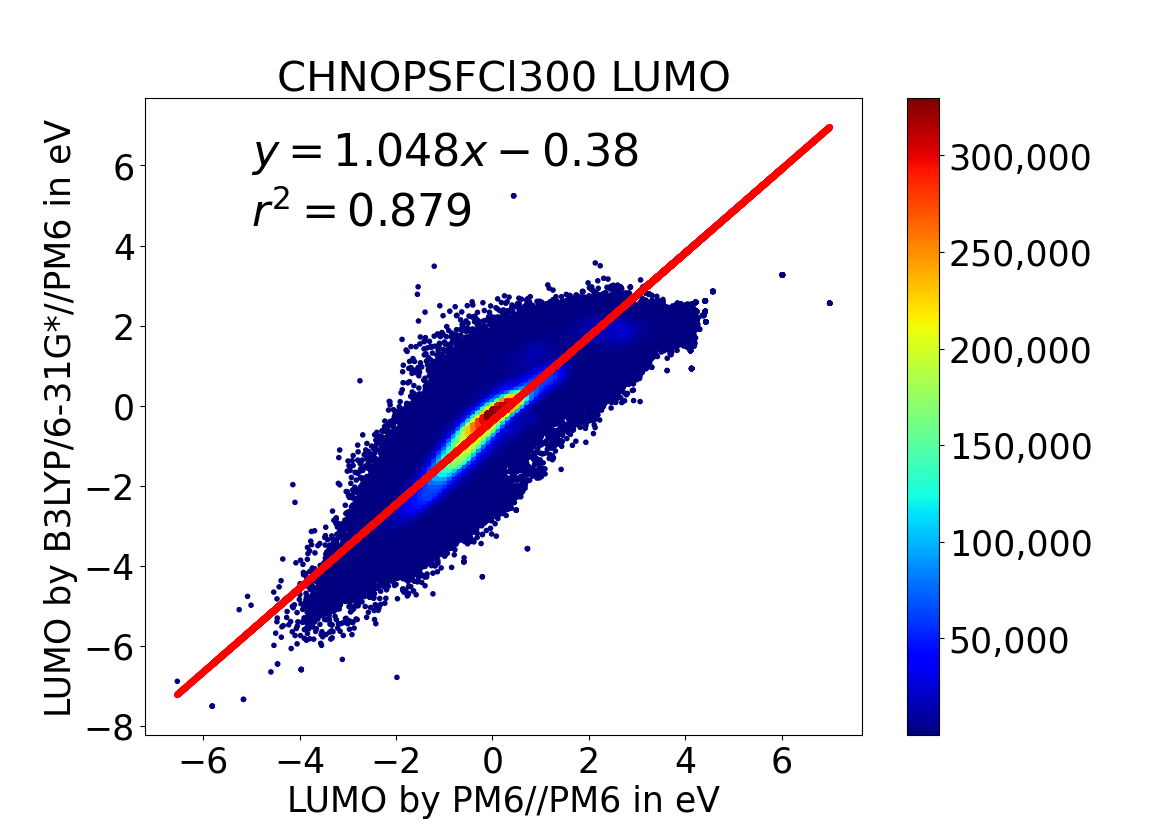} & \includegraphics[width=0.45\textwidth]{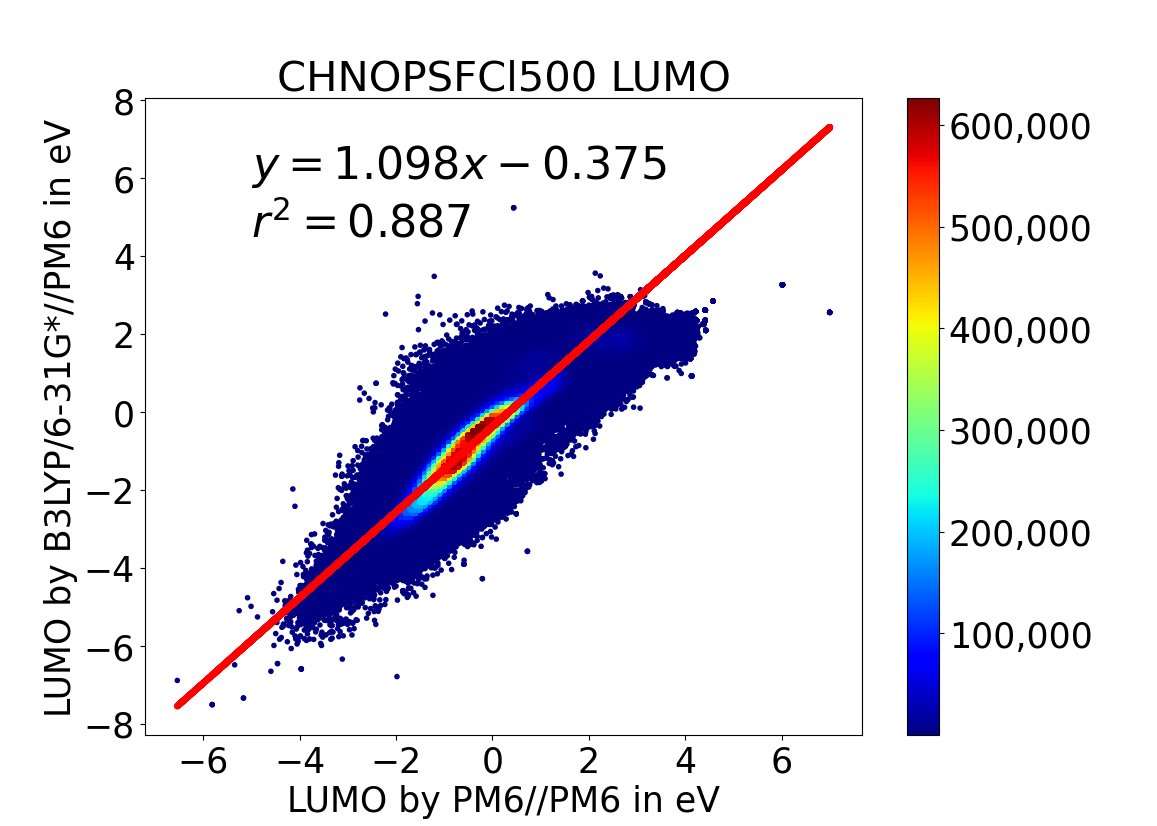} \\
      \includegraphics[width=0.45\textwidth]{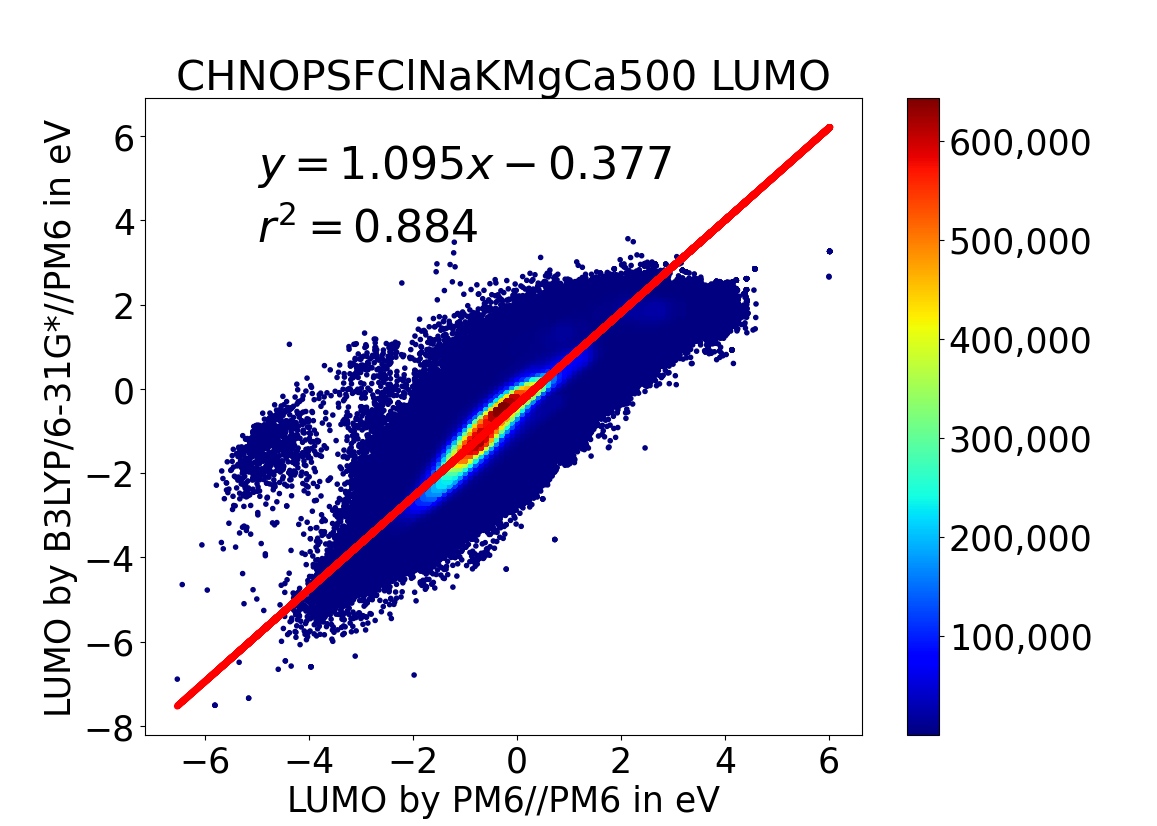} & 
      \includegraphics[width=0.45\textwidth]{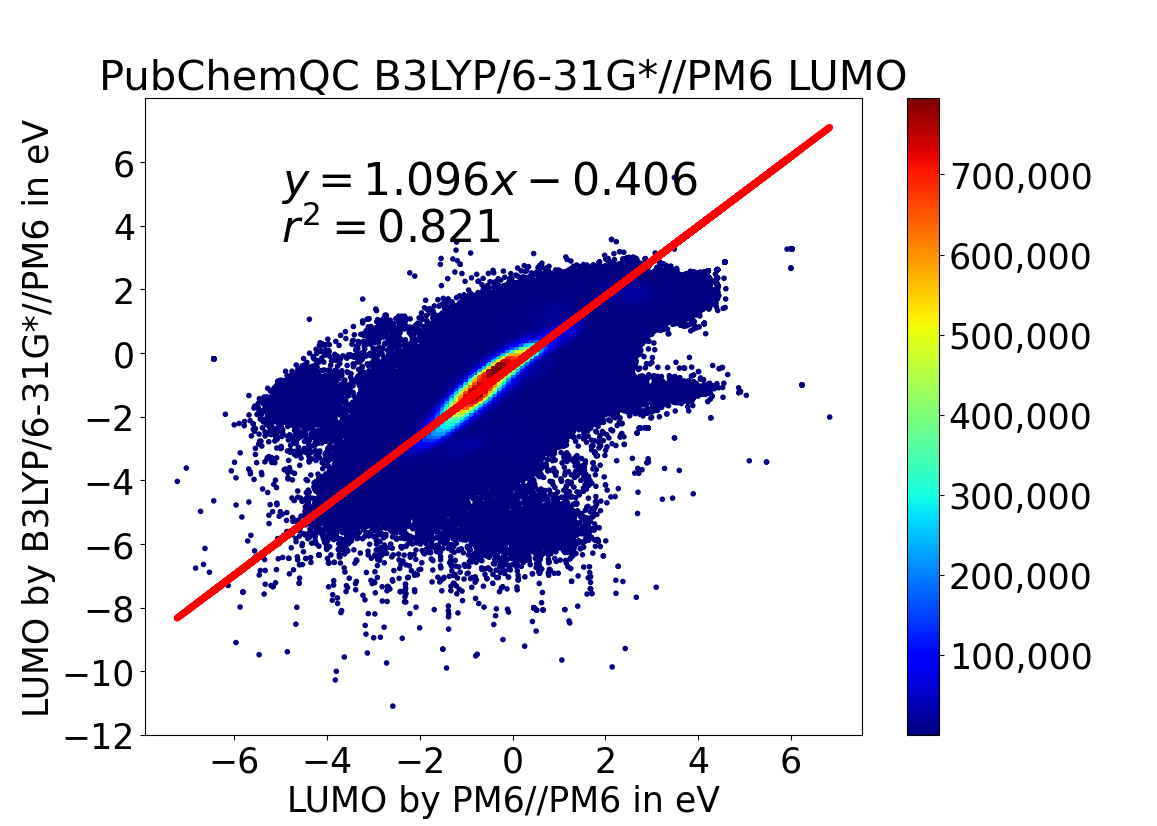} \\
    \end{tabular}
    \label{graph:lumo}
\end{figure}

The juxtaposition of LUMO energies employing the \dbname{} and PM6/\!/PM6 techniques is depicted in Fig.~\ref{graph:lumo}, showcasing a two-dimensional histogram and heatmap that elucidate the correlation of LUMO energies with a bin size of 0.1 eV each.  These energies were ascertained utilizing both PM6/\!/PM6 and \dbname{} methodologies for the entire collection and all subsets, encompassing CHON300, CHON500, CHNOPSFCl300, CHNOPSFCl500, CHNOPSFClNaKMgCa500, and the comprehensive set.

For the subset, except for CHNOPSFClNaKMgCa500 subset, we plotted all the molecules. In contrast, for the CHNOPSFClNaKMgCa500 and the complete molecular set, molecules with physically and chemically implausible values, similar to the case of HOMO, were excluded from the analysis.

The horizontal axis signifies LUMO energies PM6/\!/PM6 method, while the vertical axis exhibits the frequency distribution of HOMO values derived from B3LYP/6-31G*/\!/PM6. Moreover, a linear regression analysis was performed to assess the correlations.

The CHON300 dataset, illustrated in the upper left corner, consists of C, H, O, and N elements without salt and has a molecular weight under 300. A significant peak is observed at approximately $0.9$ eV, $0.5$ eV for PM6 versus \dbname{}. Furthermore, the coefficient of determination is 0.905, indicating an outstanding correlation based on the linear regression analysis. For CHON500, CHNOPSFCl300, CHNOPSFCl500, CHNOPSFClNaKMgCa500 subsets, and the entire set, the coefficients of determination are 0.911, 0.879, 0.887, 0.884, and 0.821 respectively.

Similar to the HOMO energy case, outliers emerged in all subsets, yet their morphology predominantly persisted. The positions and configurations of peak values demonstrated remarkable consistency across CHON and CHNOPS subsets. Nevertheless, the peak amplitude escalated in correspondence with the number of molecules in each subset. Outliers manifested with greater frequency when delineating the CHNOPSFClNaKMgCa500 subset and all molecules, presumably attributable to the augmented number of atomic species. This observation is not unexpected, as calculations typically become more intricate with the presence of heavier atoms; few molecules containing heavy atoms are found in the PubChem database. As a result, the peak values and coefficients of determination maintained a stable pattern. However, a comprehensive investigation remains indispensable.

\begin{figure}[h!]
\caption{The same figure as in Fig.~\ref{graph:lumo}, viewed
from a 45-degree angle. The plot demonstrates an almost linear relationship, suggesting that the number of molecules with substantial differences in lumo values, depending on the calculation method, as observed in the two-dimensional plots, is minimal.}
    \label{graph:lumo_3d}
    \centering
    \begin{tabular}{c c}
      \includegraphics[width=0.45\textwidth]{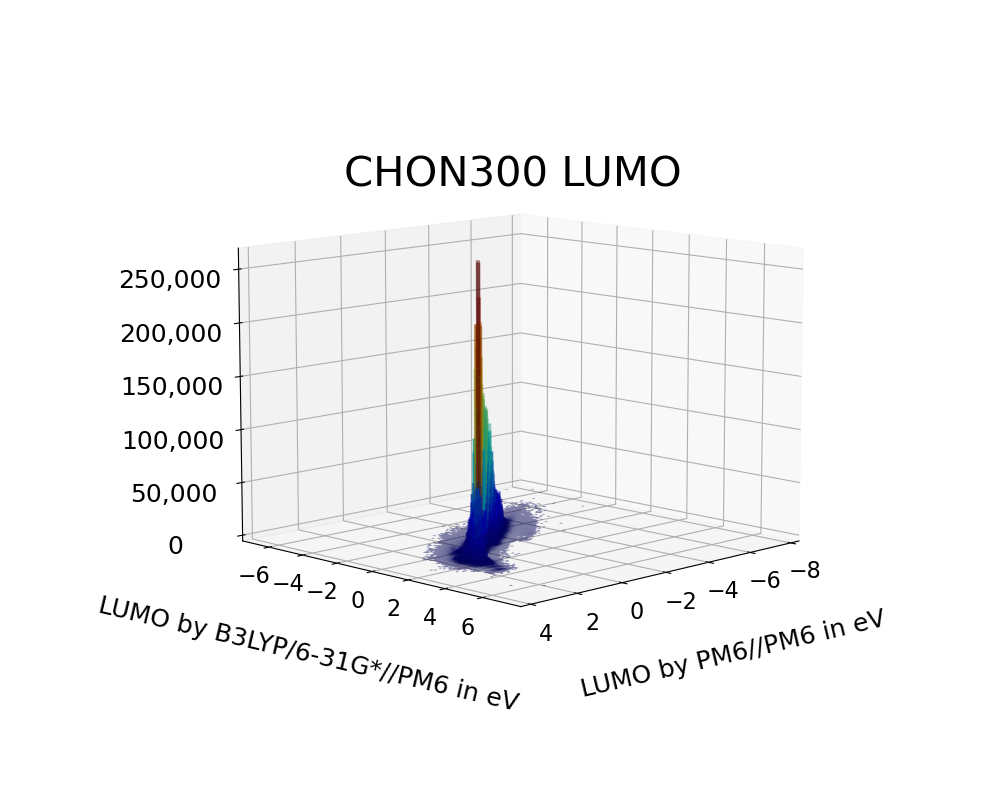} & \includegraphics[width=0.45\textwidth]{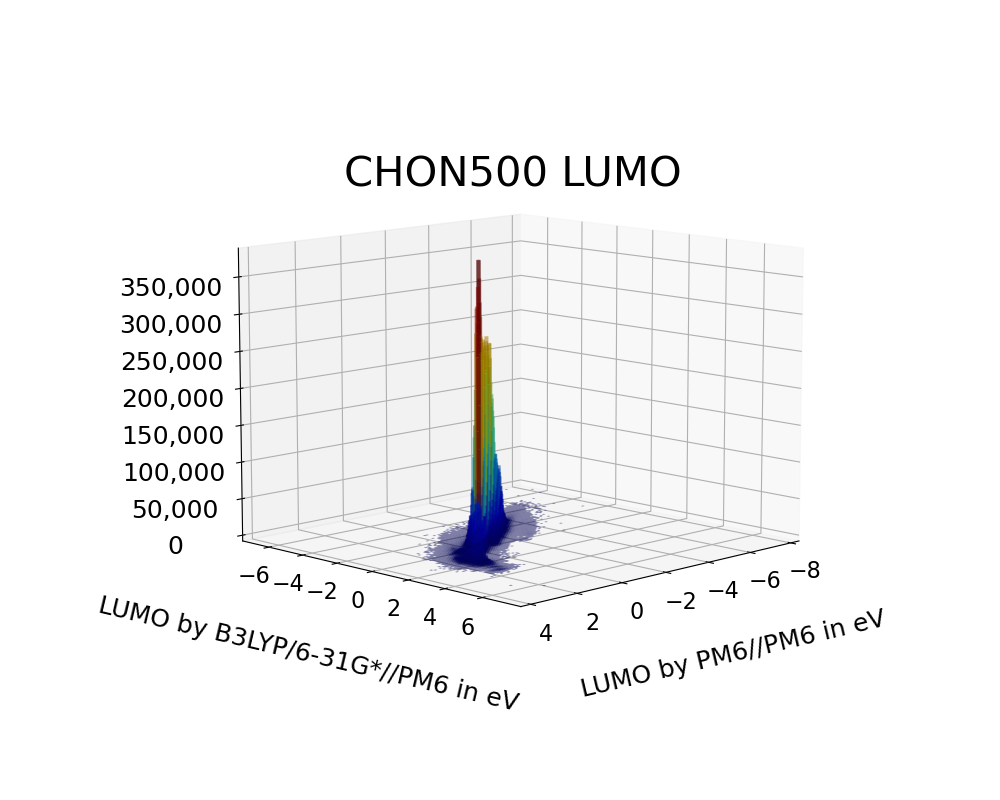} \\ 
      \includegraphics[width=0.45\textwidth]{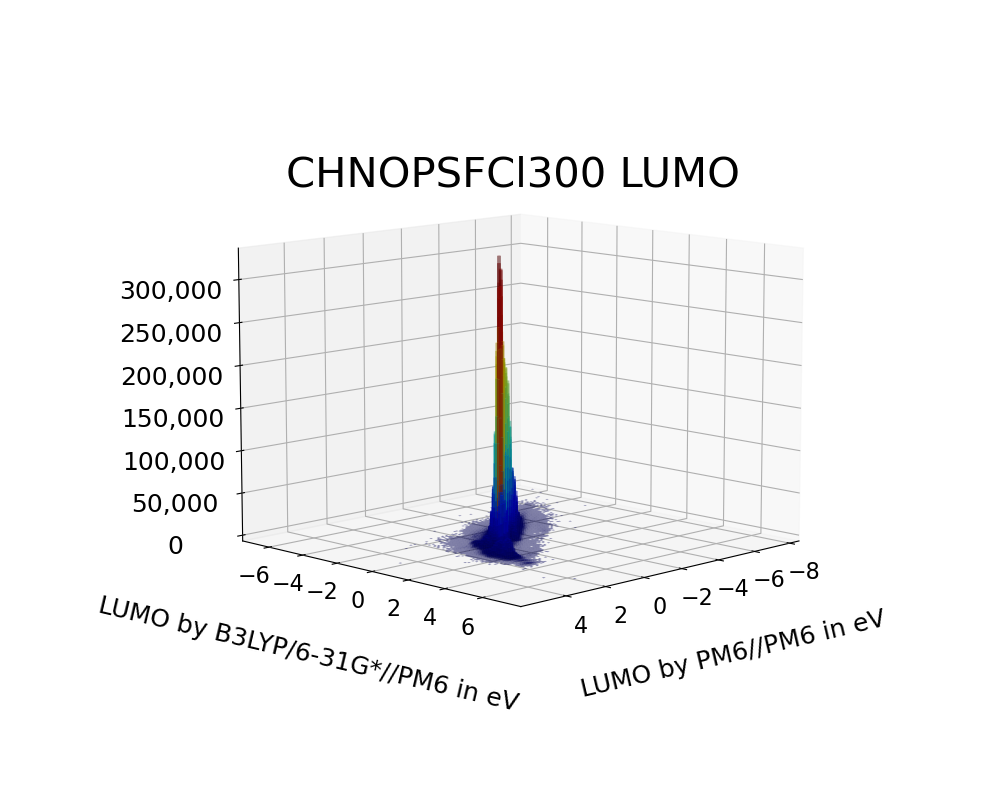} & \includegraphics[width=0.45\textwidth]{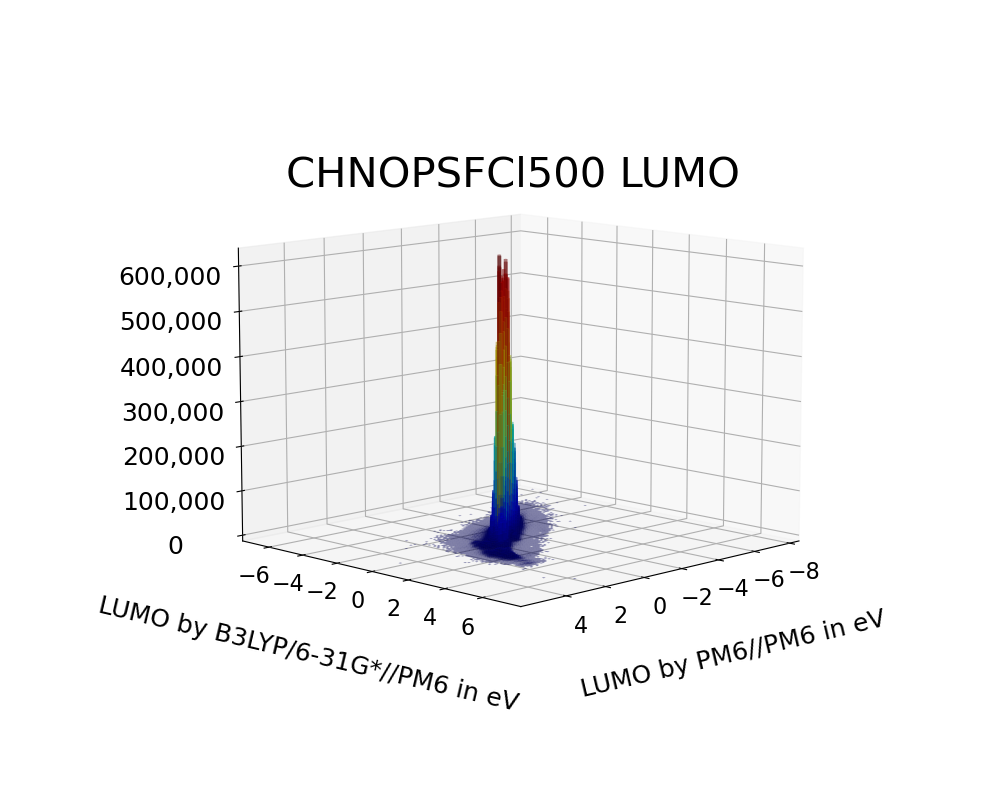} \\
     \includegraphics[width=0.45\textwidth]{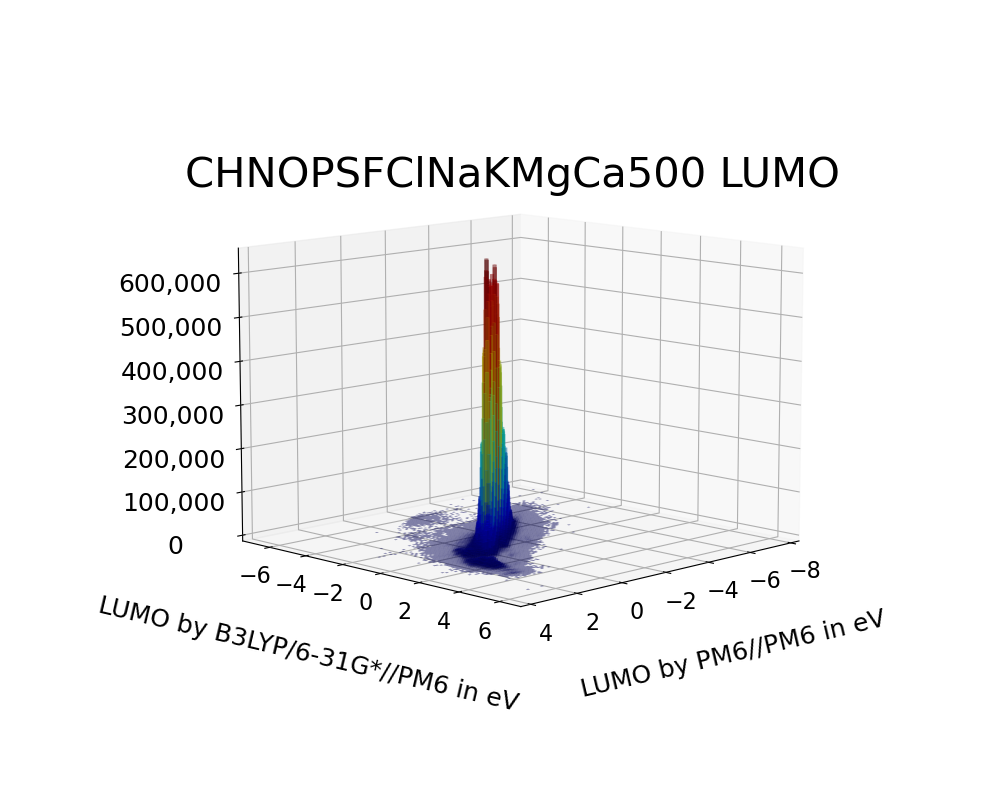} & 
      \includegraphics[width=0.45\textwidth]{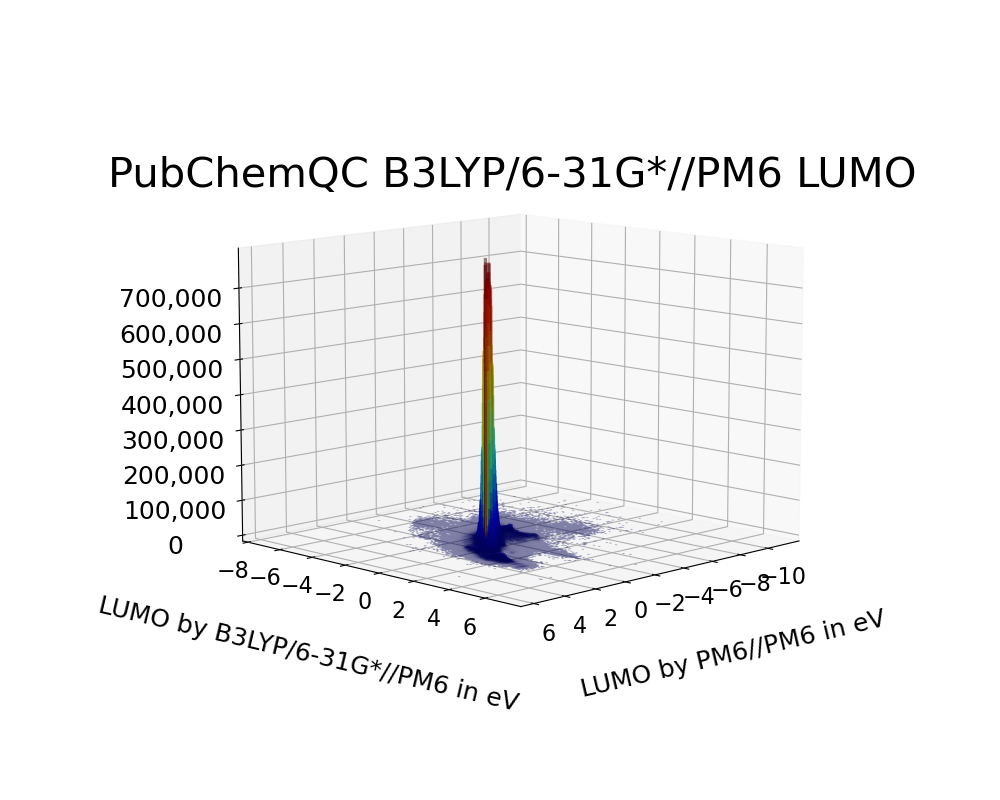} \\
    \end{tabular}
\end{figure}

Fig.~\ref{graph:lumo_3d} displays the same data as in Fig.~\ref{graph:lumo} but is viewed from a 45-degree angle. It is evident that the number of outliers is minimal, and the peaks are distinct for all sets.

\clearpage

\subsection{Comparison of HOMO-LUMO energy gap}
\begin{figure}[h!]
\caption{The figure displays a 2D histogram and heatmap of correlated LUMO energies determined using PM6/\!/PM6 and \dbname{} approaches. The horizontal axis represents PM6/\!/PM6 HOMO-LUMO energy gaps, while the vertical axis illustrates the frequency distribution of HOMO-LUMO energy gap values derived from B3LYP/6-31G* for matching structures. It is significant that the data point dispersion increases with the atom count of the molecule. The coefficients of determination are within the range of $0.803$ to $0.892$.}
    \centering
    \begin{tabular}{c c}
      \includegraphics[width=0.45\textwidth]{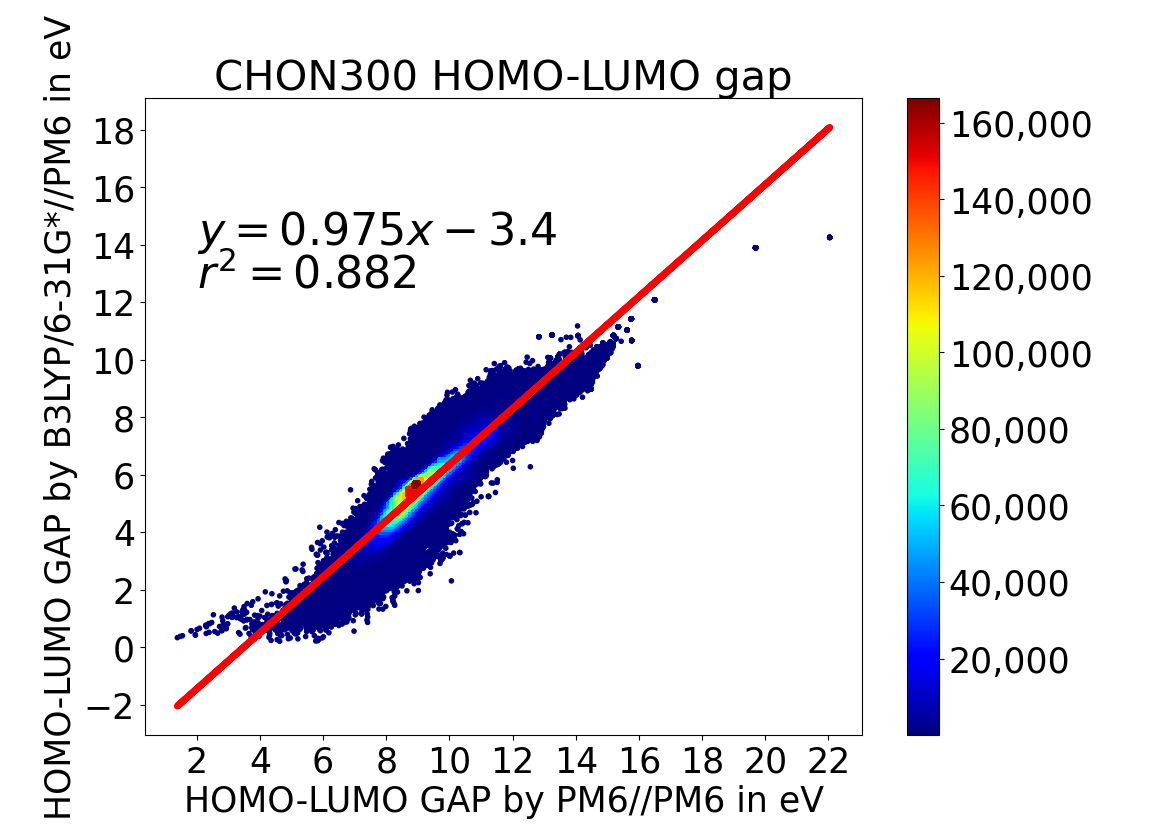} & \includegraphics[width=0.45\textwidth]{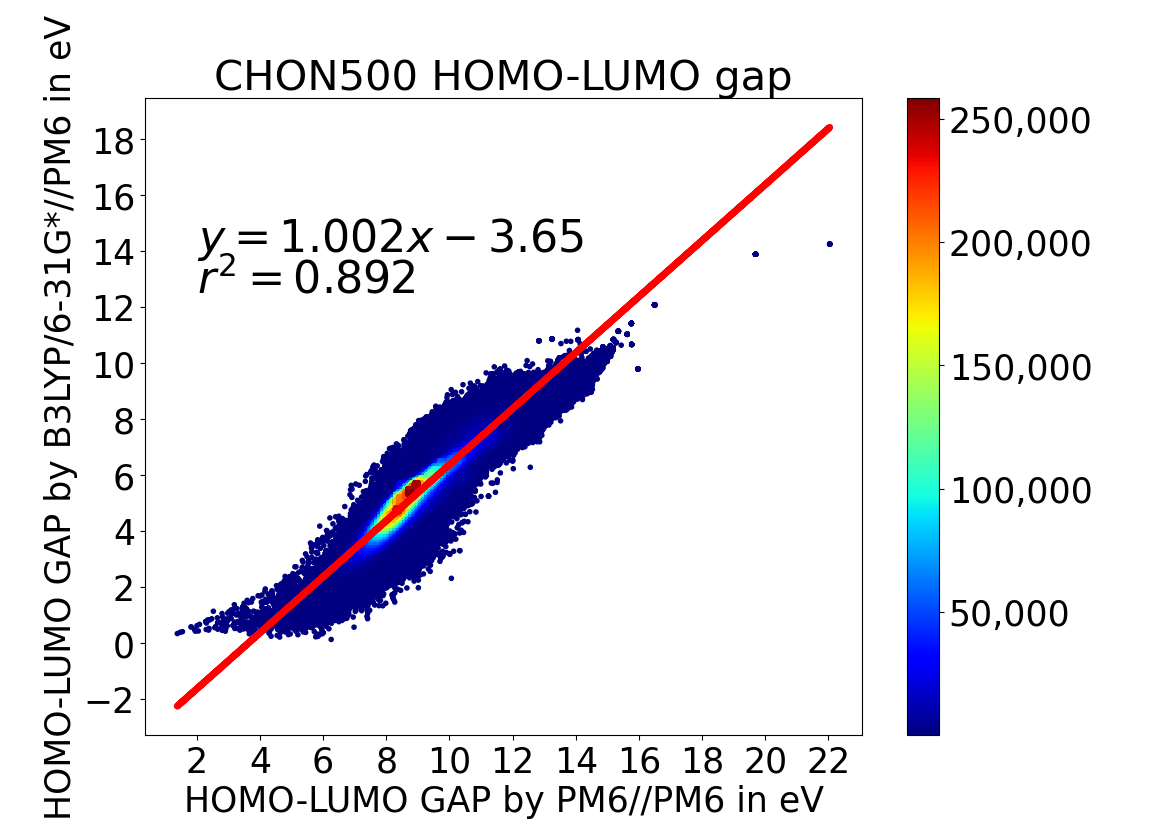} \\ 
      \includegraphics[width=0.45\textwidth]{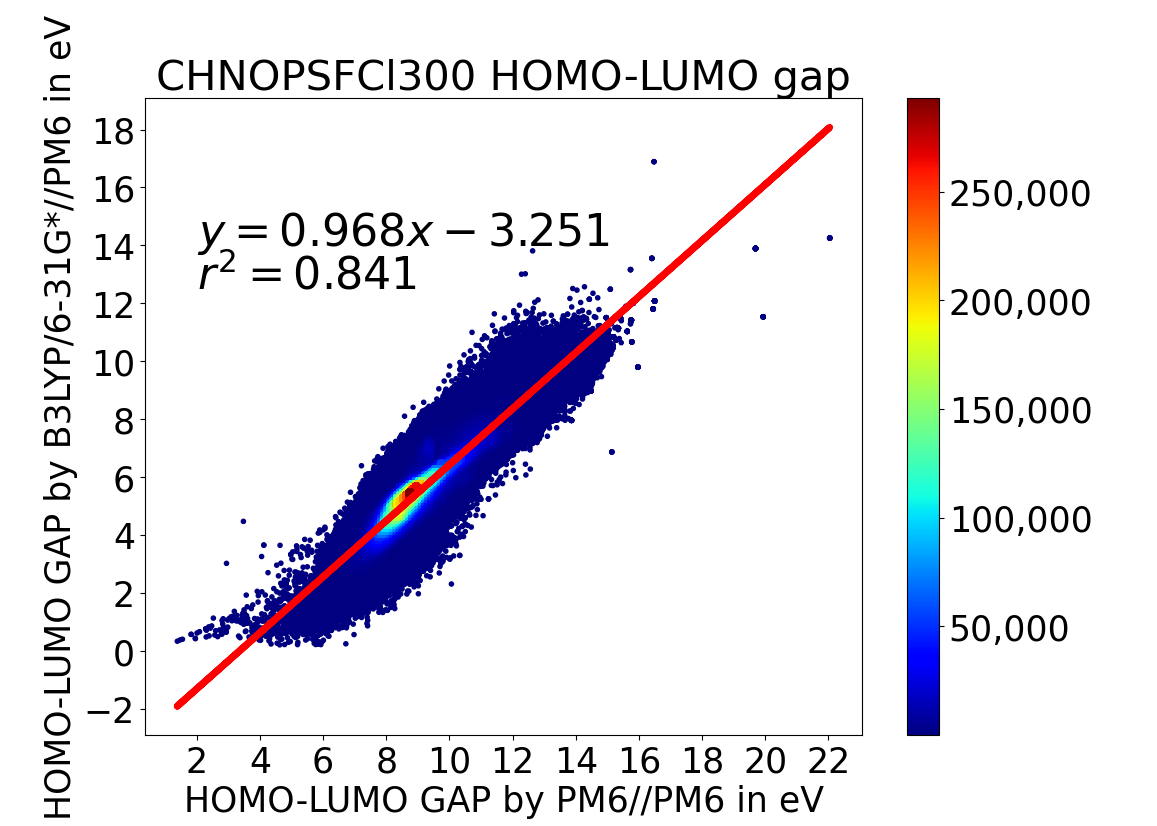} & \includegraphics[width=0.45\textwidth]{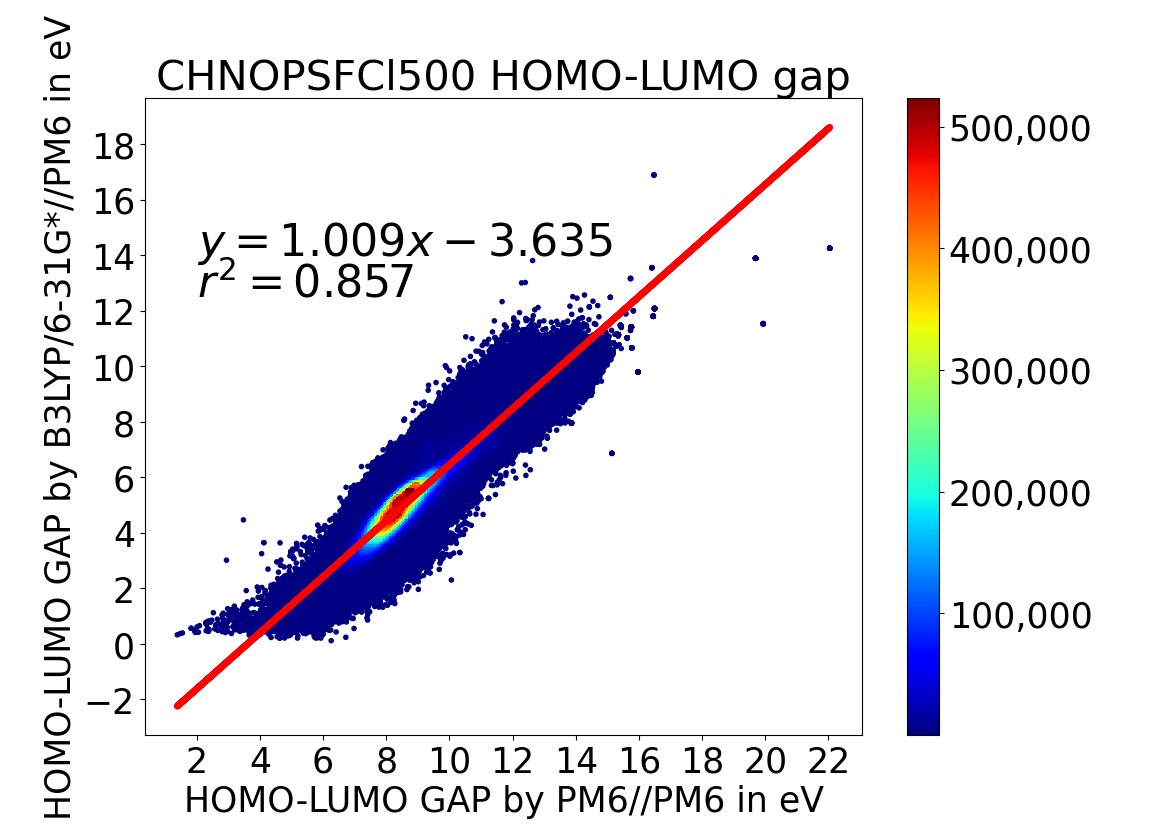} \\
      \includegraphics[width=0.45\textwidth]{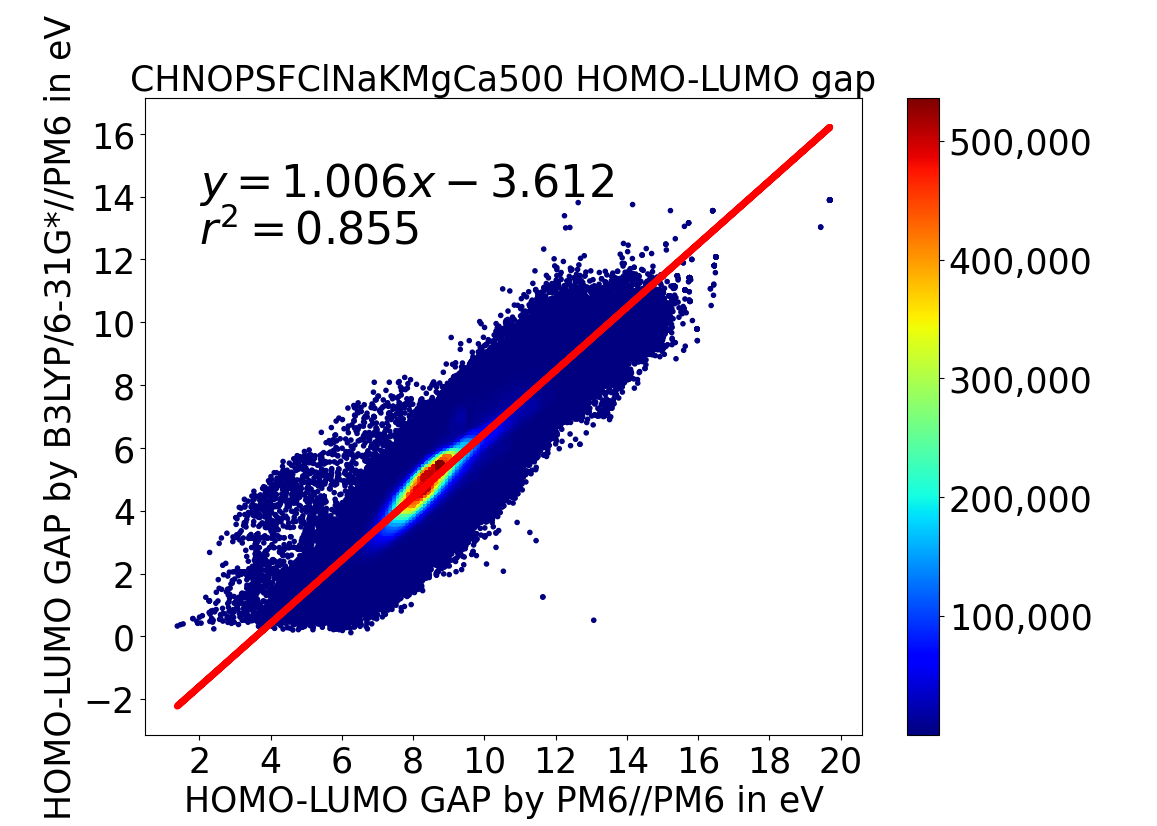} & 
      \includegraphics[width=0.45\textwidth]{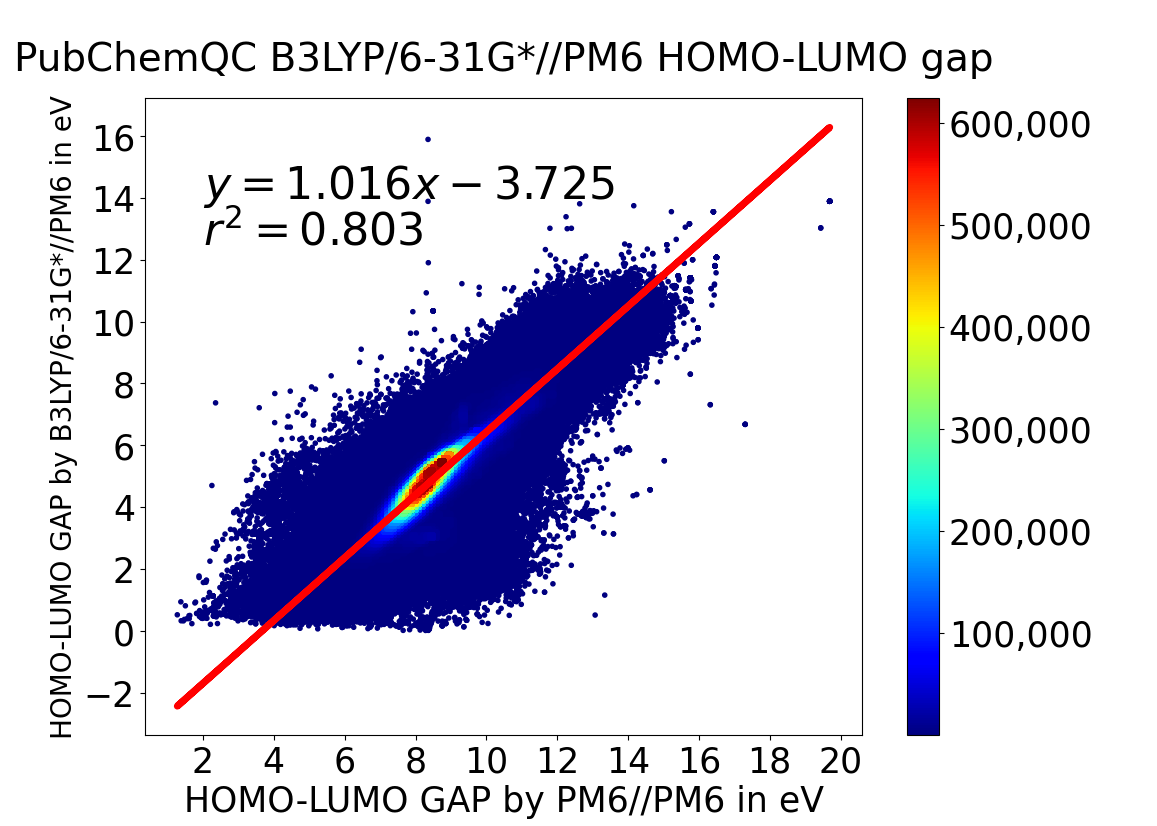} \\
    \end{tabular}
    \label{graph:homolumogap}
\end{figure}

A comparative examination of HOMO-LUMO energy gap between the \dbname{} and PM6/\!/PM6 approaches is presented in Fig.~\ref{graph:homolumogap}, displaying a two-dimensional histogram and heatmap that elucidate the correlation of HOMO-LUMO energy gap with each bin size of $0.1$ eV. These energies were ascertained utilizing both PM6/\!/PM6 and \dbname{} techniques for the entire collection and all subsets, including CHON300, CHON500, CHNOPSFCl300, CHNOPSFCl500, CHNOPSFClNaKMgCa500, and the comprehensive set.

In the case of subsets except for CHNOPSFClNaKMgCa500, all molecules were plotted. On the other hand, for CHNOPSFClNaKMgCa500 subset and the complete molecular set, molecules with physically and chemically improbable values, similar to the case of HOMO, were excluded.

The horizontal axis denotes the PM6/\!/PM6 HOMO-LUMO energy gap, whereas the vertical axis exhibits the frequency distribution of HOMO-LUMO energy gap values procured from B3LYP/6-31G*//PM6. Furthermore, a linear regression analysis was conducted to assess the correlations.

The CHON300 dataset, shown in the upper left corner, is composed of C, H, O, and N elements without salt and has a molecular weight below 300. A noticeable peak is observed at roughly $-9$ eV, $-6$ eV for PM6/\!/PM6 compared to \dbname{}. Moreover, the coefficient of determination is 0.882, reflecting an exceptionally high correlation as determined by the linear regression analysis. For CHON500, CHNOPSFCl300, CHNOPSFCl500 subsets, and the entire collection, the coefficients of determination amount to 0.892, 0.841, 0.857, 0.885, and 0.803, respectively. Compared to HOMO and LUMO energies, coefficients of determination for HOMO-LUMO energy gaps exhibit marginally inferior performance. Nonetheless, the outcome is anticipated, and the phenomenon of ``lucky cancellation'' does not appear to transpire.

While outliers were present in all subsets, their overall shape remained largely unchanged. The locations and shapes of the peak values were almost identical across all subsets. However, the peak height increased with the number of molecules in each subset. When plotting all molecules, outliers became more frequent, which is likely due to the growing number of atom species. This is expected, as calculations tend to be more complex with heavier atoms; few molecules containing heavy atoms are found in PubChem. As a result, the peak values and coefficients of determination remained relatively stable. In any event, a thorough examination is warranted.

\begin{figure}[h!]
\caption{The same figure as in Fig.~\ref{graph:homolumogap}, viewed
from a 45-degree angle. The plot demonstrates an almost linear relationship, suggesting that the number of molecules with substantial differences in HOMO-LUMO energy gap values, depending on the calculation method, as observed in the two-dimensional plots, is minimal.}
    \label{graph:homolumogap_3d}
    \centering
    \begin{tabular}{c c}
      \includegraphics[width=0.45\textwidth]{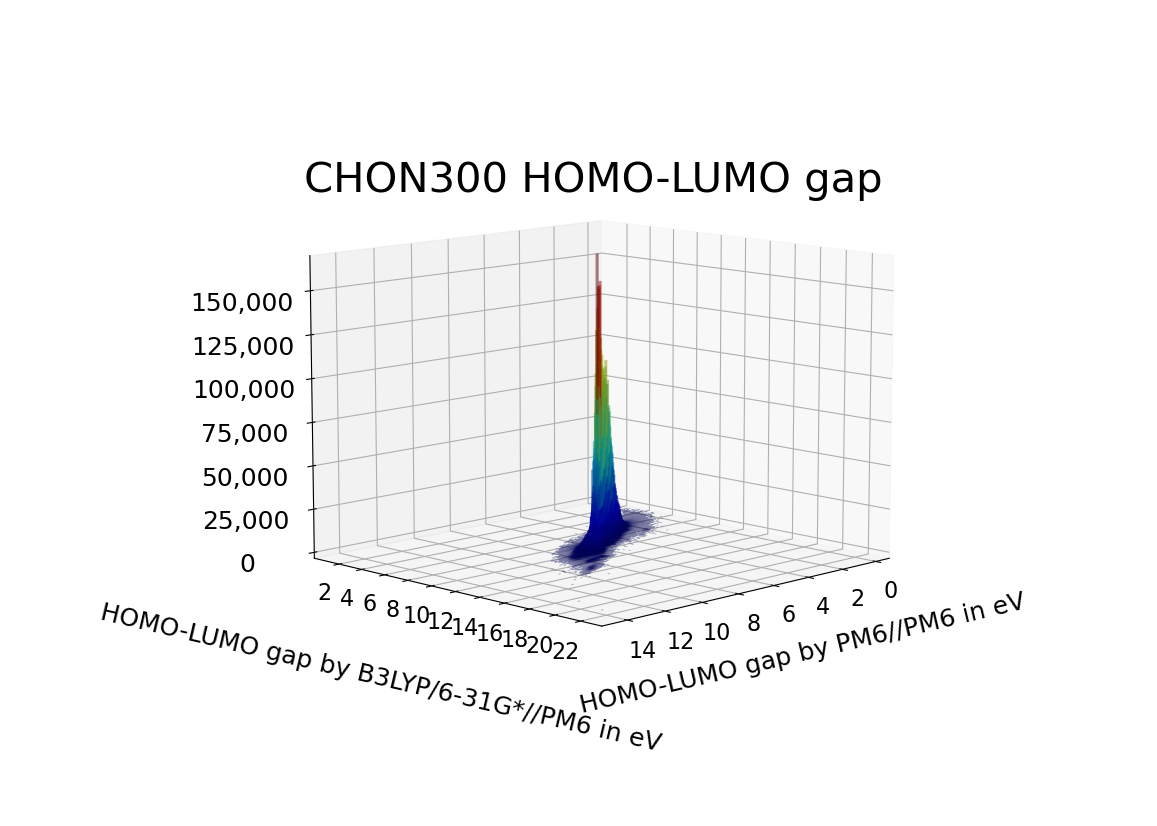} & \includegraphics[width=0.45\textwidth]{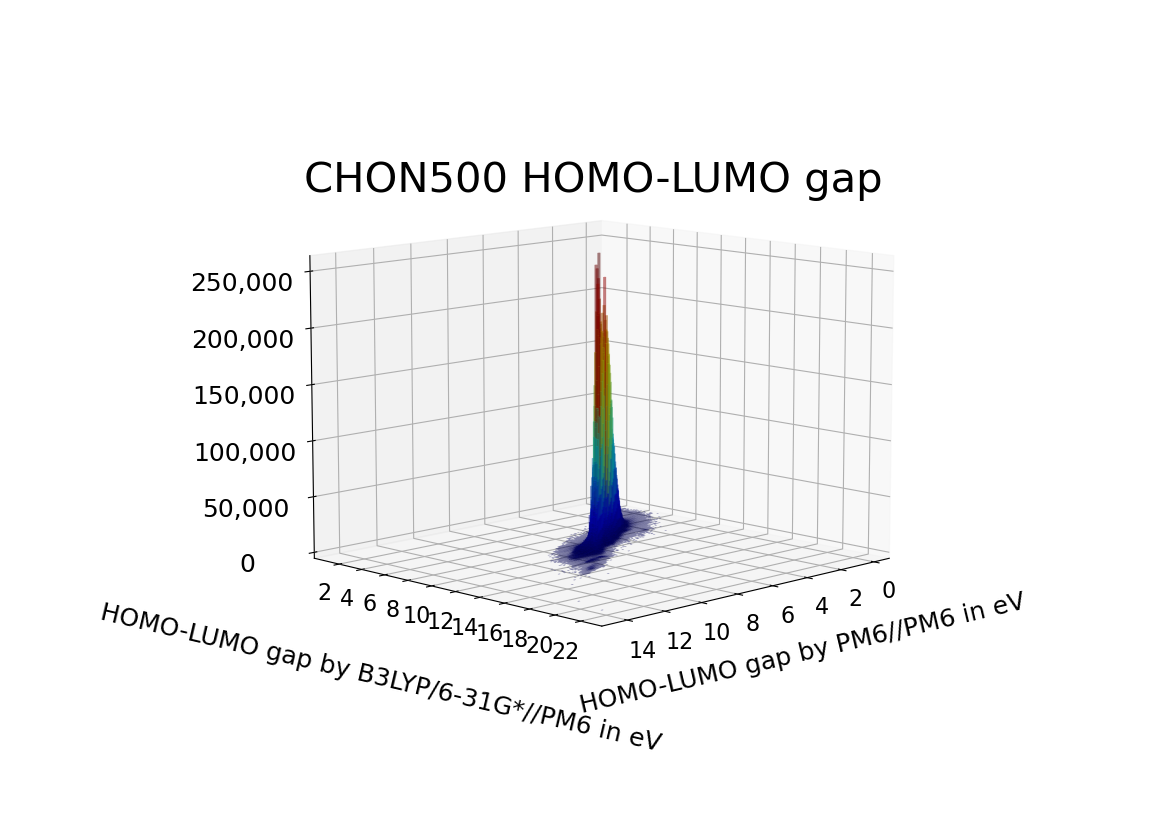} \\ 
      \includegraphics[width=0.45\textwidth]{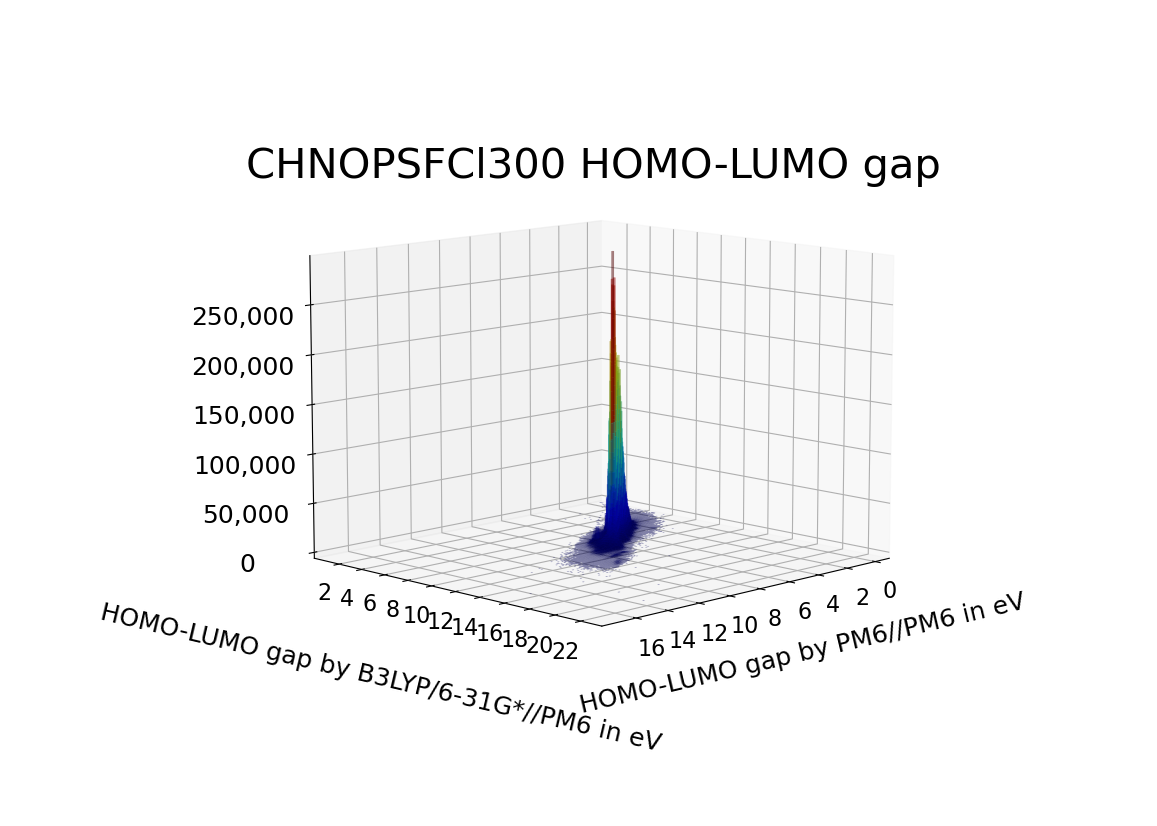} & \includegraphics[width=0.45\textwidth]{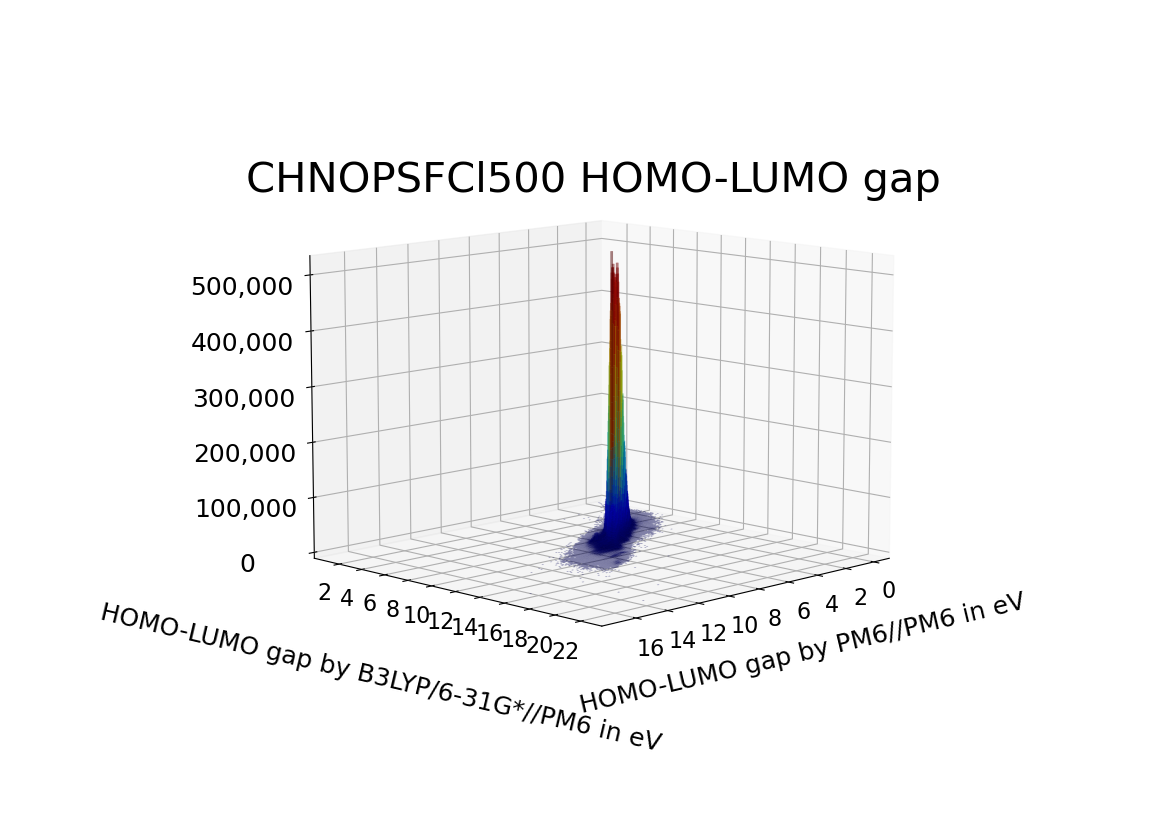} \\
      \includegraphics[width=0.45\textwidth]{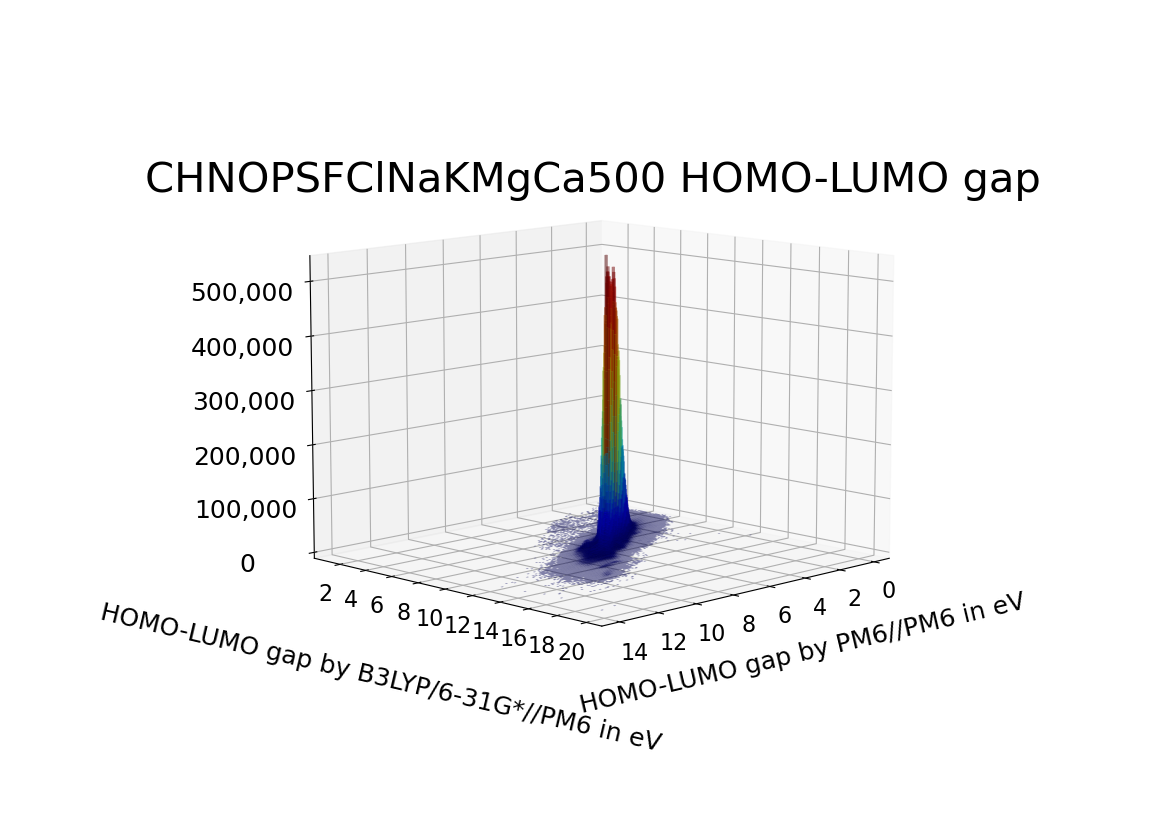} & 
      \includegraphics[width=0.45\textwidth]{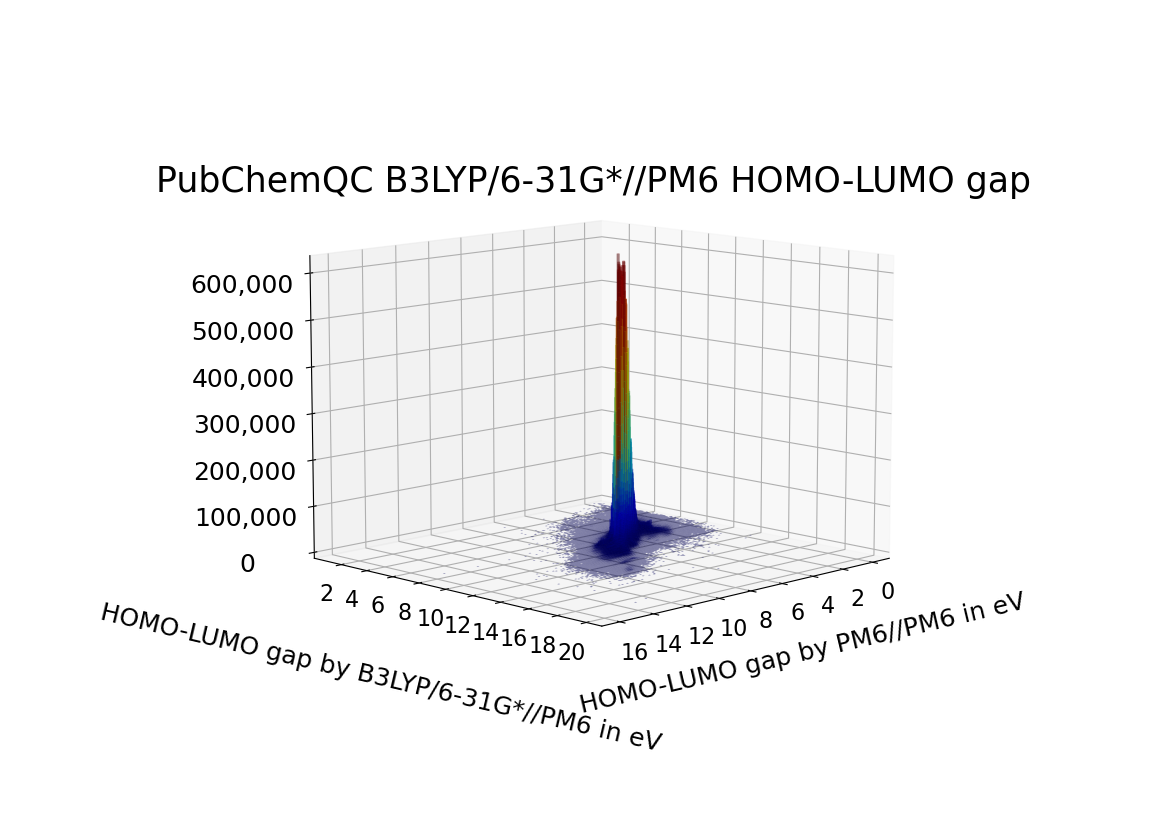} \\
    \end{tabular}
\end{figure}

Fig.\ref{graph:homolumogap_3d} presents the same data as in Fig.\ref{graph:homolumogap} but from a 45-degree angle perspective. It clearly reveals that the number of outliers is quite small, and the peaks are well-defined for all sets.

\clearpage

\subsection{Comparison of dipole moments}
\begin{figure}[h!]
\caption{The figure presents a 2D histogram and heatmap of correlated dipole moment calculated using PM6/\!/PM6 and \dbname{} methods. The horizontal axis signifies PM6/\!/PM6 dipole moment, while the vertical axis exhibits the frequency distribution of B3LYP/6-31G*/\!/PM6 dipole moment. It is noteworthy that the data point dispersion escalates with the molecule's atom count. The coefficients of determination are 0.929 to 0.945.}
    \centering
    \begin{tabular}{c c}
      \includegraphics[width=0.45\textwidth]{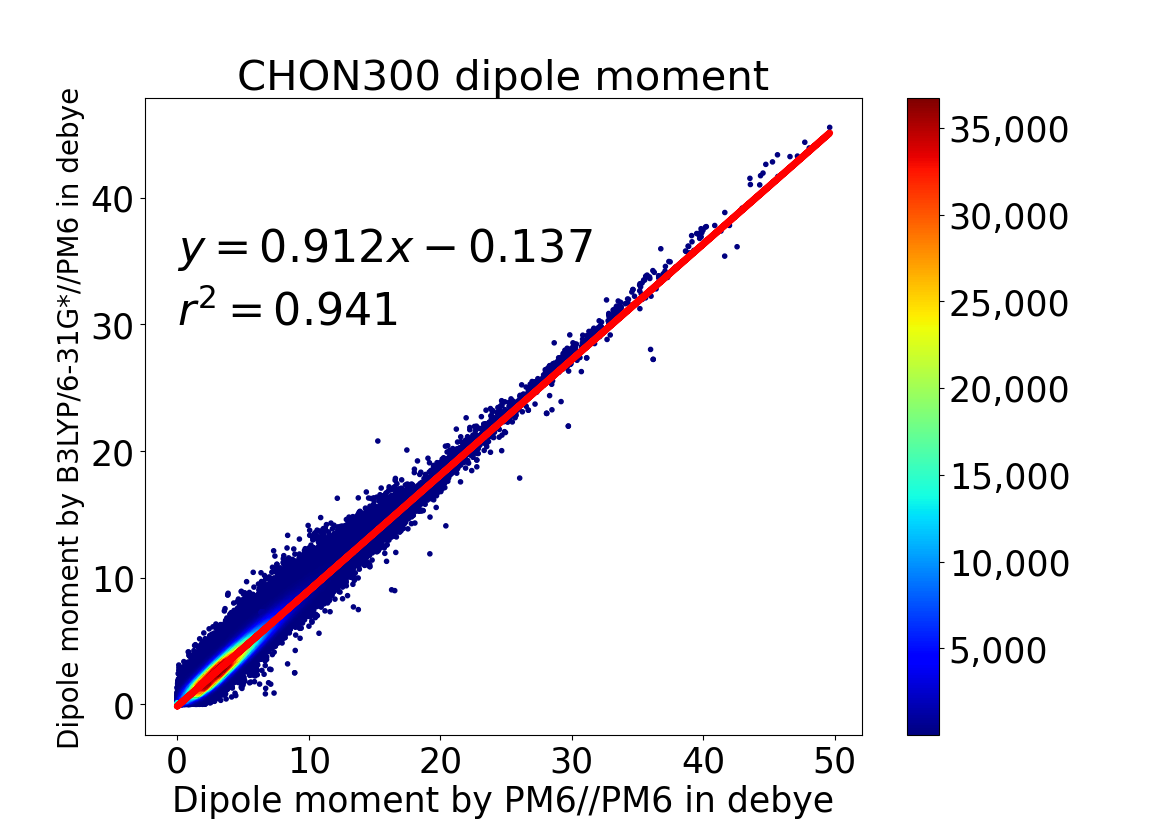} & \includegraphics[width=0.45\textwidth]{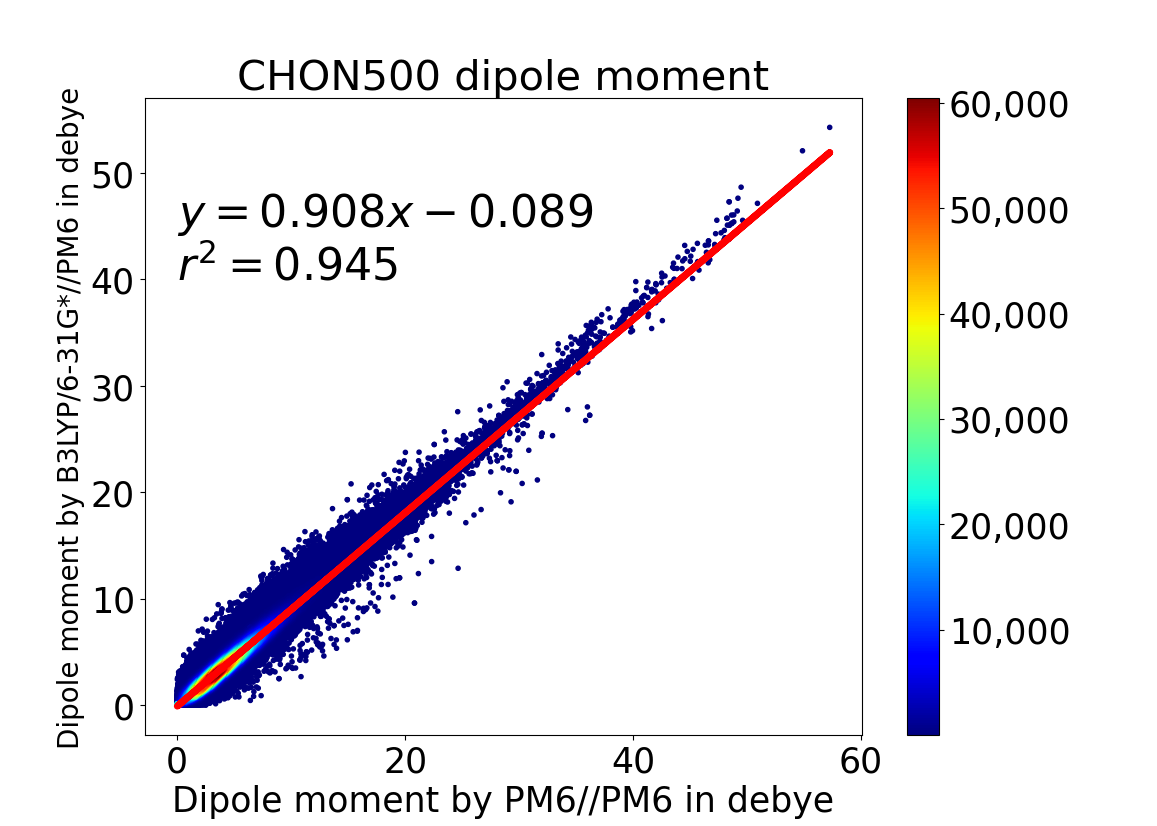} \\ 
      \includegraphics[width=0.45\textwidth]{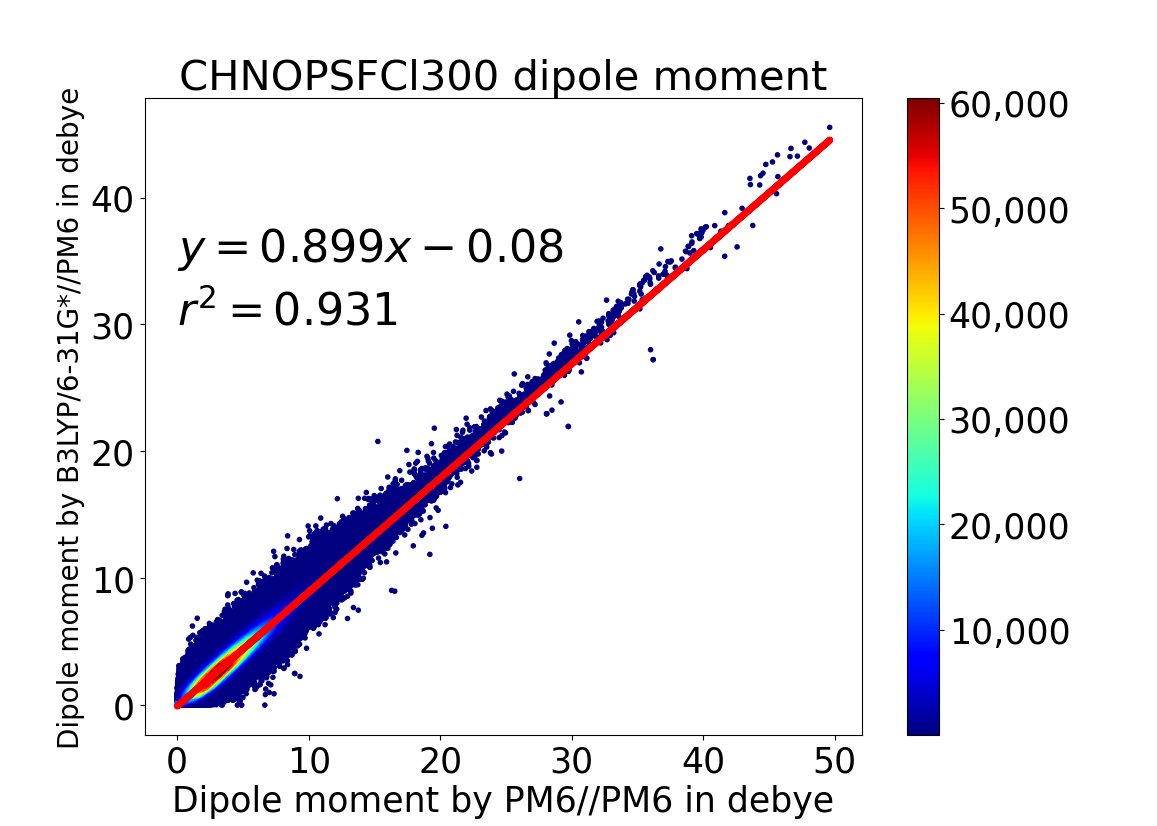} & \includegraphics[width=0.45\textwidth]{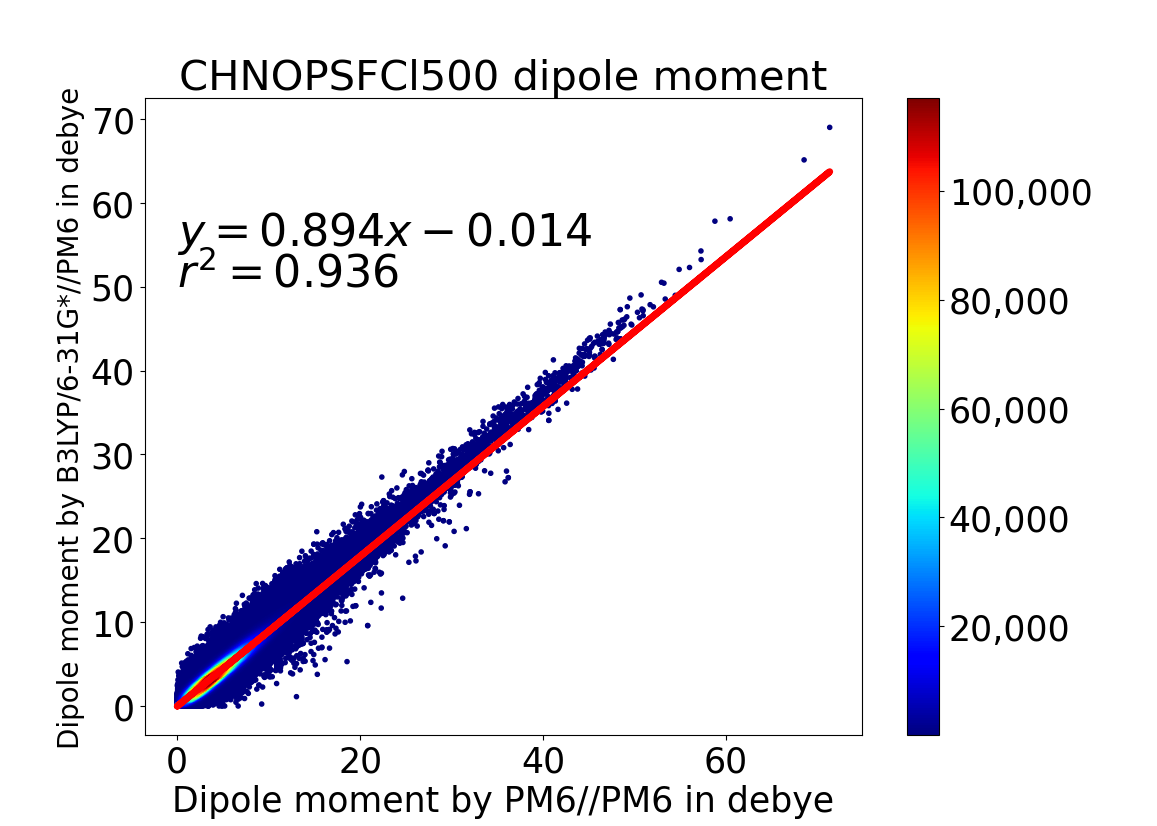} \\
      \includegraphics[width=0.45\textwidth]{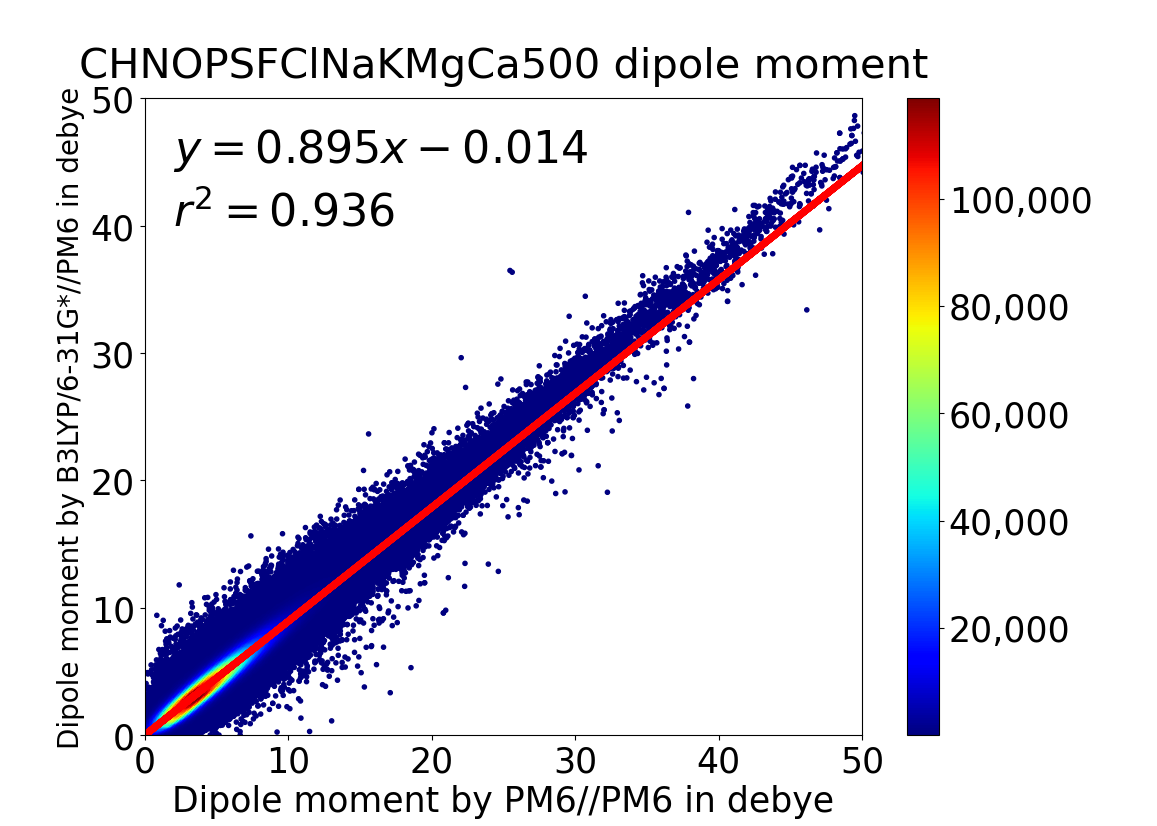} & 
      \includegraphics[width=0.45\textwidth]{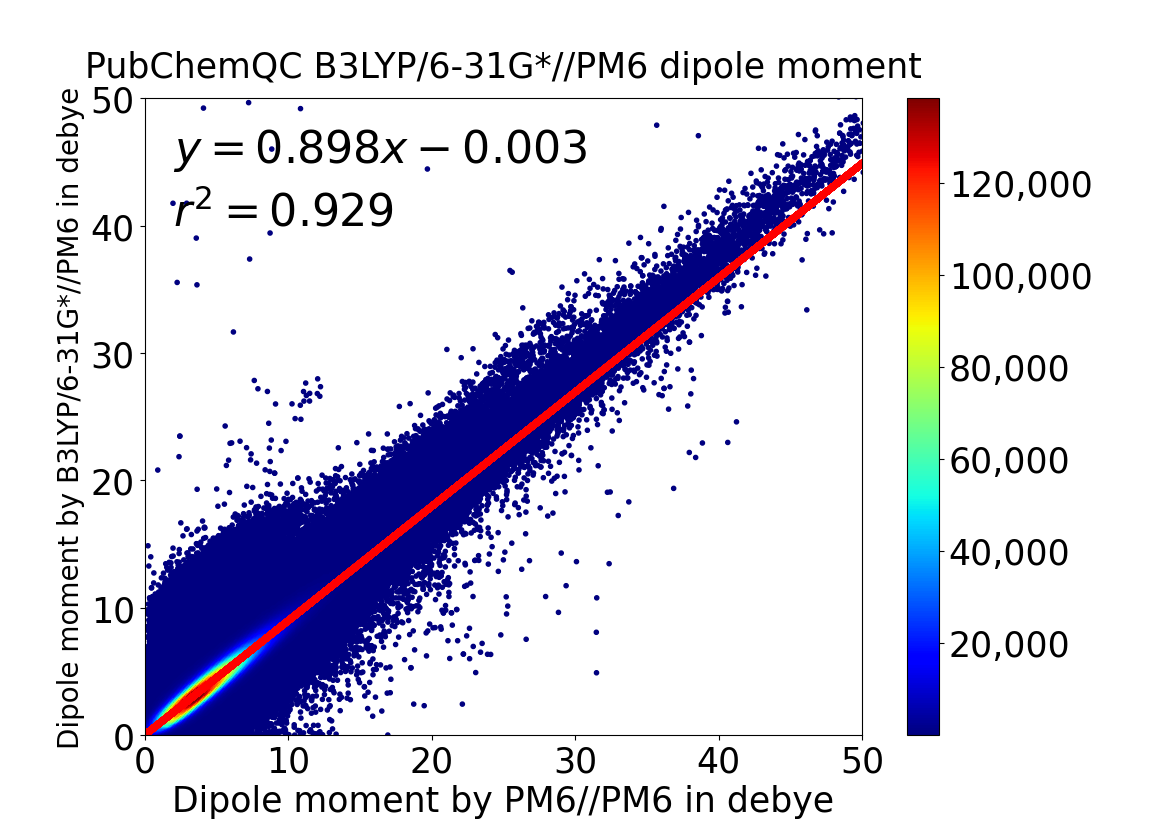} \\
    \end{tabular}
    \label{graph:dipole}
\end{figure}

The juxtaposition of dipole moment between the \dbname{} and PM6/\!/PM6 methods is presented in Fig.~\ref{graph:dipole}, showcasing a two-dimensional histogram and heatmap that illustrate the correlation of dipole moments with a bin size of $0.1$ debye each. These energies were ascertained utilizing both PM6/\!/PM6 and \dbname{} techniques for the entire collection and all subsets, encompassing CHON300, CHON500, CHNOPSFCl300, CHNOPSFCl500, CHNOPSFClNaKMgCa500, and the comprehensive set. We did not plot dipole moments surpassing $50$ debye for the CHNOPSFClNaKMgCa500 subset and the \dbname{} set. Moreover, molecules exhibiting a dipole moment greater than 10 debye are considered unrealistic.

For subsets, all molecules are plotted except for those in the CHNOPSFClNaKMgCa500 subset. In contrast, for the CHNOPSFClNaKMgCa500 subset and the entire molecular set, molecules exhibiting physically and chemically implausible values, similar to the case of HOMO, have been excluded. 

The horizontal axis denotes the dipole moment optimized using the PM6/\!/PM6 method, while the vertical axis displays the frequency distribution of dipole moment values by \dbname{}. In addition, a linear regression analysis was conducted to evaluate the correlations.

For all instances, dipole moments exhibit a strong correlation between PM6/\!/PM6 and \dbname{}. Given that the same molecular geometries were employed for both calculations, this outcome is anticipated, as the dipole moment is predominantly influenced by molecular geometry rather than the disparities in calculation methods. (A notable exception is carbon monoxide, though the direction of the dipole moment is not considered). For the CHON300, CHON500, CHNOPSFCl300, CHNOPSFCl500 subsets, and the entire collection, the coefficients of determination amount to 0.941, 0.945, 0.931, 0.936, 0.936, and 0.929, respectively.

\begin{figure}[h!]
\caption{The same figure as in Fig.~\ref{graph:dipole}, observed from a 45-degree angle. We solely plot instances where dipole moments are less than 10 debye, as determined by PM6/\!/PM6 or \dbname{}, in order to facilitate an effortless examination of the tendency. The plot exhibits an almost linear relationship, implying that the quantity of molecules with considerable discrepancies in dipole values, contingent upon the calculation method, as discerned in the two-dimensional plots, is minimal.}
    \label{graph:dipole_3d}
    \centering
    \begin{tabular}{c c}
      \includegraphics[width=0.45\textwidth]{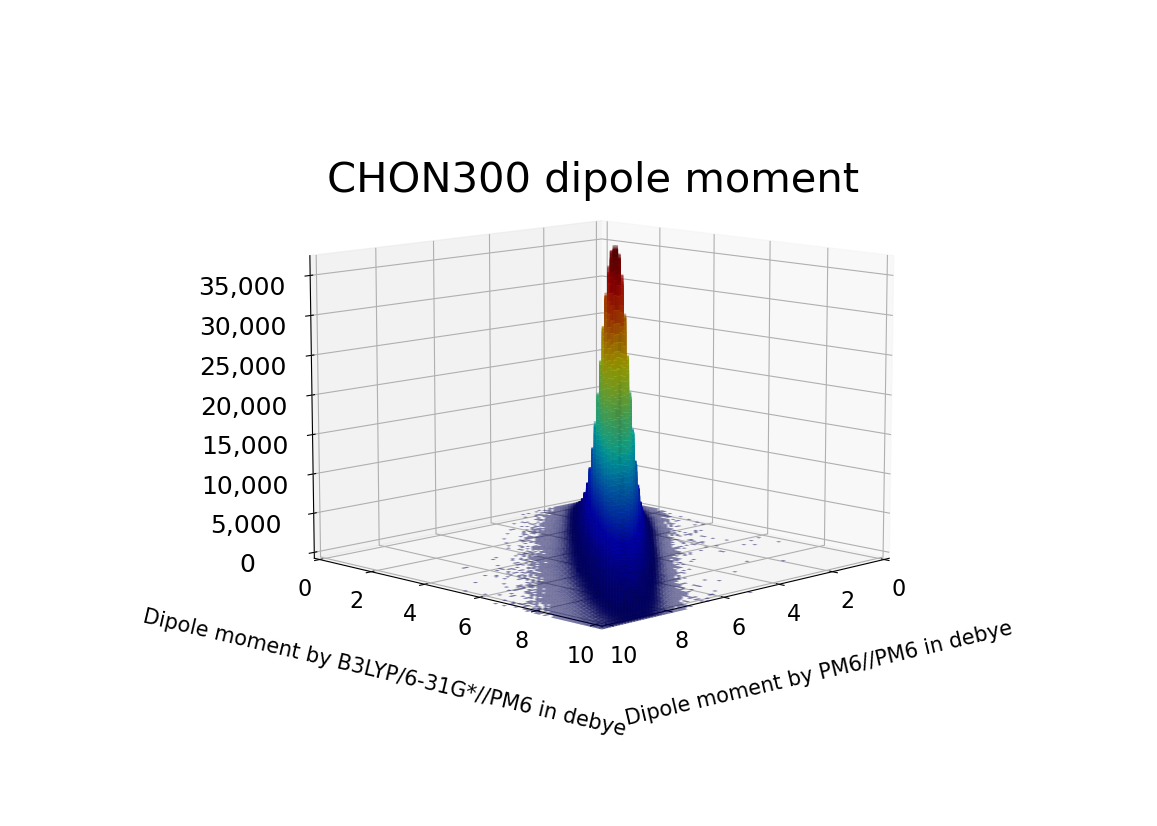} & \includegraphics[width=0.45\textwidth]{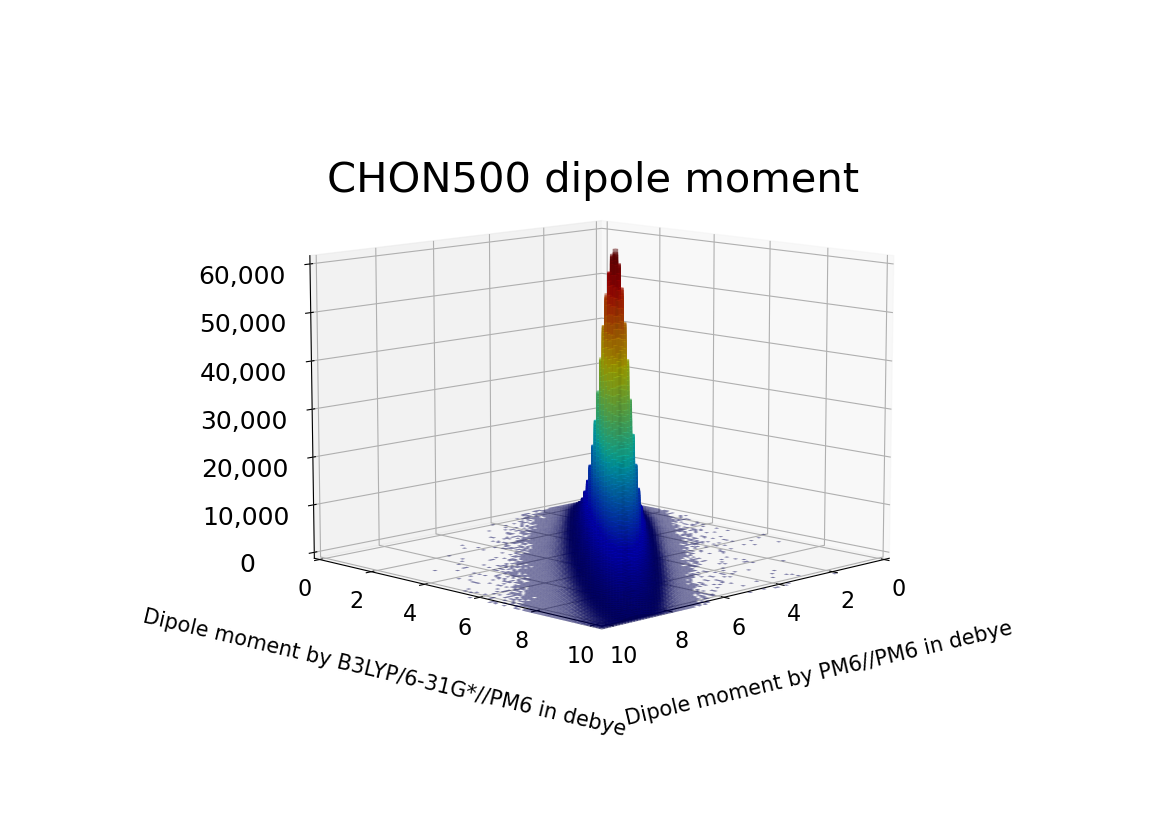} \\ 
      \includegraphics[width=0.45\textwidth]{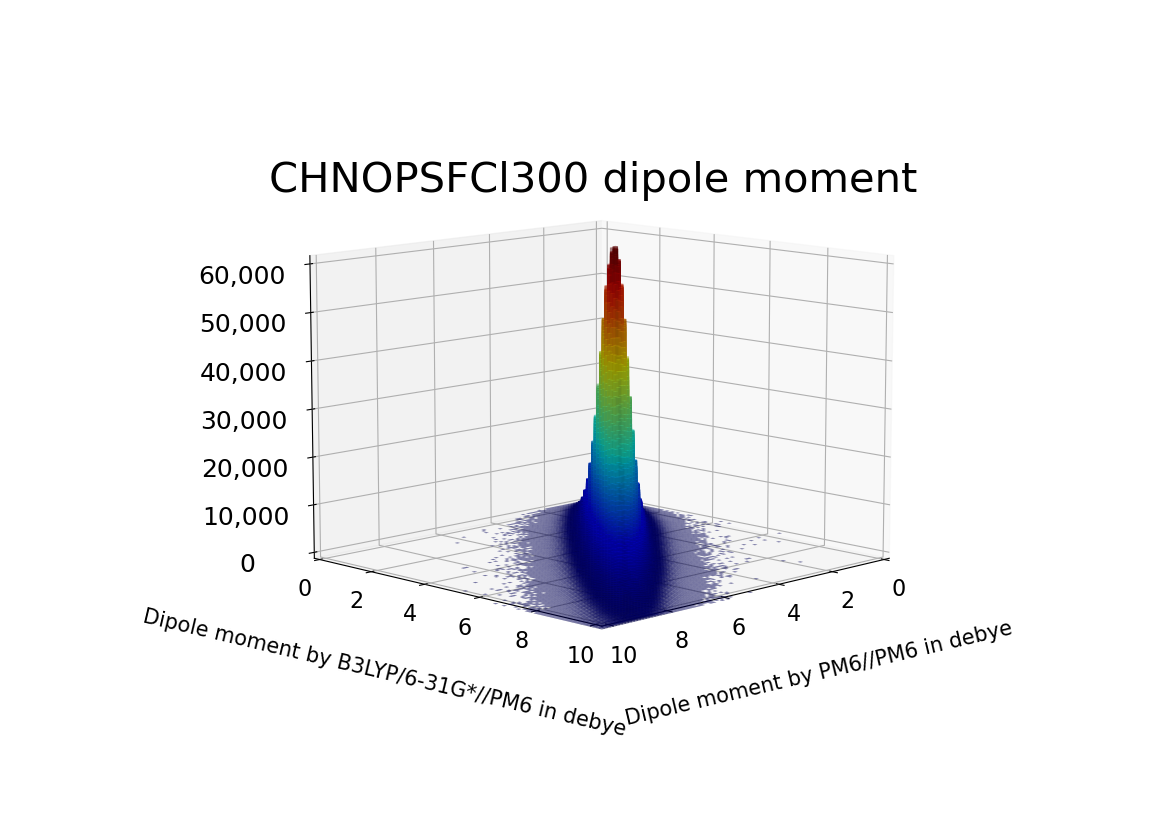} & \includegraphics[width=0.45\textwidth]{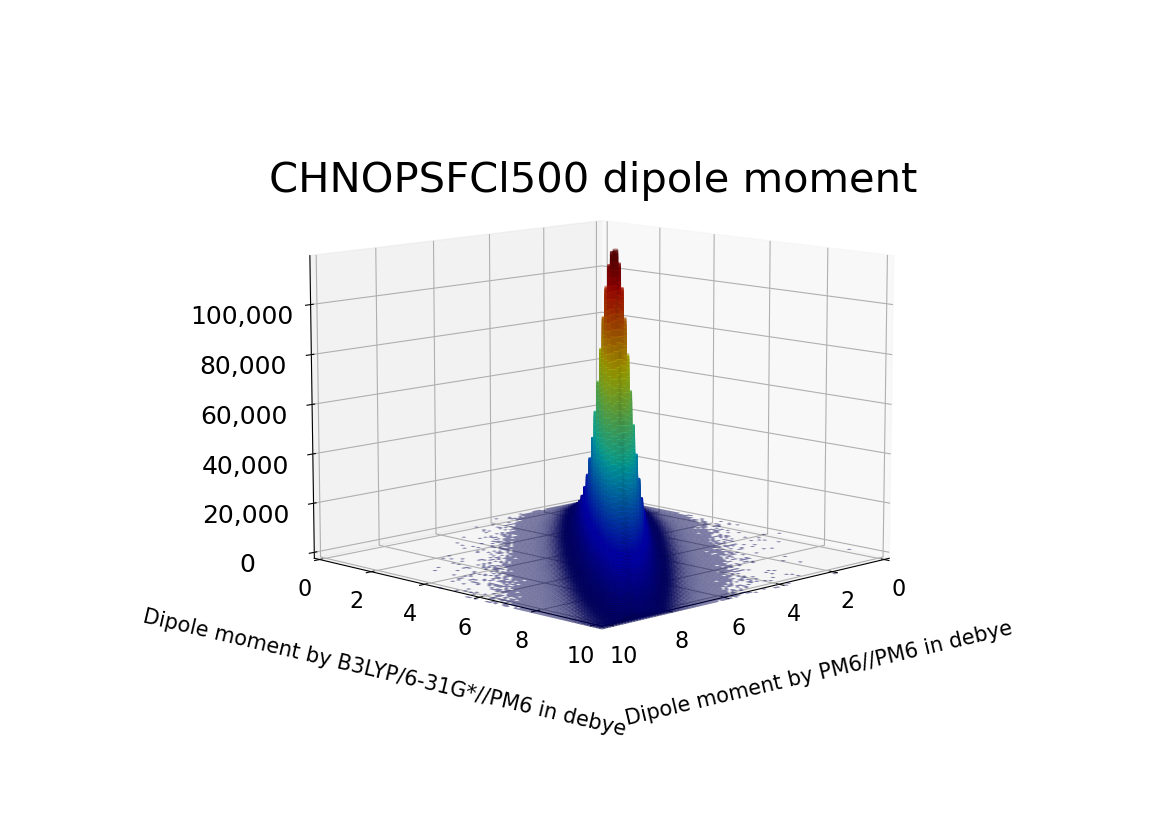} \\
      \includegraphics[width=0.45\textwidth]{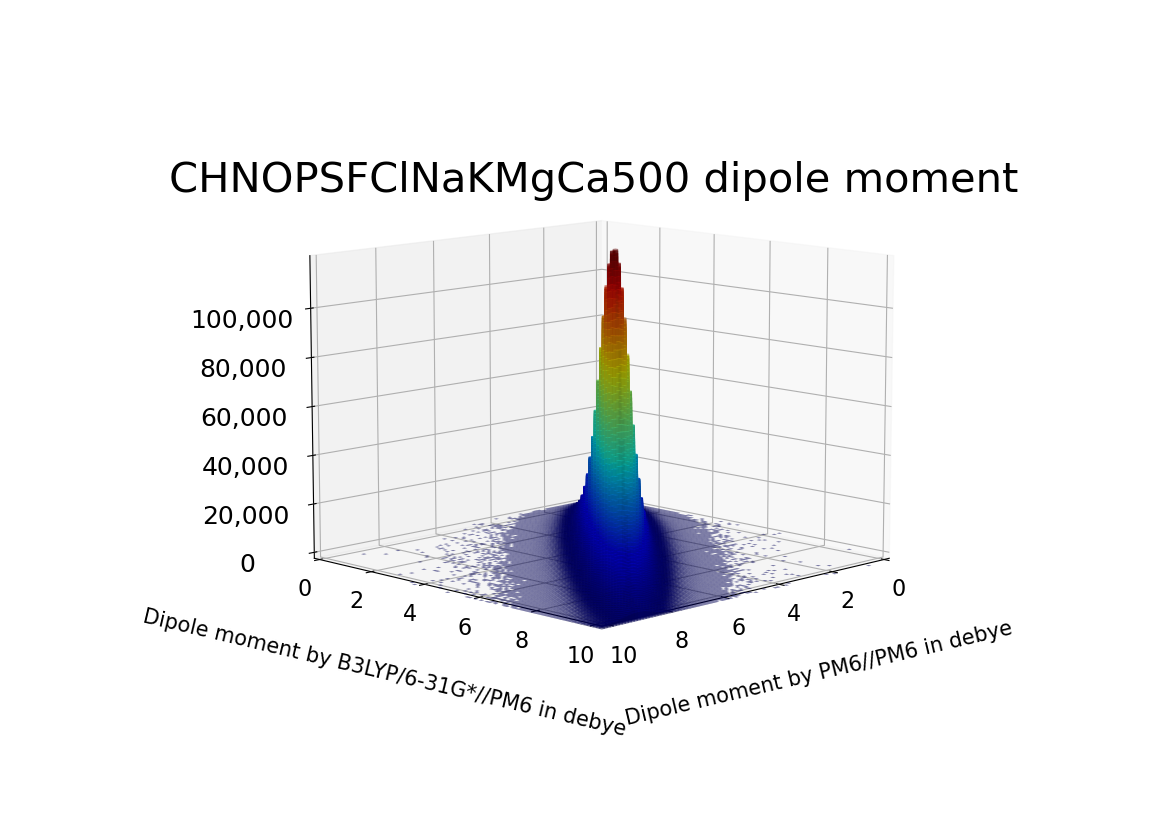} & 
      \includegraphics[width=0.45\textwidth]{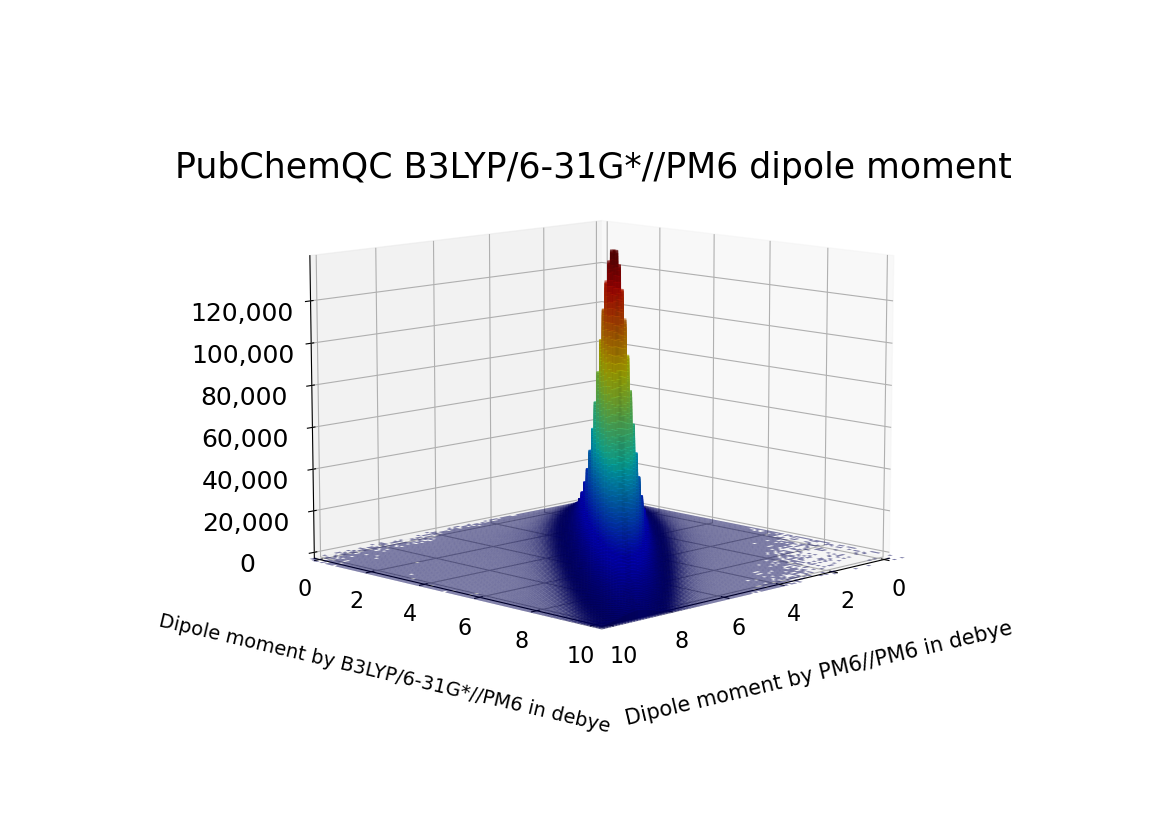} \\
    \end{tabular}
\end{figure}

Fig.\ref{graph:dipole_3d} illustrates the same data as in Fig.\ref{graph:dipole}, but from a 45-degree angle. It distinctly shows that the number of outliers is minimal, and the peaks are sharply defined for all sets.

\clearpage

\subsection{Discussion}
The results of our study can be compared with those of Aires-de-Sousa \etal{}~\cite{doi:10.1021/acs.jcim.6b00340}, who investigated 111,725 organic molecules comprising C, H, B, N, O, F, Si, P, S, Cl, Se, and Br. In their research, the geometry of the molecules was optimized with PM6 or PM7, and the electronic structure was calculated using B3LYP/6-31G* (DFT). The determination of the correlation ($R^2$) for HOMO between PM7 and DFT was 0.78 using a random subset of 84 molecules from the validation data set, while our \dbname{} obtained a determination of coefficient, calculated between PM6 and DFT, ranging from 0.877 to 0.907. For LUMO, their calculation yielded 0.895, whereas \dbname{} determined the coefficient, calculated between PM6 and DFT, ranging from 0.821 to 0.905. For the HOMO-LUMO energy gap, their result was 0.652, whereas our results calculated a determination of coefficient, ranging from 0.803 to 0.882. 

Despite analyzing 86 million molecules and employing disparate methodologies (PM6 and PM7), the current investigation has revealed consistent and slightly improved correlations for HOMO, LUMO, and HOMO-LUMO energy gaps between PM6/\!/PM6 and \dbname{} results. The application of machine learning techniques, as demonstrated by Aires-de-Sousa \etal{}~\cite{doi:10.1021/acs.jcim.6b00340}, facilitates the accurate prediction of HOMO, LUMO, and HOMO-LUMO energy gaps, leveraging the PubChemQC \dbname{} dataset, which already exhibits strong correlations with B3LYP/6-31G* results. Furthermore, the adoption of a linear regression formula for the HOMO-LUMO energy gap in eV is viable: 
Additionally, the implementation of a linear regression formula for the HOMO-LUMO energy gap in eV proves feasible: 
\[ E_{\text{(HOMO-LUMO energy gap by B3LYP/6-31G*/\!/PM6)}} = 1.016 E_{\text{(HOMO-LUMO energy gap by PM6/\!/PM6)}} - 3.725. 
\]
Analogous formulas have already been demonstrated and can be devised for other electronic properties, thereby expanding the range of computational methods available for predicting and understanding molecular characteristics.

Nevertheless, it remains uncertain whether the equation maintains its efficacy when applied to PM6-calculated electronic structures with arbitrary molecular geometries, particularly in cases where the molecular structure has not undergone optimization through the PM6 method.

Generally, the objective is to predict molecular properties when the molecule is optimized using a more advanced level of theory, such as $\omega$B97M-V/cc-pVTZ. However, geometry optimization and conformer searches are exceedingly time-consuming processes. It is worth noting that properties such as the HOMO-LUMO energy gap are obtained automatically during the geometry optimization procedure.

Consequently, the ability to predict electronic structures solely based on SMILES representations is a highly desirable approach.

Recently, Hu \etal{}~\cite{2103.09430}, a contest was held to estimate the HOMO-LUMO energy gaps of 3.8 million molecules using previous datasets~\cite{PubChemQC2017}, using 4 hours, using a single GPU/TPU/IPU and single CPU~\cite{OGB-LSC2002-rules}. With only SMILES as the molecular information, they estimated the HOMO-LUMO energy gap using the B3LYP/6-31G*/\!/B3LYP/6-31G* method. The most accurate model, GIN-virtual, achieved a mean absolute error (MAE) of 0.11 eV. Even with only 10\% of the training data, GIN-virtual obtained an MAE of 0.15 eV. In 2022, the Open Graph Benchmark Large-Scale Challenge (OGB-LSC) took place, in which molecular geometries were added to the reference data. The WeLoveGraph team's GPS++ was declared the winner with an MAE of 0.072 eV for the HOMO-LUMO energy gap estimation~\cite{2212.02229}. Fascinatingly, this result manifests a notable contrast when compared to the conventional chemical accuracy standards~\cite{RevModPhys.71.1267}, which is designated as 1 kcal/mol or equivalently, 0.0432 eV, as the obtained outcome is approximately twice the standard value.


In contrast to Aires-de-Sousa \etal{}'s result, which exhibited an MAE of 0.206 eV, the WeLoveGraph team's findings were three times more accurate for 350 wider chemical space. Although a direct comparison might not be entirely informative, it underscores the importance of molecular diversity and optimized molecular geometries within the training set.

\section{How to use PubChemQC \dbname{}}
\label{sec:howtouse}

\subsection{Description of input/output data} 
We provide access to over 50 TB of compressed archives that contain both input and output files generated by the GAMESS. Within the PubChem Compounds database, there are 91,679,247 molecules and 121,494,125 CIDs. It should be noted that some CIDs have become obsolete. The archive filenames are split by CID, with each file containing a range of 25,000 CIDs, such as 1 to 25,000, 25,001 to 50,000, and so on. Consequently, the filenames have the format {\tt Compound\_000XXXXXX\_000YYYYYY.tar}, where {\tt XXXXXX} and {\tt YYYYYY} represent the starting and ending CIDs in increments of 25,000. In total, there are 4,856 files. Additionally, we provide md5sum and sha256sum for each tar file to ensure data integrity.

To illustrate the integrity-checking process, consider the following example. After downloading all the files, the following files would be present: {\tt Compound\_000000001\_000025000.tar}, {\tt Compound\_000000001\_000025000.tar.md5sum}, and \\
{\tt Compound\_000000001\_000025000.tar.sha256sum}. \\
The integrity of {\tt Compound\_000000001\_000025000.tar} can be verified using md5sum and sha256sum as follows:\\
\\
\begin{myverb}
$ md5sum Compound_000000001_000025000.tar 
f9f4b9875e4c3ea5eecc8a8187036422 Compound_000000001_000025000.tar 
$ cat Compound_000000001_000025000.tar.md5sum 
f9f4b9875e4c3ea5eecc8a8187036422 Compound_000000001_000025000.tar 
$ cat Compound_000000001_000025000.tar.sha256sum
e4a112d2acc33115a5cf98e10f6fe8513be8744560317ffcc213f5901cfeea7a Compound_000000001_000025000.tar
$ cat Compound_000000001_000025000.tar.sha256sum
e4a112d2acc33115a5cf98e10f6fe8513be8744560317ffcc213f5901cfeea7a Compound_000000001_000025000.tar
\end{myverb}

Now we extract the {\tt Compound\_000000001\_000025000.tar} as follows:\\

\begin{myverb}
$ tar xvf Compound_000000001_000025000.tar
Compound_000000001_000025000/000000001/000000001.B3LYP@PM6.S0.inp
Compound_000000001_000025000/000000001/000000001.B3LYP@PM6.S0.out.xz
Compound_000000001_000025000/000000003/000000003.B3LYP@PM6.S0.inp
Compound_000000001_000025000/000000003/000000003.B3LYP@PM6.S0.out.xz
Compound_000000001_000025000/000000004/000000004.B3LYP@PM6.S0.inp
Compound_000000001_000025000/000000004/000000004.B3LYP@PM6.S0.out.xz
Compound_000000001_000025000/000000005/000000005.B3LYP@PM6.S0.inp
...
Compound_000000001_000025000/000024998/000024998.B3LYP@PM6.S0.out.xz
Compound_000000001_000025000/000024999/000024999.B3LYP@PM6.S0.inp
Compound_000000001_000025000/000024999/000024999.B3LYP@PM6.S0.out.xz
\end{myverb}
\\

After extracting the {\tt Compound\_000000001\_000025000.tar} file, we can see many input files with the filename extension ``.inp.'' These are the GAMESS input files for the corresponding CIDs in the archive. Similarly, output files with the filename extension ``.out.xz'' contain the GAMESS output data compressed by ``xz'' for the corresponding CIDs. The typical header of input files are follows:\\

\begin{myverb}
 $CONTRL RUNTYP=ENERGY DFTTYP=B3LYPV1R $END
 $SYSTEM MWORDS=200 $END
 $SCF DIRSCF=.T. FDIFF=.T. $END
 $BASIS GBASIS=N31 NGAUSS=6 NDFUNC=1 $END

 $DATA
PUBCHEM 0000XXXXX B3LYP 6-31G(d) at PM6 optimized geometry
...
 $END
\end{myverb}
\\
Indeed, from the output files, we can extract valuable information about the molecules, including but not limited to:

\begin{itemize} 
\item Total energy calculated using the B3LYP method and its components (one electron and two electron energy)
\item Information about the basis set used 
\item Energies of all molecular orbitals 
\item The molecular orbital (MO) coefficients 
\item Populations in each atomic orbital, analyzed using both Mulliken and Löwdin methods \item Bond order and valence analysis 
\item The dipole moment of the molecule 
\end{itemize}
The information provides a comprehensive understanding of the electronic properties and behavior of the molecules under study.
\\
We can also obtain information using {\tt cclib}~\cite{cclib} as follows:\\

\begin{myverb}
$ ccget --list 0000XXXXX.B3LYP@PM6.S0.out
Attempting to read 0000XXXXX.B3LYP@PM6.S0.out
cclib can parse the following attributes from 0000XXXXX.B3LYP@PM6.S0.out:
  aonames
  atombasis
  atomcharges
  atomcoords
  atomnos
  charge
  coreelectrons
  gbasis
  homos
  metadata
  mocoeffs
  moenergies
  moments
  mosyms
  mult
  natom
  nbasis
  nmo
  scfenergies
  scftargets
  scfvalues
\end{myverb}
\\
Additionally, cclib allows you to convert the output file into a more accessible format, such as {\tt JSON}~\cite{rfc8259}, as shown in the following example:\\
\begin{myverb}
$ ccwrite json 0000XXXXX.B3LYP@PM6.S0.out
\end{myverb}
\\
The conversion to {\tt JSON} format streamlines subsequent analysis and data processing, thereby simplifying integration with diverse tools and workflows. Nevertheless, it is important to note that the extraction of only partial data is possible from the output file; for instance, the resulting {\tt JSON} file does not encompass single-electron and double-electron energy values.

We can acquire the three-dimensional coordinates of molecules using Open Babel~\cite{OpenBabel}. Let us demonstrate how to accomplish this with an example, specifically, {\tt 000000004.B3LYP@PM6.S0.inp} as follows: \\

\begin{myverb}
$ obabel -igamin 000000004.B3LYP@PM6.S0.inp -o xyz
14
PUBCHEM 000000004 B3LYP 6-31G(d) at PM6 optimized geometry
C          0.85391        0.09871       -0.00694
C          2.37832        0.02676       -0.05044
C          2.91538       -1.03693       -1.04053
N          2.46323       -0.71745       -2.42461
O          2.91936        1.30992       -0.37318
H          0.40326       -0.85848        0.26711
H          0.43305        0.41215       -0.96943
H          0.53147        0.84787        0.72872
H          2.79780       -0.15591        0.96766
H          2.62354       -2.04989       -0.69557
H          4.02883       -0.99557       -1.01803
H          1.48775       -0.97228       -2.54708
H          2.99913       -1.25300       -3.10068
H          2.77975        1.50805       -1.32926
1 molecule converted
\end{myverb}
\\
Note that we did not perform geometry optimization, allowing us to use the input file directly. The molecular geometries can be obtained by utilizing the databases with Docker integration.

\subsection{Running Databases with Docker Integration}

Besides the main database, we offer a Docker~\cite{merkel2014docker}-compatible version to facilitate seamless deployment. Our analysis was conducted using {\tt PostgreSQL}~\cite{10.5555/548095} and {\tt PostgREST}~\cite{PostgREST} as the principal tools. All data is maintained in the {\tt JSON} format. Employing {\tt PostgreSQL} queries allows for expeditious data extraction, while {\tt PostgREST} queries produce {\tt JSON} output, which can be efficiently interpreted with a {\tt JSON} parser.

Essential alterations for the Docker version entailed: The output files were processed using cclib, and selective trimming of {\tt atombasis} and {\tt mocoeffs} was performed owing to their considerable dimensions.

Despite these modifications, the resulting database remains sizable, and we have divided it into five distinct sub-datasets:

The first subset consists of molecules containing C, H, O, and N~\cite{CHON}, with a molecular weight of less than 300, excluding salts. The second subset encompasses molecules composed of C, H, O, and N, with a molecular weight below 500, and also omits salts.

The third subset includes molecules featuring C, H, N, O, P, S~\cite{10.5555/1198994}, F, and Cl, with a molecular weight under 300, and without salts. The fourth subset comprises molecules containing C, H, N, O, P, S, F, and Cl, with a molecular weight less than 500, and excludes salts. Finally, the fifth subset consists of molecules containing C, H, N, O, P, S, F, Cl, Na, K, Mg, and Ca, with a molecular weight below 500~\cite{}. The collection represents the essential elements present in the human body~\cite{ZORODDU2019120}, except for fluorine.

It is of significance to explicitly mention that no curation procedure was implemented on the \dbname{} dataset. As a result, there may be instances where certain molecules have unusual HOMO, LUMO energies and dipole moments. It is recommended that users manually filter such molecules according to their individual requirements.

In summary, we offer the following databases along with their respective sha256sums:
\begin{itemize}
\item {\tt b3lyp\_pm6\_CHNOPSFCl300noSalt\_ver1.0.1-postgrest-docker-compose.tar.xz}, \\ sha256sum: {\tt bfcb5916715eeb93d536e20bff111921ca4f05d5b5b82488d6f46f0d94d1dbfd}
\item {\tt b3lyp\_pm6\_CHNOPSFCl500noSalt\_ver1.0.1-postgrest-docker-compose.tar.xz}, \\ sha256sum: {\tt e95c4b5dbe0fc7630d552bb4ec9f49e6457a99940640f6d3039b5059a9d0c706}
\item {\tt b3lyp\_pm6\_CHNOPSFClNaKMgCa500\_ver1.0.1-postgrest-docker-compose.tar.xz}, \\ sha256sum: {\tt b212db50e0da2afd7c361ed9d9114ec74f245e8efa60ae448f207161d097c26d}
\item {\tt b3lyp\_pm6\_CHON300noSalt\_ver1.0.1-postgrest-docker-compose.tar.xz}, \\ sha256sum: {\tt 51bf899a24ff790ebfff21c796917a8e1c1ffcdc12887303b82d5f739f355623}
\item {\tt b3lyp\_pm6\_CHON500noSalt\_ver1.0.1-postgrest-docker-compose.tar.xz}, \\ sha256sum: {\tt 56227542c97c1eda516012721c4ce20abde9a3f64a396c126ab73d1df6cff225}
\item {\tt b3lyp\_pm6\_ver1.0.1-postgrest-docker-compose.tar.xz}, \\ sha256sum: {\tt 
5e941a7e9482795060e59f91b0049bdf0e9764af07bde036a4aa6a077cf8acae}
\end{itemize}

Next, we will explore some applications of the database. To do so, let’s begin by expanding the smallest size,\\ {\tt b3lyp\_pm6\_CHNOPSFCl300noSalt\_ver1.0.1-postgrest-docker-compose.tar.xz}.

\subsubsection{Check, Extract, and Launch the Database}
First, we check, extract, and launch the database. To accomplish this, input the following command in a Linux terminal: \\
\\
\begin{myverb}
$ sha256sum b3lyp_pm6_CHON300noSalt_ver1.0.1-postgrest-docker-compose.tar.xz
51bf899a24ff790ebfff21c796917a8e1c1ffcdc12887303b82d5f739f355623  b3lyp_pm6_CHON300noSalt_ver1.0.1-postgrest-docker-compose.tar.xz
$ cat b3lyp_pm6_CHON300noSalt_ver1.0.1-postgrest-docker-compose.tar.xz.sha256sum
51bf899a24ff790ebfff21c796917a8e1c1ffcdc12887303b82d5f739f355623  b3lyp_pm6_CHON300noSalt_ver1.0.1-postgrest-docker-compose.tar.xz
\end{myverb}
\\

\begin{myverb}
$ tar xzJ b3lyp_pm6_CHON300noSalt_ver1.0.1-postgrest-docker-compose.tar.xz
$ cd b3lyp_pm6_CHON300noSalt-postgrest-docker-compose
$ docker-compose up -d --build
Creating network "b3lyp_pm6_chon300nosalt-postgrest-docker-compose_default" with the default driver
Creating b3lyp_pm6_chon300nosalt-postgrest-docker-compose_db_1 ... done
Creating b3lyp_pm6_chon300nosalt-postgrest-docker-compose_server_1 ... done
\end{myverb}
\\
If successful, the output message should be similar to the one provided above.

\subsubsection{Verify the table's existence and enumerate the molecules in the database}
To determine the number of registered molecules, assume the user is in the \\
{\tt b3lyp\_pm6\_CHON300noSalt-postgrest-docker-compose} subdirectory. The following command checks for the presence of the table:\\

\begin{myverb}
$ docker exec -i $(docker-compose ps -q db) psql -U pgrest db -c "\dt"
                 List of relations
 Schema |          Name           | Type  | Owner
--------+-------------------------+-------+--------
 public | b3lyp_pm6_chon300nosalt | table | pgrest
(1 row)
   
\end{myverb}
\\
Subsequently, we enumerate the molecules present in the database by:\\

\begin{myverb}
$ docker exec -i $(docker-compose ps -q db) psql -U pgrest db -c \
"SELECT count(*) FROM b3lyp_pm6_chon300nosalt"
  count
----------
 17308321
(1 row)
\end{myverb}
\\
The number 17308321 is the same as shown in the Table~\ref{tab:statistics}.

\subsection{Structures of {\tt JSON} with Docker Integration}
In the subsection, we elucidate the structure of {\tt JSON} by employing a water molecule as an example. The PubChem CID of the molecule is 962, and we utilize it to avoid the abstract nature of the dataset structure.

\subsubsection{Complete Dataset for a Water Molecule}
The dataset's top-level key is composed of three elements: "{\tt cid}", "{\tt state}", and "{\tt data}". The "{\tt cid}" element signifies the PubChem CID, whereas the "{\tt state}" element specifies the electronic state as either "{\tt S0}" or "{\tt D0}" The "{\tt data}" element incorporates all pertinent information regarding the molecule. As an example, to obtain complete details about a water molecule, one could employ the following query command:\\

\begin{myverb}
$ curl "http://localhost:3000/b3lyp_pm6_chon300nosalt?cid=eq.962" | python -m json.tool | less
[
    {
        "cid": 962,
        "state": "S0",
        "data": {
            "pubchem": {
                "cid": 962,
                "InChI": "InChI=1S/H2O/h1H2",
                "charge": 0,
                "version": "20160829",
                "B3LYP@PM6": {

...
                "multiplicity": 1,
                "Isomeric SMILES": "O",
                "molecular weight": 18.01528,
                "molecular formula": "H2O"
            }
        }
    }
]
\end{myverb}

\subsubsection{InChI and SMILES for a Water Molecule}

It is important to note that our dataset contains two InChI representations and two SMILES representations, respectively, for a water molecule. One originates from the original PubChem Compound, while the other is recalculated using PM6 optimized molecular geometries by Open Babel. These two InChI and SMILES representations may differ, and it is recommended to use the recalculated ones when considering the electronic structure and/or the 3D coordinates of the molecule's atoms.

The "{\tt data->pubchem->InChI}" and "{\tt data->pubchem->Isomeric SMILES}" field contains the original InChI and Isomeric SMILES derived from the PubChem dataset:\\

\begin{myverb}
$ curl -s "http://localhost:3000/b3lyp_pm6_chon300nosalt?\
cid=eq.962&select=data->pubchem->InChI"
[{"InChI":"InChI=1S/H2O/h1H2"}]
$ curl -s "http://localhost:3000/b3lyp_pm6_chon300nosalt?\
cid=eq.962&select=data->pubchem->Isomeric%20SMILES"
[{"Isomeric SMILES":"O"}]
\end{myverb}

It should be noted that in order to extract the "Isomeric SMILES," one needs to use "{\tt Isomeric\%20SMILES} " instead of the "{\tt Isomeric SMILES}".

InChI and SMILES by Open Babel are \\
"{\tt data->pubchem->B3LYP@PM6->openbabel->InChI}" and \\
"{\tt data->pubchem->B3LYP@PM6->openbabel->Canonical\%20SMILES}".
The queries for those keys are following:\\

\begin{myverb}
$ curl -s "http://localhost:3000/b3lyp_pm6_chon300nosalt?\
cid=eq.962&\
select=data->pubchem->\"B3LYP@PM6\"->openbabel->InChI"
[{"InChI":"InChI=1S/H2O/h1H2"}]
$ curl -s "http://localhost:3000/b3lyp_pm6_chon300nosalt?\
cid=eq.962&\
select=data->pubchem->\"B3LYP@PM6\"->openbabel->Canonical%20SMILES"
[{"Canonical SMILES":"O"}]
\end{myverb}

\subsection{Retrieving Atomic Coordinates through Query and reconstruct .xyz file}

To obtain the atomic coordinates, you can access the "{\tt pubchem->B3LYP@PM6->atoms->coords->3D}" field. Then, the three-dimensional coordinates are provided in the form of a list labeled "3D." Each element within the list corresponds to the x, y, and z coordinates of the respective atoms in \AA{}ngstrom. Below is an example demonstrating how to retrieve the coordinates for a water molecule:
\\

\begin{myverb}
$ curl -s "http://localhost:3000/b3lyp_pm6_chon300nosalt?\
cid=eq.962&select=data->pubchem->\"B3LYP@PM6\"->atoms->coords->3d" \
| python -m json.tool
[
    {
        "3d": [
            1.00389992727911,
            0.098379992886939,
            0.016349998815803797,
            1.9528298585118447,
            0.10610999233477432,
            0.0022299998635043703,
            0.7027599490669395,
            0.5304499615746316,
            -0.773189943956068
        ]
    }
]

\end{myverb}

To retrieve the atomic numbers corresponding to the obtained atomic coordinates for the aforementioned water molecule, you can access the "{\tt pubchem->B3LYP@PM6->atoms->elements->number}" field. The atomic numbers are provided as a list, arranged in a natural number sequence. Here is an example of how to retrieve the atomic numbers for the water molecule:\\

\begin{myverb}
$ curl -s "http://localhost:3000/b3lyp_pm6_chon300nosalt?\
cid=eq.962&select=data->pubchem->\"B3LYP@PM6\"->atoms->elements->number" | \
python -m json.tool
[
    {
        "number": [
            8,
            1,
            1
        ]
    }
]
\end{myverb}

Subsequently, the acquired atomic coordinates and corresponding atomic numbers are combined to generate the resulting ".xyz" file.\\

\begin{myverb}
3
water molecule PM6
O          1.00390        0.09838        0.01635
H          1.95283        0.10611        0.00223
H          0.70276        0.53045       -0.77319
\end{myverb}

Note that the geometry mentioned in the previous context was not optimized using the B3LYP/6-31G* method. Instead, the PM6 method was employed for the optimization of the geometry.

\subsection{Retrieving the Electronic structure}
The "{\tt pubchem->B3LYP@PM6->properties}" field encompasses several properties, such as the charge, HOMO, LUMO, HOMO-LUMO energy gap, orbital energies, the number of basis, the dipole moment and the orbital numbers of HOMO and LUMO. Additionally, it includes Mulliken populations and L\"owdin populations, among others.

\begin{myverb}
$ curl -s "http://localhost:3000/b3lyp_pm6_chon300nosalt?cid=eq.962&\
select=data->pubchem->\"B3LYP@PM6\"->properties" | python -m json.tool
[
    {
        "properties": {
            "charge": 0.0,
            "energy": {
...
                "alpha": {
                    "gap": 9.790656340990001,
                    "homo": -7.926676465065,
                    "lumo": 1.8639798759250006
                },
                "total": -2079.166848697776
            },
            "orbitals": {
                "homos": [
...
                "mulliken": [
                    -0.784985,
                    0.392494,
                    0.392491
                ]
            },
            "total dipole moment": 2.0276952463259854
        }
    }
]
\end{myverb}

\subsection{Custom queries}

\subsubsection{Query the entries of molecules with PubChem CIDs}
The following command shows the entries of molecules whose PubChem CIDs are between 950 and 1000:\\
\\
\begin{myverb}
$ curl 'http://localhost:3000/b3lyp_pm6_chon300nosalt?and=(cid.gte.950,cid.lte.1000)'
\end{myverb}

\subsubsection{Querying a Molecule based on HOMO-LUMO Energy gap}

The following command generates a table that displays the PubChem Compound Identifiers (CIDs), electronic states, and HOMO-LUMO energy gaps for molecules with CIDs ranging from 950 to 1000 and HOMO-LUMO energy gaps ranging from 1.1 to 5.5:\\

\begin{myverb}
$ curl "http://localhost:3000/b3lyp_pm6_chon300nosalt"\
"?select=*&"\
"and=(cid.gte.950,cid.lte.1000)&"\
"and=(data->pubchem->\"B3LYP@PM6\"->properties->energy->alpha->gap.gte.1.1,"\
"data->pubchem->\"B3LYP@PM6\"->properties->energy->alpha->gap.lte.5.5)"
\end{myverb}

The query below produces a subtable comprising of three columns: "{\tt cid}", "{\tt molecular formula}", "{\tt Canonical SMILES}", and "{\tt gap}", wherein the CIDs fall within the range of 950 to 1000, and the HOMO-LUMO energy gaps fall within the range of 1.1 to 5.5:\\

\begin{myverb}
$ curl "http://localhost:3000/b3lyp_pm6_chon300nosalt"\
"?select=cid,"\
"data->pubchem->molecular%20formula,"\
"data->pubchem->\"B3LYP@PM6\"->openbabel->Canonical%20SMILES,"\
"data->pubchem->\"B3LYP@PM6\"->properties->energy->alpha->gap&"\
"and=(cid.gte.950,cid.lte.1000)&"\
"and=(data->pubchem->\"B3LYP@PM6\"->properties->energy->alpha->gap.gte.1.1,"\
"data->pubchem->\"B3LYP@PM6\"->properties->energy->alpha->gap.lte.5.5)"
\end{myverb}

\subsubsection{Querying a Molecule based on HOMO-LUMO energy gaps and Dipole moments}

The subsequent command generates a table presenting the PubChem Compound Identifiers (CIDs), electronic states, and HOMO-LUMO energy gaps for molecules with CIDs in the range of 950 to 1000, HOMO-LUMO energy gaps between 1.1 and 5.5, and dipole moments between 0.9 and 2.5: \\

\begin{myverb}
$ curl "http://localhost:3000/b3lyp_pm6_chon300nosalt"\
"?select=*&"\
"and=(cid.gte.950,cid.lte.1000)&"\
"and=(data->pubchem->\"B3LYP@PM6\"->properties->energy->alpha->gap.gte.1.1,"\
"data->pubchem->\"B3LYP@PM6\"->properties->energy->alpha->gap.lte.5.5)&"\
"and=(data->pubchem->\"B3LYP@PM6\"->properties->total%20dipole%20moment.gte.0.9,"\
"data->pubchem->\"B3LYP@PM6\"->properties->total%20dipole%20moment.lte.2.5)"
\end{myverb}

The following query generates a subtable containing four columns: "{\tt cid}", "{\tt molecular formula}", "{\tt Canonical SMILES}", and "{\tt gap}", with the CIDs ranging from 950 to 1000, HOMO-LUMO energy gaps between 1.1 and 5.5, and dipole moments between 0.9 and 2.5:\\

\begin{myverb}
$ curl "http://localhost:3000/b3lyp_pm6_chon300nosalt"\
"?select=cid,"\
"data->pubchem->molecular%20formula,"\
"data->pubchem->\"B3LYP@PM6\"->openbabel->Canonical%20SMILES,"\
"data->pubchem->\"B3LYP@PM6\"->properties->energy->alpha->gap,"\
"data->pubchem->\"B3LYP@PM6\"->properties->total%20dipole%20moment&"\
"and=(cid.gte.950,cid.lte.1000)&"\
"and=(data->pubchem->\"B3LYP@PM6\"->properties->energy->alpha->gap.gte.1.1,"\
"data->pubchem->\"B3LYP@PM6\"->properties->energy->alpha->gap.lte.5.5)&"\
"and=(data->pubchem->\"B3LYP@PM6\"->properties->total%20dipole%20moment.gte.0.9,"\
"data->pubchem->\"B3LYP@PM6\"->properties->total%20dipole%20moment.lte.2.5)" | python -m json.tool
[
    {
        "cid": 964,
        "molecular formula": "C3H4O4",
        "Canonical SMILES": "OCC(=O)C(=O)O",
        "gap": 4.76199238375,
        "total dipole moment": 2.3771959354939596
    },
    {
        "cid": 976,
        "molecular formula": "C5H6O3",
        "Canonical SMILES": "OC(=O)C(=O)CC=C",
        "gap": 4.71301189066,
        "total dipole moment": 2.3151423409445044
    },
    {
        "cid": 997,
        "molecular formula": "C9H8O3",
        "Canonical SMILES": "O=C(C(=O)O)Cc1ccccc1",
        "gap": 4.38103299305,
        "total dipole moment": 2.3833796625863033
    }
]
\end{myverb}

\subsubsection{Shutdown the Database}
To initiate the shutdown process for the database, please execute the following command: \\

\begin{myverb}
$ cd b3lyp_pm6_CHON300noSalt-postgrest-docker-compose
$ docker-compose down
\end{myverb}

If the database shutdown process is completed successfully, a message will be displayed as follows:\\

\begin{myverb}
Stopping b3lyp_pm6_chon300nosalt-postgrest-docker-compose_server_1 ... done
Stopping b3lyp_pm6_chon300nosalt-postgrest-docker-compose_db_1     ... done
Removing b3lyp_pm6_chon300nosalt-postgrest-docker-compose_server_1 ... done
Removing b3lyp_pm6_chon300nosalt-postgrest-docker-compose_db_1     ... done
Removing network b3lyp_pm6_chon300nosalt-postgrest-docker-compose_default
\end{myverb}



 
\section{Future work}
\label{sec:futurework}


Our PubChemQC \dbname{} dataset offers numerous research directions to explore. In this section, we outline several possible approaches for future work. (i) Prediction or recommendation systems for novel molecules: Gebauer \etal{}~\cite{Gebauer22:_invers} recently proposed a method for designing molecular structures using QM9 quantum chemical data. Our dataset, which includes a more diverse range of molecules than QM9, could significantly improve the predictability of such methods. Additionally, the learning process in these methods often involves the total energy of molecules, which can be readily obtained from the PubCheQC \dbname{} dataset. (ii) Creation of a more accurate dataset considering conformers and employing advanced levels of theory: although we used Open Babel to generate initial molecular geometries, we did not account for conformers, resulting in molecules that may not correspond to the lowest energy conformer. Acquiring the lowest energy conformers of molecules is a challenging task. The CREST~\cite{C9CP06869D} program provides a heuristic approach to partially address this issue. We are currently calculating molecules in PubChem using CREST to obtain more accurate molecular conformers, which we will use as an initial guess. We plan to calculate the optimized geometries and their properties using $\omega$B97X-D\cite{B810189B} or $\omega$B97M-V\cite{b97m-v} with improved basis sets. These advancements could lead to more accurate and reliable molecular design and property prediction computational models. (iii) In-depth analysis: our dataset enables researchers to investigate more detailed properties based on the results using optimized molecular geometries from PubChemQC \dbname{}. This can lead to a more in-depth understanding of structure-property relationships. For instance, we possess HOMO and LUMO data for all molecules, allowing us to predict their reactivity. By utilizing the HOMO, LUMO energies, and HOMO-LUMO energy gaps, it is possible to develop high-throughput screening methods for advancements in OLED technology. (iv) Developing enhanced databases with Docker integration: Our current databases integrated with Docker do not encompass molecular orbital and basis set information. Developing databases that incorporate such data is an enormous challenge, as the compressed size of the data reaches 50 TB. One potential approach is to develop a smaller subset by randomly selecting molecules and limiting the variety of elements considered. (v) Development of a web interface: this would allow for easier access to the dataset and visualization and analysis tools for researchers in the field.

\begin{acknowledgement}
The HOKUSAI facility was utilized for a portion of the calculations. Generating such large data sets would have been difficult without RIKEN's rich computer resources. The Japan Society supported this research for the Promotion of Science (JSPS KAKENHI Grant no. 18H03206). We thank Dr. Masatomo Hashimoto and Dr. Tomomi Shimazaki for engaging in insightful discussions with us.
\end{acknowledgement}

\bibliography{main}

\end{document}